\documentclass[a4paper,11pt]{article}
\pdfoutput=1

\usepackage{jinstpub} 
\usepackage{siunitx}
\usepackage{lineno}
\usepackage[perpage]{footmisc}
\usepackage[utf8]{inputenc}

\title{ICARUS at the Fermilab Short-Baseline Neutrino Program - Initial Operation}

\author[a]{P.~Abratenko}
\author[b]{A.~Aduszkiewicz}
\author[c]{F.~Akbar}
\author[d]{M.~Artero~Pons}
\author[e]{J.~Asaadi}
\author[f,1]{M.~Aslin\note{Now at University of Wisconsin, Madison, USA}}
\author[g,2]{M.~Babicz\note{Also at INP-Polish Acad. Sci, Krakow,Poland. Now at University of Zurich, Switzerland}}
\author[f]{W.F.~Badgett } 
\author[f]{L.F.~Bagby}
\author[d]{B.~Baibussinov}
% ...

\affiliation[a]{Tufts University, Medford, MA 02155, USA}
\affiliation[b]{University of Houston, Houston, TX 77204, USA}
\affiliation[c]{University of Rochester, Rochester, NY 14627, USA}
\affiliation[d]{INFN Sezione di Padova and University of Padova, Padova, Italy}
\affiliation[e]{University of Texas at Arlington, Arlington, TX 76019, USA}
\affiliation[f]{Fermi National Accelerator Laboratory, Batavia, IL 60510, USA}
\affiliation[g]{CERN, European Organization for Nuclear Research 1211 Gen\`eve 23, Switzerland, CERN}
% ...

\emailAdd{alessandro.menegolli@unipv.it}

\abstract{The ICARUS collaboration employed the 760-ton T600 detector in a successful three-year physics run at the underground LNGS laboratory studying neutrino oscillations with the CERN Neutrino to Gran Sasso beam (CNGS) and searching for atmospheric neutrino interactions. ICARUS performed a sensitive search for LSND-like anomalous $\nu_e$ appearance in the CNGS beam, which contributed to the constraints on the allowed parameters to a narrow region around 1 eV$^2$, where all the experimental results can be coherently accommodated at 90\% C.L.. After a significant overhaul at CERN, the T600 detector has been installed at Fermilab. In 2020, cryogenic commissioning began with detector cool down, liquid argon filling and recirculation. ICARUS has started operations and successfully completed its commissioning phase, collecting the first neutrino events from the Booster Neutrino Beam (BNB) and the Neutrinos at the Main Injector (NuMI) beam off-axis, which were used to test the ICARUS event selection, reconstruction and analysis algorithms. The first goal of the ICARUS data taking will then be a study to either confirm or refute the claim by Neutrino-4 short baseline reactor experiment both in the $\nu_\mu$ channel with the BNB and in the $\nu_e$ with NuMI. ICARUS will also address other fundamental studies such as neutrino cross sections with the NuMI beam and a number of Beyond Standard Model searches. After the first year of operations, ICARUS will commence its search for evidence of a sterile neutrino jointly with the Short Baseline Near Detector, within the Short-Baseline Neutrino program.}


\begin{document}

\maketitle

\twocolumn

\section{Introduction}\label{sec:introduction}

The Liquid Argon Time Projection Chamber (LAr-TPC) is a continuously sensitive and self triggering detector that can provide excellent 3D imaging and calorimetric reconstruction of any ionizing event. First proposed by C. Rubbia in 1977 \cite{crubbia}, this detection technique allows a detailed study of neutrino interactions, spanning a wide energy spectrum (from a few keV to several hundreds of GeV), as demonstrated by the first large scale experiment performed by the ICARUS Collaboration at the LNGS underground laboratory.

Several experiments, in particular the Liquid Scintillator Neutrino Detector (LSND)~\cite{lsnd} and MiniBooNE~\cite{miniboone}, have reported anomalous signals that may imply the presence of additional (mass-squared difference $\Delta m^2_{new}$ $\sim$ \SI{1}{eV^2}) flavor oscillations at small distances pointing toward the possible existence of nonstandard heavier sterile neutrino(s). A sensitive search for a possible $\nu_e$ excess related to the LSND anomaly in the CNGS $\nu_\mu$ beam has already been performed using the neutrino events collected in the ICARUS-T600 detector during the Gran Sasso run. A total of 2,650 CNGS neutrino interactions, identified in 7.9$\cdot$10$^{19}$ POT (Protons On Target) exposure, have been studied to identify the $\nu_e$ interactions. Globally, 7 electron-like events have been observed to be compared to 8.5$\pm$1.1 expected from the intrinsic beam contamination and standard 3-flavor oscillations. This study constrained the LSND signal to a narrow parameter region at sin$^2 2\theta \sim 0.005$, $\Delta m^2$ $<$ \SI{1}{eV^2}, which requires further investigation \cite{lngs_nue}. 

The primary goal of the Short-Baseline Neutrino (SBN) program at Fermilab is to further investigate the possibility of sterile neutrinos in the {\em O}(1 eV) mass range and provide the required clarification of the LSND anomaly. It is based on three LAr-TPC detectors (ICARUS-T600, with 476 tons active mass, MicroBooNE with 89 tons active mass and SBND with 112 tons active mass) exposed at shallow depth to the $\sim$ 0.8 GeV Booster Neutrino Beam (BNB) at different distances from the target (600 m, 470 m and 110 m respectively) \cite{sbn_proposal,sbn_2019}.

The detection technique used will provide an unambiguous identification of neutrino interactions,  measurement of their energy and a strong mitigation of possible sources of background. Performing this study with almost identical detectors at various distances from the neutrino source allows identification of any variation of the spectra, which is a clear signature of neutrino oscillations.

In particular, SBN will allow for a very sensitive search for $\nu_\mu \rightarrow \nu_e$ appearance signals, covering the LSND 99\% C.L. allowed region at $\sim 5\sigma$ C.L.~\cite{sbn_proposal,sbn_2019}. 
The high correlations between the event samples of the three LAr-TPC’s and the huge event statistics at the near detector will also allow for a simultaneous sensitive search in the $\nu_\mu$ disappearance channel.

During data taking at Fermilab, the 760-ton T600 detector is also exposed to the off-axis neutrinos from the Neutrinos at the Main Injector (NuMI) beam, where most of events are in the 0 -- 3 GeV energy range, with an enriched component of electron neutrinos (few \%). The analysis of these events will provide useful information related to detection efficiencies and neutrino cross-sections at energies relevant to the future long baseline experiment with the multi-kiloton DUNE LAr-TPC detector. 

In addition to the LSND anomaly, ICARUS will test the oscillation signal reported by the Neutrino-4 experiment \cite{neutrino4} both in the $\nu_\mu$ and $\nu_e$ channels with the BNB and NuMI beams, respectively.

This paper is organized as follows: in Section~\ref{sec:T600} the ICARUS-T600 detector is described with a particular emphasis on its achievements during three years data taking at the INFN LNGS underground laboratories in Italy; in Section~\ref{sec:overhauling}, the ICARUS-T600 overhauling activities, most of which were carried out at CERN in the Neutrino Platform framework~\cite{NP01}, are shown; the new Cosmic Ray Tagger (CRT) detector, used to mitigate the cosmic ray background due to operating ICARUS at shallow depth, is detailed in Section~\ref{sec:CRT}. In Section~\ref{sec:installation}, the first operations of ICARUS at Fermilab, in particular the installation of the cryogenic plant, TPC electronics, scintillation light detection system and CRT are described. A successful commissioning phase followed soon after as described in Section~\ref{sec:commissioning}. Finally, the procedure for the selection, reconstruction, and analysis of the first collected BNB and NuMI off-axis neutrino events is introduced in Section~\ref{sec:reco}. 

\section{The ICARUS-T600 detector}\label{sec:T600}

The ICARUS-T600, with a total active mass of 476 ton, is the first large-scale operating LAr-TPC detector~\cite{t600_pavia}: it consists of two large and identical adjacent modules with internal dimensions 3.6 $\times$ 3.9 $\times$ 19.6~m$^3$, filled with a total of 760 tons of ultra-pure liquid argon. Each module houses two LAr-TPCs separated by a common cathode with a maximum drift distance of 1.5~m, equivalent to $\sim$~1~ms drift time for the nominal 500~V/cm electric drift field. The cathode is built up by an array of nine
panels made of punched stainless-steel, allowing for a 58\% optical
transparency between the two drift regions. The anode is made of three parallel wire planes positioned 3~mm apart, where the stainless-steel \SI{100}{\micro m} wires are oriented on each plane at a different angle with respect to the horizontal direction: 0$^{\circ}$ (Induction 1), +60$^{\circ}$ (Induction 2) and -60$^{\circ}$ (Collection). In total, 53,248 wires with a 3~mm pitch and length up to 9 m are installed in the detector. By appropriate voltage biasing, the first two planes (Induction 1 and Induction 2) provide a nondestructive charge measurement, whereas the ionization charge is fully collected by the last Collection plane.
Photo-Multiplier Tubes (PMTs) are located behind the wire planes to collect the scintillation light produced by charged particles in LAr and used for the trigger of the detector.

In 2013, ICARUS concluded a very successful 3-year long run in the Gran Sasso underground laboratory~\cite{ica_lngs}, demonstrating the feasibility of the LAr-TPC technology at the kiloton scale in a deep underground environment and paving the way to the construction of the next generation of experiments dedicated to study neutrino oscillation physics such as DUNE.
During the data taking, the liquid argon was kept at an exceptionally high purity level ($<$ 50 ppt of O$_2$ equivalent contaminants) reaching in 2013 a 16 ms lifetime corresponding to 20 ppt O$_2$ equivalent LAr contamination \cite{lngs_purity}, demonstrating the possibility to build larger LAr-TPC detectors with drift distances up to 5 m. 

The detector has been exposed to the CNGS neutrino beam and to cosmic rays, recording events that demonstrate high-level performance and the physical potential of this detection technique: the detector showed a remarkable $e/\gamma$ separation and particle identification exploiting the measurement of $dE/dx$ versus range \cite{lngs_reco}. The momentum of escaping muons has been measured by studying the multiple Coulomb scattering with $\sim$ 15\% average resolution in the 0.4 -- 4~GeV/c energy range, which is relevant for the next generation neutrino experiments \cite{lngs_mcs}. 

Events related to cosmic rays have been studied to identify atmospheric neutrino interactions: 6 $\nu_\mu$CC and 8 $\nu_e$CC events in a 0.43~kton$\cdot$y exposure have been identified and reconstructed, demonstrating that the automatic search for the $\nu_e$CC in the sub-GeV range of interest for the future short and long baseline neutrino experiments is feasible \cite{lngs_atm}. 

\section{The overhaul of ICARUS-T600}\label{sec:overhauling}

The ICARUS-T600 detector at Fermilab takes data at shallow depth, shielded by a $\sim$ 3-meter concrete overburden: neutrino interactions must be recognized among the $\sim$~11~cosmic muons that are expected to cross the detector randomly in the 1~ms drift time during each triggered event. High-energy photons produced by cosmic rays can become a serious background source for the $\nu_e$ search since the electrons produced via Compton scattering and pair production can mimic $\nu_e$CC events. 

\begin{figure}[ht]
    \centering
    \includegraphics[width=0.45\textwidth]{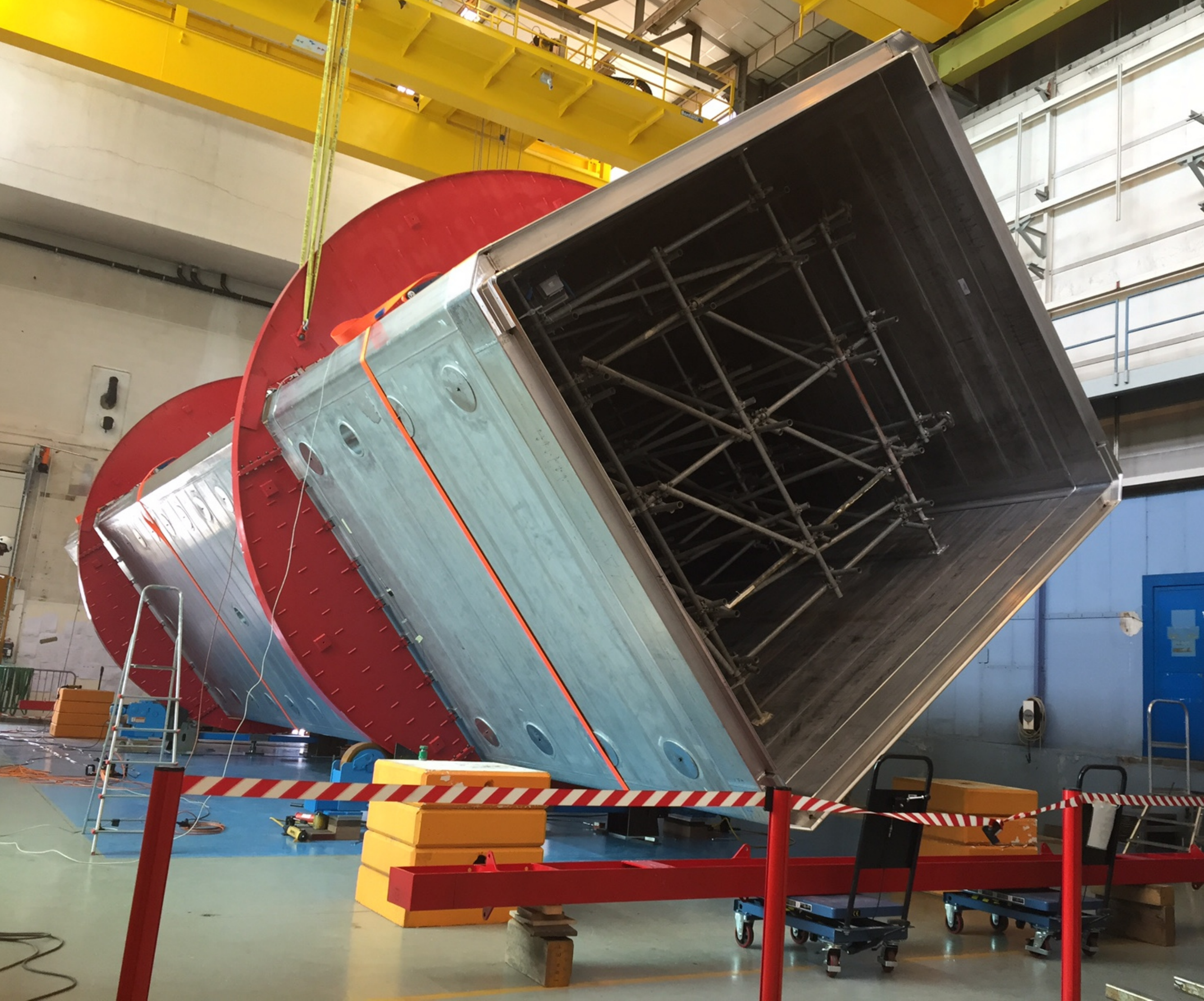}
    \caption{One of the two new ICARUS cryostats during its assembly at a CERN workshop.}
    \label{fig:newcryo}
\end{figure}

In order to prepare the detector for SBN data taking, the T600 underwent an intensive overhaul at CERN in the Neutrino Platform framework (WA104/NP01 project) before being shipped to the USA in 2017, introducing several technology developments while maintaining the achieved performance at Gran Sasso. The refurbishing mainly consisted of: the realization of new cold vessels (Fig.~\ref{fig:newcryo}) with purely passive insulation; an update of the cryogenics and of the LAr purification equipment; flattening of the TPC cathode (the punched hole stainless-steel panels underwent a thermal treatment improving the planarity to a few mm); the implementation of new, higher performance TPC read-out electronics; the upgrade of the LAr light detection system.

\subsection{The TPC electronics}\label{subsec:TPCele}

The electronics used at LNGS was based on flange modularity, each flange serving 576 TPC wire-channels. The analogue front-end was a Radeka type amplifier, using a custom BiCMOS chip to integrate the cascode stage with two different filtering, one for Collection and Induction~1, another for Induction~2 with the aim to produce in all the cases a unipolar signal. 
This solution, however, showed strong limitations in the Induction 2 signals in the case of dense showers. Analog signals were converted to digital via multiplexers by 10-bit ADCs with sampling rate of 400 ns. The analogue circuits were housed in a custom crate, connected to the flange by flat cables, with 18 boards (32 channels per board). 
Analogue boards had a digital link to corresponding digital modules hosted in VME crates that contained memory buffers and performed lossless data compression and data transmission through a VME bus. Both crates were housed in a rack next to the flange. 

One of the largest tasks of the overhauling program was the design of new electronics for the 53,248 wire-channels that would be compatible with higher data rates foreseen at shallow depth operation at FNAL. The new electronics adopts the same modularity and architecture but takes advantage of newer technology that allows for integrating both the analogue and the digital electronics on the same board on a custom crate mounted onto the flange~\cite{ica_electronics}. 

\begin{figure}[ht]
    \centering
    \includegraphics[width=0.23\textwidth]{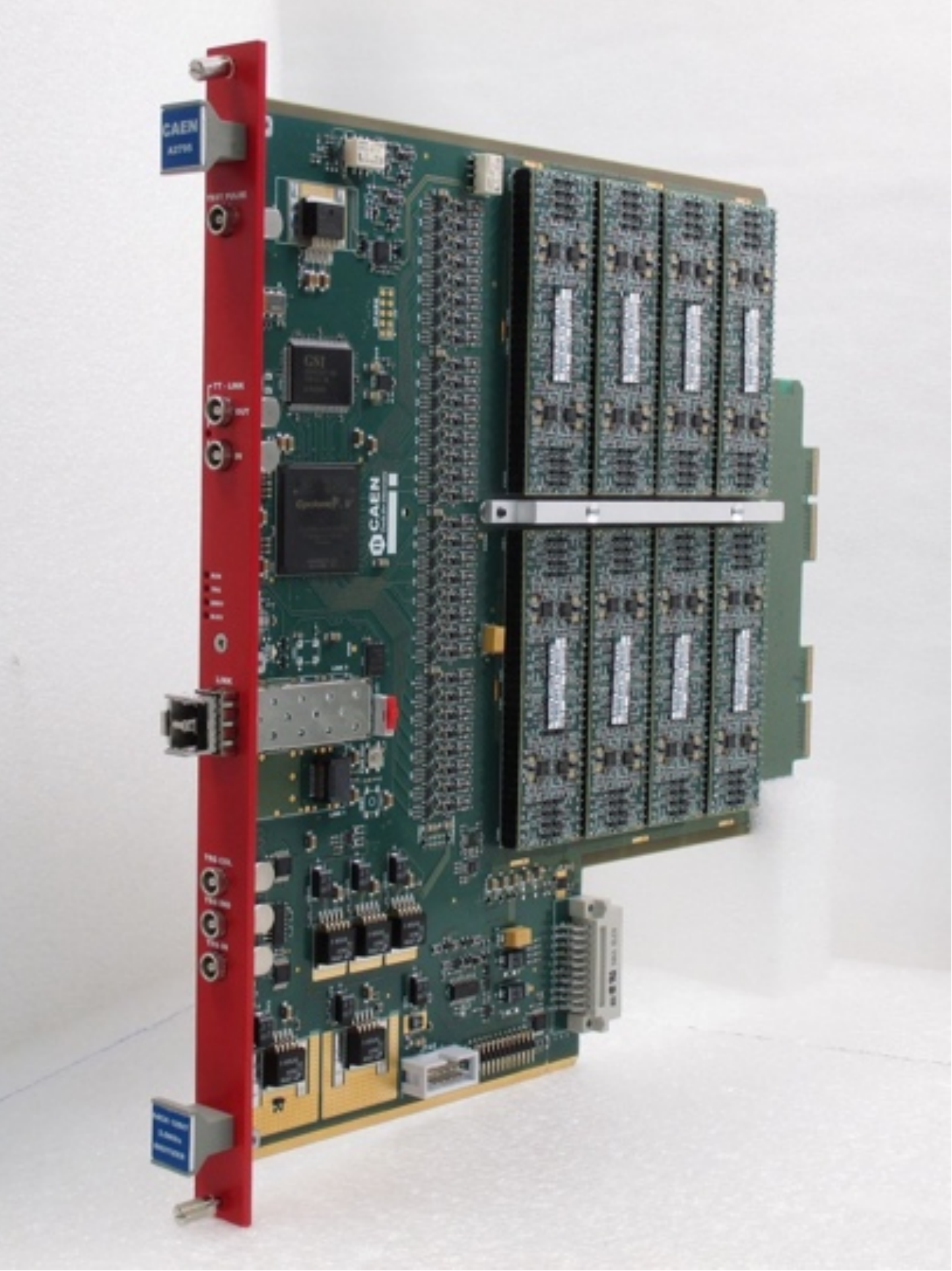}
    \includegraphics[width=0.23\textwidth]{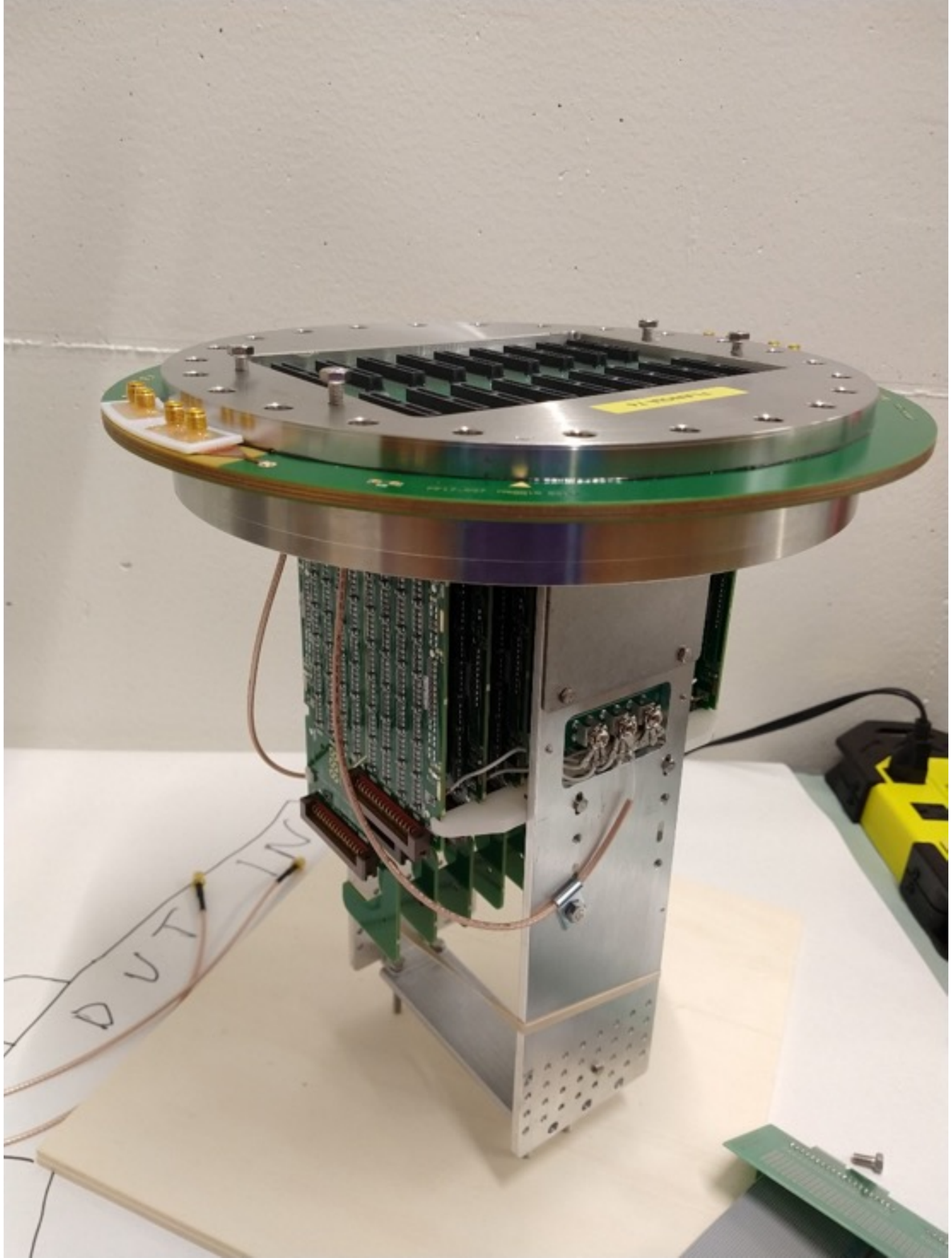}
    \includegraphics[width=0.47\textwidth]{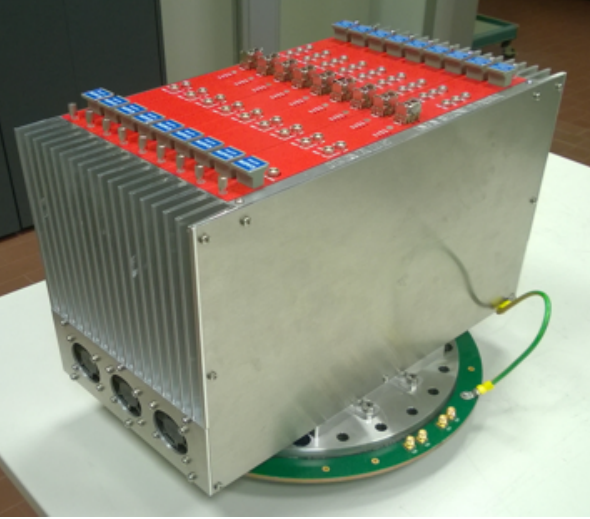}
    \caption{A2795 custom board housing 64 amplifiers (far end), AD converter, digital control, and optical link (top-left). An assembled feed-through with nine DBBs and the biasing cables (top-right). A mini-crate populated by the nine A2795 boards installed on a feed-through flange (bottom).} 
    \label{fig:tpc_board}
\end{figure}

New packaging for the BiCMOS custom cascode allowed the design of a small piggyback module with 8 amplifiers and to house 8 of these modules on a single board serving 64 channels, see Fig.~\ref{fig:tpc_board} (top-left). The digital part is also
completely contained in the same board. Moreover, all the amplifiers now have the same filtering, preserving the bipolar structure of Induction 2 signals without distortion. Each amplifier is followed by a serial 12-bit ADC avoiding the cumbersome signal multiplexing. The digital part is based essentially on a large powerful FPGA allowing the possibility to use  different signal treatments if required from running experience.
The VME standard was abandoned in favor of a serial optical link, allowing for gigabit bandwidth data transmission compatible with shallow depth data rates. 

TPC wire signals are fed into the front-end amplifiers by means of Decoupling Biasing Boards (DBBs). The DBB has two functions: biasing of each wire and conveying, with blocking capacitors, the signals to the amplifiers. The DBBs work in argon gas and can withstand up to 400~V input biasing. The flange CF250 is realized with a G10 multi-layer solid PCB, about 6 mm thick with three internal layers of copper to guarantee the required stiffness. SMD external connectors provide receptacles for the A2795 boards, while another set of SMD connectors in correspondence (inner side) provide receptacles for DBBs, see Fig.~\ref{fig:tpc_board} (top-right). Finally, nine electronic A2795 boards are hosted by a mini-crate which is installed on a feed-through CF250 flange, see Fig.~\ref{fig:tpc_board} (bottom). 

\subsection{The scintillation light detection system}\label{subsec:newPMT}

A new light detection system that is sensitive to the photons produced by the LAr scintillation is a fundamental feature for the T600 operation at shallow depth (contributing to the rejection of the cosmic background). The light detection system complements the 3D track reconstruction, unambiguously providing the absolute timing for each track and identifying the interactions occurring in the BNB and NuMI spill gates.

\begin{figure}[ht]
    \centering
    \includegraphics[width=0.45\textwidth]{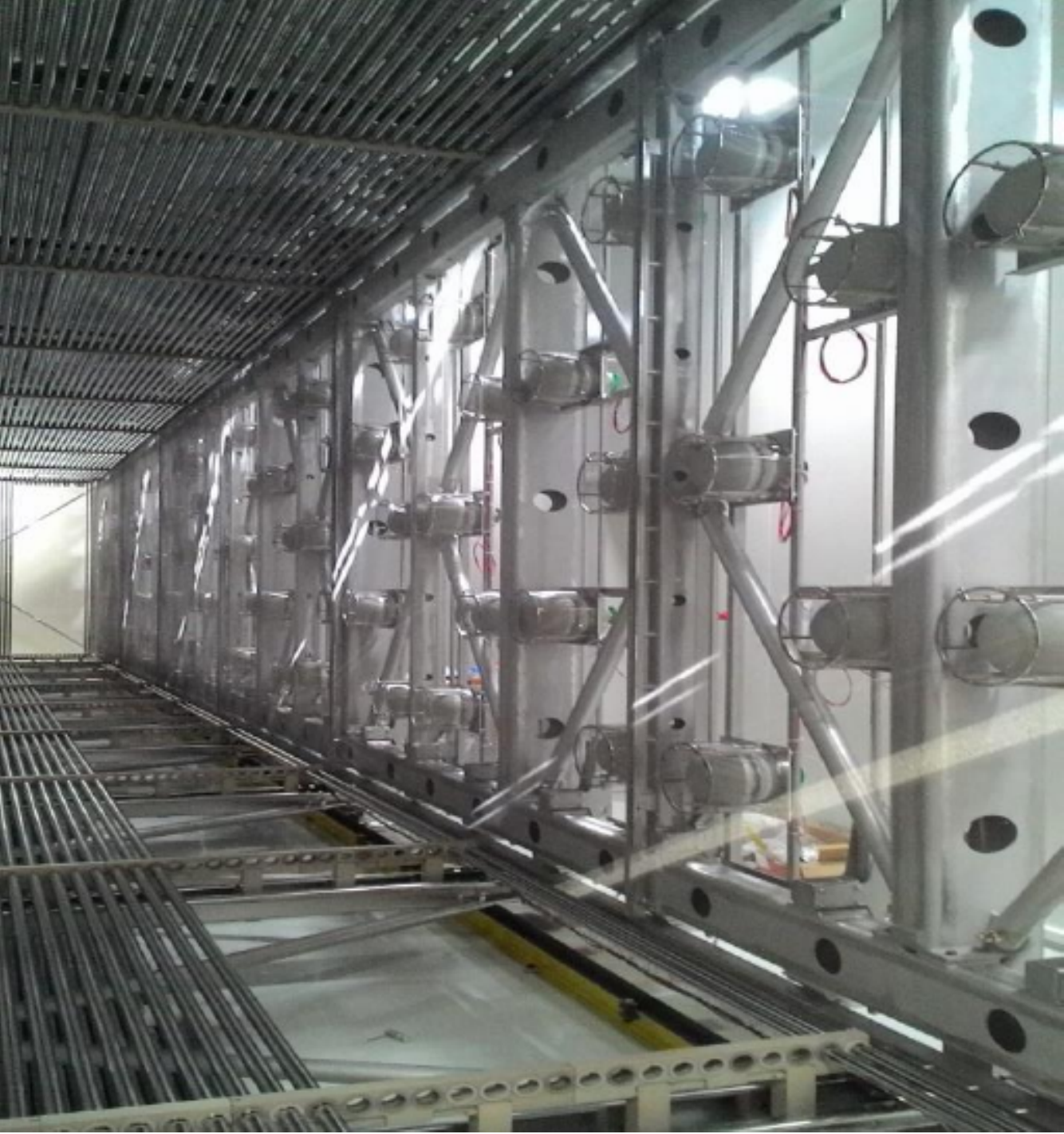}
    \caption{The new ICARUS PMTs mounted behind the wires of one TPC.}
    \label{ica_pmt}
\end{figure}

The ICARUS-T600 light detection system  consists of 360 8" Hamamatsu R5912-MOD PMTs deployed behind the 4 wire chambers, 90 PMTs per TPC~\cite{Babicz:2018svg,Ali_Mohammadzadeh_2020}, see Fig.~\ref{ica_pmt}. Since the PMT glass is not transparent to the $128$~nm wavelength scintillation light produced in liquid argon, each unit is provided with a $\approx$ 
\SI{200}{\micro g}/cm$^2$ coating of Tetra-Phenyl Butadiene (TPB), to convert the VUV photons to visible light~\cite{Bonesini:2018ubd}. 

All PMTs are mounted onto the wire chamber mechanical frames using a
supporting system, that allows the PMT to be positioned about 5~mm behind the Collection planes wires. A stainless steel grid cage is mounted around each PMT to mitigate the induction of fake signals on the nearby wire planes by the relatively large PMT signals.

The light detection setup, realized by INFN, is complemented by a laser calibration system
allowing for gain equalization, timing and monitoring of all the PMTs.
Laser pulses ($\lambda$ = 405~nm, FWHM = 60~ps), generated
by a laser diode head (Hamamatsu PLP10), are sent to each
PMT window by means of a light distribution system based on optical fibers, light splitters and an optical switch~\cite{laser}.

\section{The Cosmic Ray Tagger}\label{sec:CRT}

ICARUS-T600 based at FNAL faces more challenging experimental conditions than at LNGS: due to its shallow depth operation, identification of neutrino interactions among 11 kHz of cosmic rays is required. A $\sim$~3-meter concrete overburden was designed to almost completely remove the contribution from charged hadrons and high energy photons~\cite{ICARUSOverburden}. However, $\sim$~11 muon tracks occur per triggered event in the 1~ms TPC drift readout; 
photons associated with the muons represent a serious background for identifying $\nu_e$ candidates since electrons produced via Compton scattering/pair production can mimic a genuine $\nu_e$CC event. 

Rejecting the cosmic background, i.e. reconstructing the triggering event, requires to know precisely the timing of each track in the TPC image. Operating at FNAL, ICARUS exploits an improved light detection system with high granularity and $O$(1 ns) time resolution, and an external $\sim$ 4$\pi$ high coverage Cosmic Ray Tagger (CRT). 
The primary function of the CRT is to tag muons passing through or near the cryostats.

Timestamps associated to a particle tagged by the CRT are compared with timestamps from PMT signals, both with a few nanosecond resolution, allow the determination of whether an interaction in the TPC originated from an outside cosmic ray or from an internal interaction. 
The ICARUS CRT consists of a top, side and bottom subsystem. 
 
\begin{figure}[ht]
    \centering
    \includegraphics[width=0.45\textwidth]{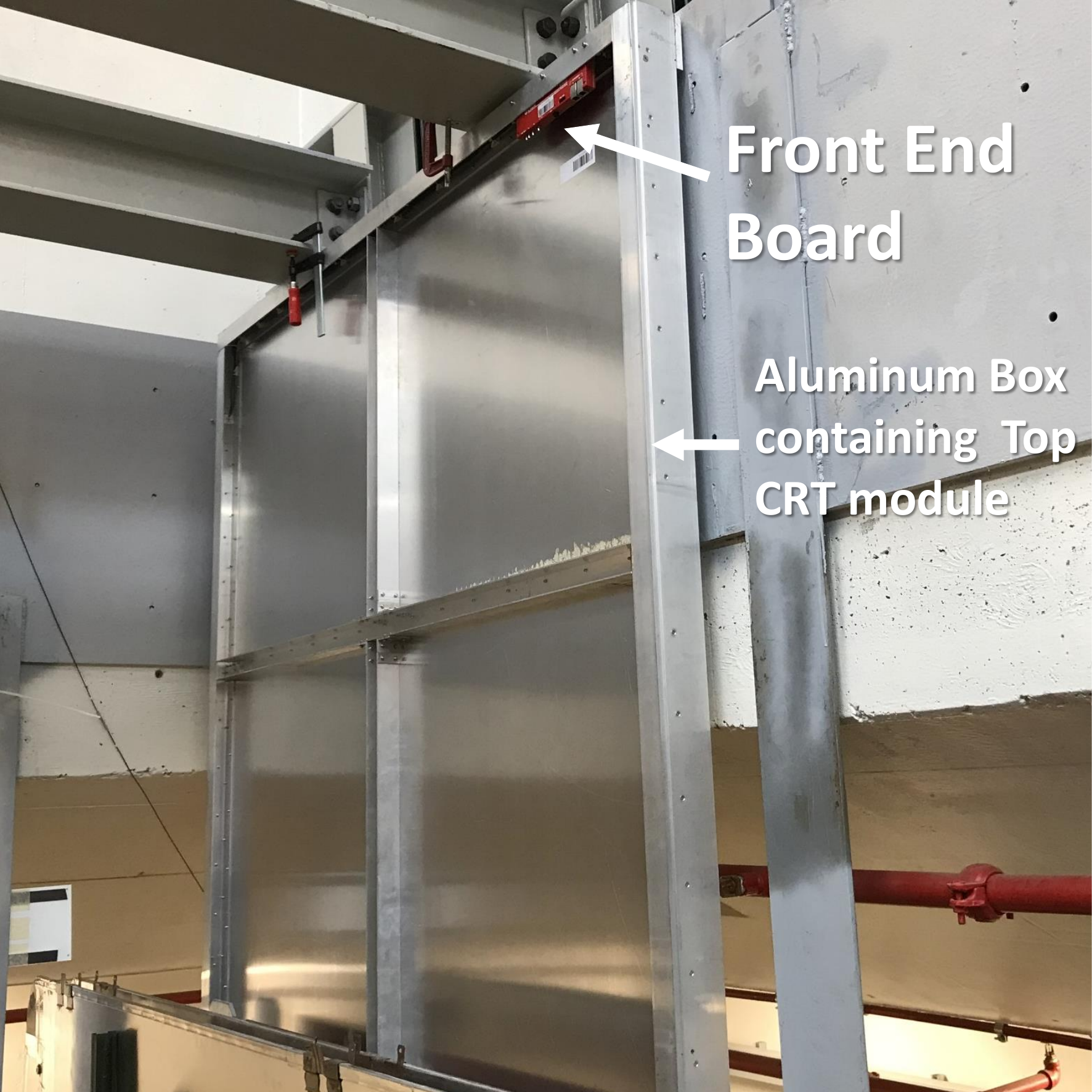}
    \caption{Picture of a vertical TOP CRT module installed in the detector hall.}
    \label{fig:TopCRTpic}
\end{figure}

The ICARUS Top CRT system is divided in 123 detector modules covering a surface of about 426~m$^2$: 84 horizontal and 39 vertical modules along the perimeter of the cryostat top surface. Its design is such that more than 80$\%$ of the cosmic muon flux is intercepted by the Top CRT. Each module is a 1.86 $\times$ 1.86~m$^2$ aluminum box containing two orthogonal layers of eight scintillator bars for position reconstruction. The bars, coated with white paint, are 23~cm wide, 184~cm long and have different thickness depending on the layer: 1~cm and 1.5~cm for the top layer and the bottom layer, respectively.
In each scintillator, the light is collected by two wave-length shifting (WLS) fibers Kuraray Y-11(200) then read out from one end by a Silicon Photo-Multiplier (SiPM), Hamamatsu S13360-1350C model. The 32~SiPM signals of one module are routed via 50 $\Omega$ micro-coaxial cables to a patch panel connected to the CAEN DT5702 Front End Board (FEB) which provides a bias voltage adjustable for each channel. The FEB triggers on the coincidence between two SiPM signals of the same bar and provides a coincidence logic between the two scintillator layers in the module. In Fig.~\ref{fig:TopCRTpic}, a picture of a vertical Top CRT module installed in the detector hall is shown. The Top CRT was a brand new detector designed and built by INFN and CERN before shipping to Fermilab in summer 2021.

The ICARUS Side CRT makes use of scintillator modules formerly used by the MINOS experiment. Each module is composed of twenty adjacent strips of 800~$\times$~4~$\times$~1~cm$^3$ Polystyrene (1.0$\%$ PPO, 0.03$\%$ POPOP) scintillator. The full Side CRT system consists of 2,710 readout channels across 93 FEBs, with 136 full and 81 cut modules in total. The scintillator is contained in a metal sheath and each strip has an embedded WLS fiber running down the middle. These fibers are collected into “snouts” at the ends of the modules, onto which the optical readout, consisting of an array of ten Hamamatsu S14160-3050HS SiPMs, is mounted onto a snout. Each SiPM reads out two fibers and corresponds to a single electronic readout channel on CAEN A1702 Front-End Boards (FEBs). A full MINOS module has two snouts, one on each end. The ICARUS Side CRT System is double layered, with an inner and outer layer of MINOS modules to apply coincidence logic between the two layers. To account for geometric constraints, some MINOS modules were cut and sealed on the cut end with mylar and tape to only have a single snout for readout. The South Side CRT wall consists of an inner and outer layer of cut modules oriented orthogonally in an X-Y configuration, with the added benefit of improved position reconstruction on the southern side of the TPCs, upstream along the BNB beam. The East and West walls utilize full length MINOS modules mounted horizontally, while the North Wall use cut modules mounted horizontally. 

The Bottom CRT consists of 14 modules divided into two daisy chains of 7 modules each, positioned underneath the warm vessel in a north and south section. These modules are refurbished veto modules from the Double Chooz reactor neutrino experiment. Each module consists of 64 Polystyrene scintillator strips, running in parallel and divided into two layers of 32 strips offset 2.5~cm from each other. Scintillation light is collected in a WLS optical fiber and read out at one end of each strip by an Hamamatsu H7546B M64 multi-anode PMT, while the other end is mirrored to maximize light collection. 

\section{First operations at FNAL}\label{sec:installation}

\begin{figure*}[ht]
    \centering
    \includegraphics[width=.44\textwidth]{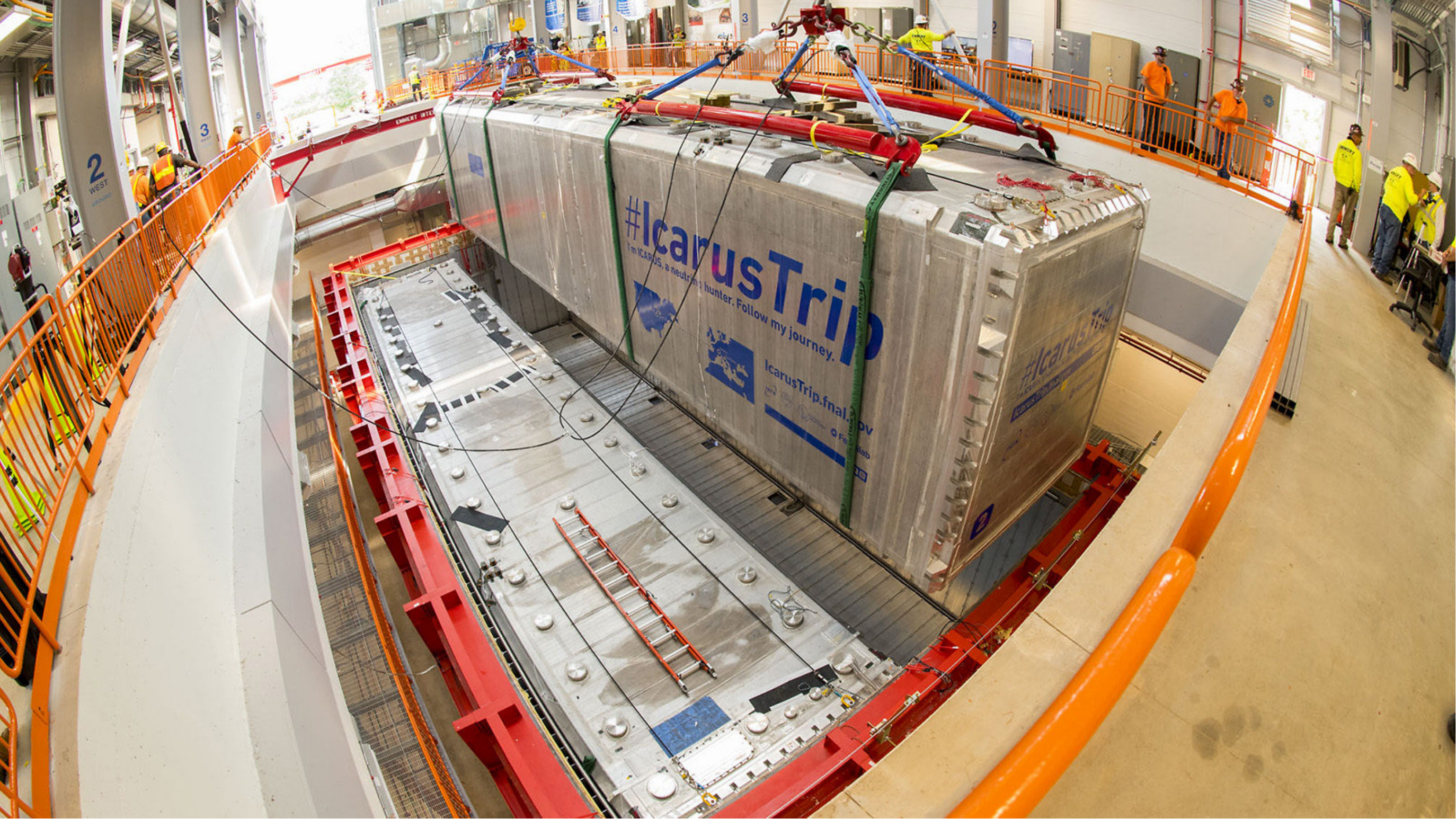}
    \includegraphics[width=.35\textwidth]{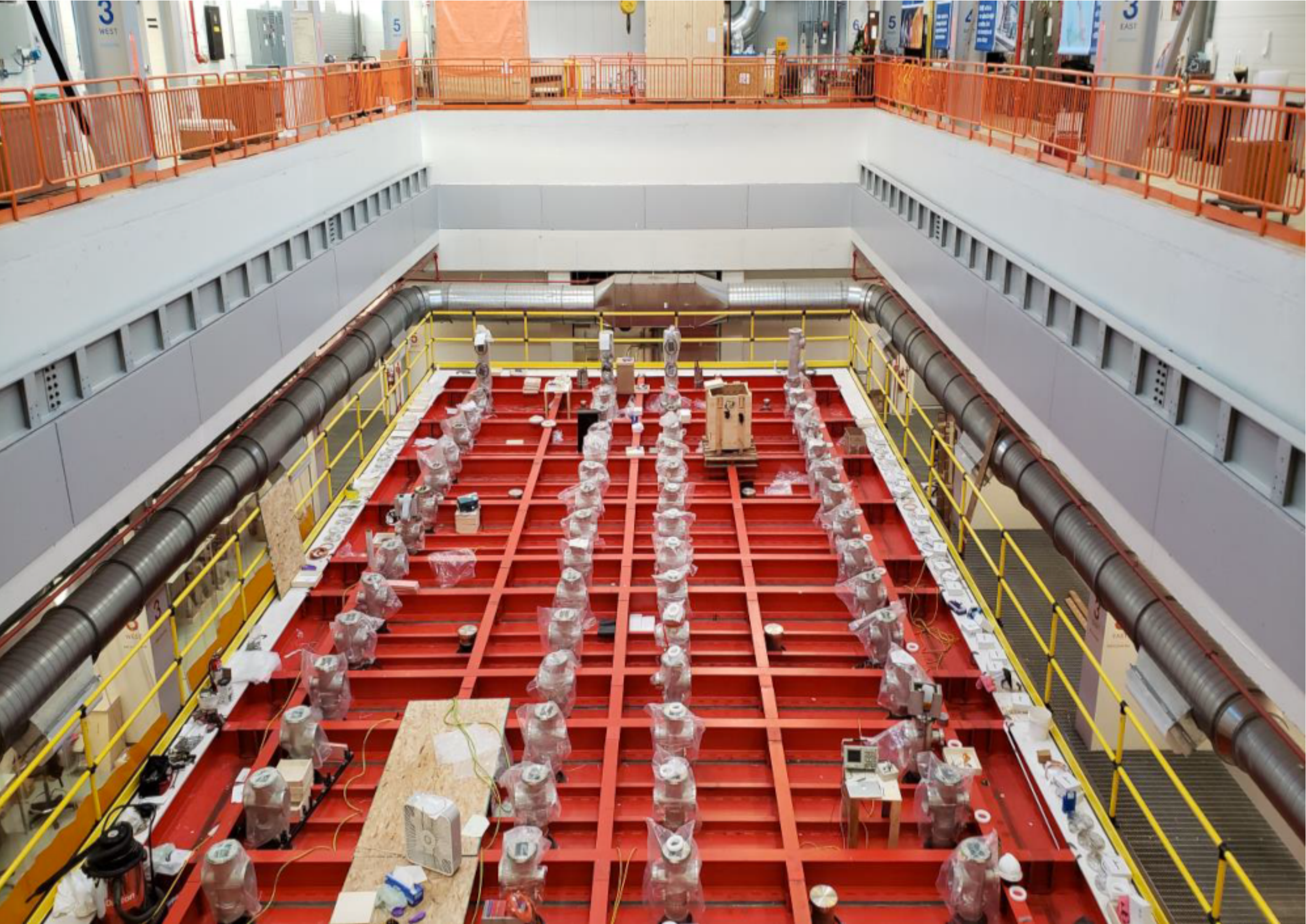}
    \includegraphics[width=.185\textwidth]{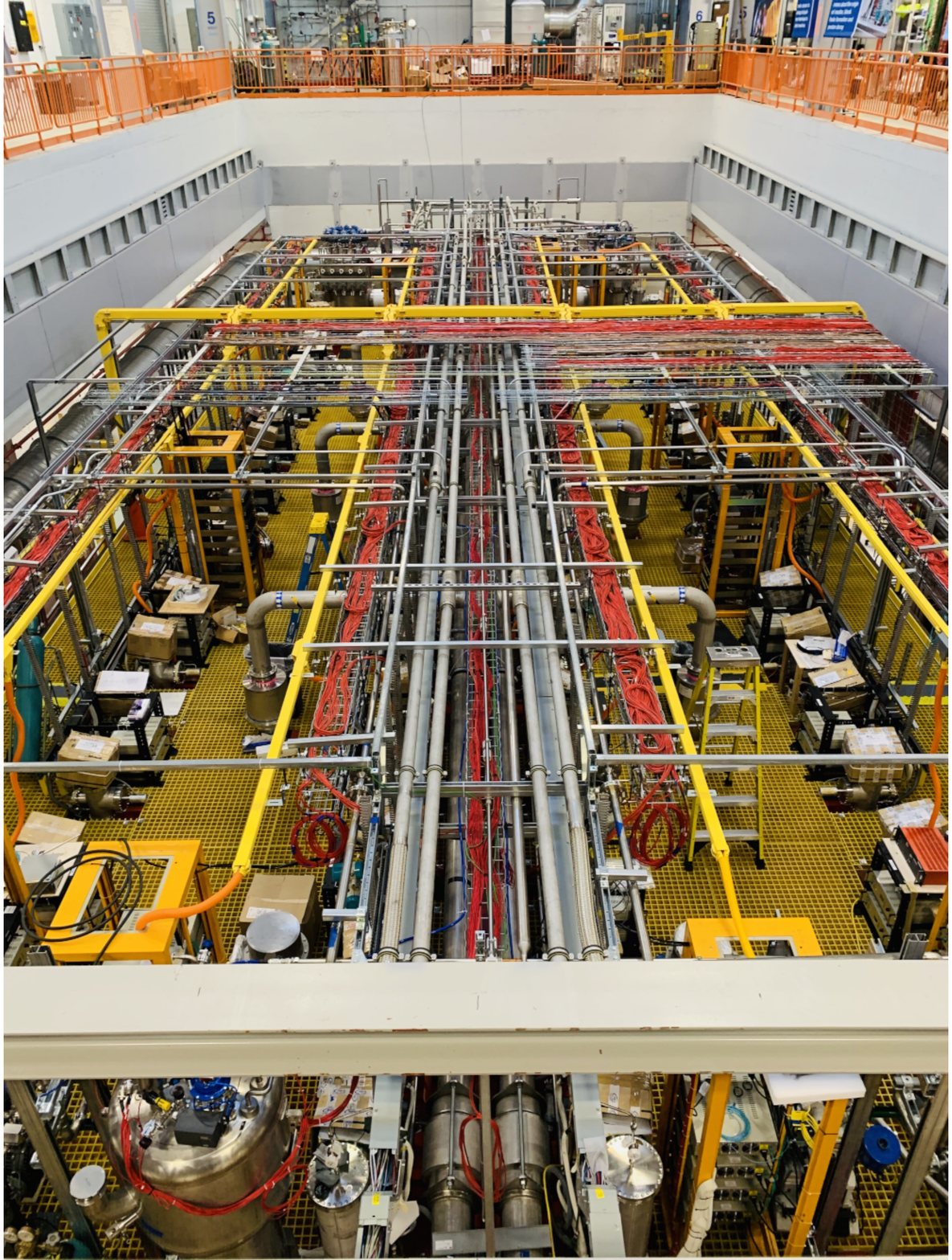}
    \caption{Deployment of the ICARUS cryostats inside the pit of the SBN Far Detector experimental hall at Fermilab in August 2018 (left). Installation of TPC, PMT and laser feed-through flanges in December 2018 (center). Status of the ICARUS detector at the beginning of data taking for commissioning (right).}
    \label{fig:installation}
\end{figure*}

Following the overhauling activities at CERN, ICARUS-T600 was shipped to Fermilab in July 2017 and the two cryostats hosting the TPCs were finally deployed in their shallow depth position in August 2018. Work began soon after to install and test all main subsystems before the cryogenic commissioning, see Fig.~\ref{fig:installation}.

\subsection{Cryogenic plant installation}\label{subsec:cryo_inst}

The ICARUS cryogenic plant was designed, built and installed at Fermilab by a collaboration of three international institutions, CERN, INFN and Fermilab to support operations of the ICARUS LAr-TPC. 
For the installation at Fermilab, the entire ICARUS-T600 cryogenic and purification system was rebuilt anew. The new design followed closely the original implementation at the LNGS with one important exception: at Fermilab, the LN$_2$ boiloff is vented to the atmosphere (open loop cooling circuit), while at LNGS the LN$_2$ boiloff was re-condensed by means of a set of cryocoolers (closed loop cooling circuit). The main components of the cryogenic and purification system are the following:

\begin{itemize}
    \item Main LAr containers (2$\times$ cold vessels): 273~m$^3$ each, containing the TPC detectors and the LAr scintillation light system.
    \item Cold shields: set of heat exchangers filled with LN$_2$, completely surrounding the main LAr containers and designed to prevent heat, coming from the thermal insulation, to reach the LAr volumes.
    \item Thermal insulation: polyurethane foam panels, $\sim$ 600~mm thick, surrounding the cold shields.
    \item Warm vessel: provides enclosure and mechanical support for the thermal insulation.
    \item LN$_2$ cooling circuits: piping, circulation pumps, regulating valves, phase separators, etc., providing LN$_2$ supply to heat exchangers serving the cold shields and the purifying units.
    \item Argon gas recirculation units (4$\times$, two per cold vessel): set of units that re-condense and purify the argon flowing from the gas phase on top of the main LAr containers.
    \item Liquid argon recirculation units (2$\times$, one per cold vessel): provide forced circulation, with a cryogenic pump, of argon coming from the cold vessels through a set of purifiers before injecting it back into the cold vessel.
    \item Cryogenic control system: to provide automation, data display, recording and alarming.
    \item LN$_2$ and LAr storage dewars and relative transfer lines.
    \item A dedicated purification unit used for the filling of the cold vessels, equipped with a regeneration system and a set of gas analyzers.
\end{itemize}  

\begin{figure*}[ht]
    \centering
    \includegraphics[width=\textwidth]{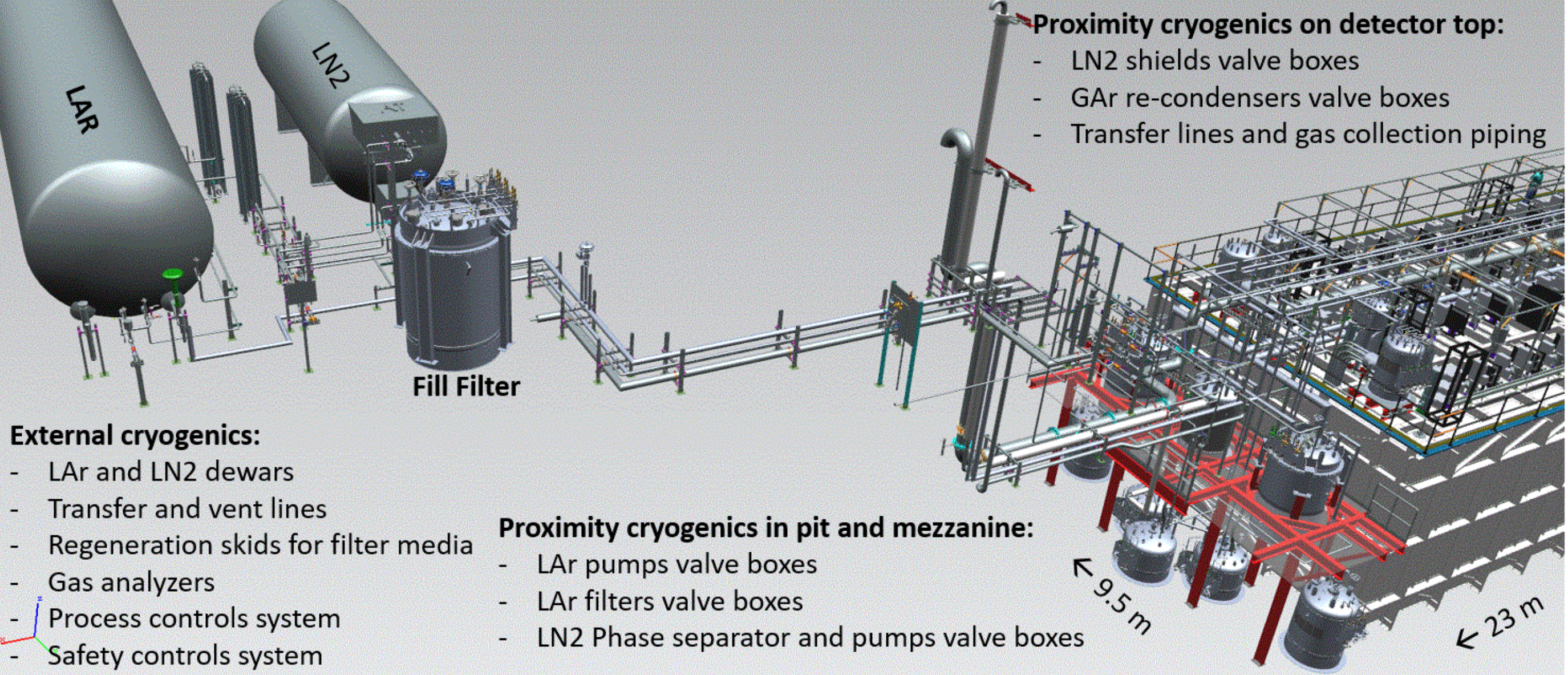}
    \caption{ICARUS cryogenic plant physical layout.}
    \label{fig:cryo_layout}
\end{figure*}

The ICARUS cryogenic plant at the SBN Far Detector Hall at Fermilab was fully designed, delivered, and installed by July 2019, with the commissioning phase started by January 2020.  The equipment included the ICARUS cryogenic plant is schematically divided into the external components supplied by Fermilab, the proximity components supplied by Demaco Holland B.V. under contract with CERN and components internal to the cryostats supplied by INFN. Fig.~\ref{fig:cryo_layout} shows the ICARUS plant physical layout.

\subsection{TPC electronics installation}\label{subsec:TPCele_inst}

\begin{figure}[ht]
    \centering
    \includegraphics[width=0.45\textwidth]{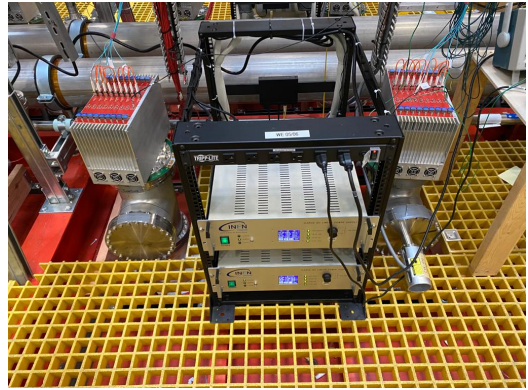}
    \caption{Two Low Voltage Power Supply (LVPS) modules powering the two adjacent mini-crates populated with nine A2795 boards, serving 576 wires each.}
    \label{fig:minicrate}
\end{figure}

Each mini-crate, housing nine A2795 boards, was mounted onto the flange on top of the chimney that contains flat cables connecting wires of the chambers to DBBs and powered by a linear power supply next to the chimney, see Fig.~\ref{fig:minicrate}. Each set of nine A2795 in a single crate are read out through two fibers that implement a CAEN proprietary protocol named CONET (Chain-able Optical NETwork). The two sets of fibers are read through an A3818 PCI Express board installed in dedicated PCs.

The full TPC electronics (96 mini-crates) is synchronized by a serial link (one cable), named TTLink, which sends clock, trigger, and commands. The TTLinks are distributed to all mini-crates by four fan-out modules with the same cable lengths to guarantee equal time delay. The TPC electronics system is fully installed and operational.

\subsection{PMT system installation}\label{subsec:PMT_inst}

Electrical connections between PMTs and electronics, located in a building alcove
adjacent to the detector, were realized by means of 360 signal cables and 360 high voltage cables. Signal cables are RG316/U, 7~m of which are deployed inside the detector and 37~m outside, the two parts connected by means of BNC-BNC feedthrough flanges. High voltage cables are 7-m long HTC-50-1-1 deployed inside the detector and 37~m  RG58/U outside; the two parts connected by means of SHV-SHV feedthrough flanges. Power supply voltages are generated and distributed by 8 CAEN A7030 boards, each with 48 channels that can provide 3~kV, housed in two CAEN SY4527 crates.

The PMT electronics are designed to allow continuous read-out, digitization and independent waveform recording of signals coming from the 360 PMTs. This operation is performed by 24 CAEN V1730B digitizers installed in 8 VME crates (3 digitizers per crate). Each module consists of a 16-channel 14-bit 500-MSa/s FLASH ADC
with 2~Vpp input dynamic range. In each board 15 channels are used for the acquisition of PMT pulses, while one channel is used for the acquisition of ancillary signals such as the beam gates
and the trigger pulses.

For each channel, an internal trigger-request logic signal is generated every time the sampled PMT pulse passes through a programmable threshold. For each couple of adjacent channels, trigger-requests are logically combined (OR, AND, Ch0, Ch1) and the result is presented in a low-voltage differential signaling (LVDS) logic output with settable duration. For triggering purposes, an OR logic between neighboring PMTs is adopted.
A total of 192 LVDS lines (8 lines per digitizer) are connected to the ICARUS trigger system for exploiting the scintillation light information for trigger purposes.

The PMT electronics are complemented by a common 62.5~MHz clock distribution system,
an external trigger network, an external time-stamp reset network, and 24 optical link interfaces based on the CAEN CONET2 protocol.

\subsection{Cosmic Ray Tagger installation}\label{subsec:CRT_inst}

\begin{figure*}[ht]
    \centering
    \includegraphics[width=0.37\textwidth]{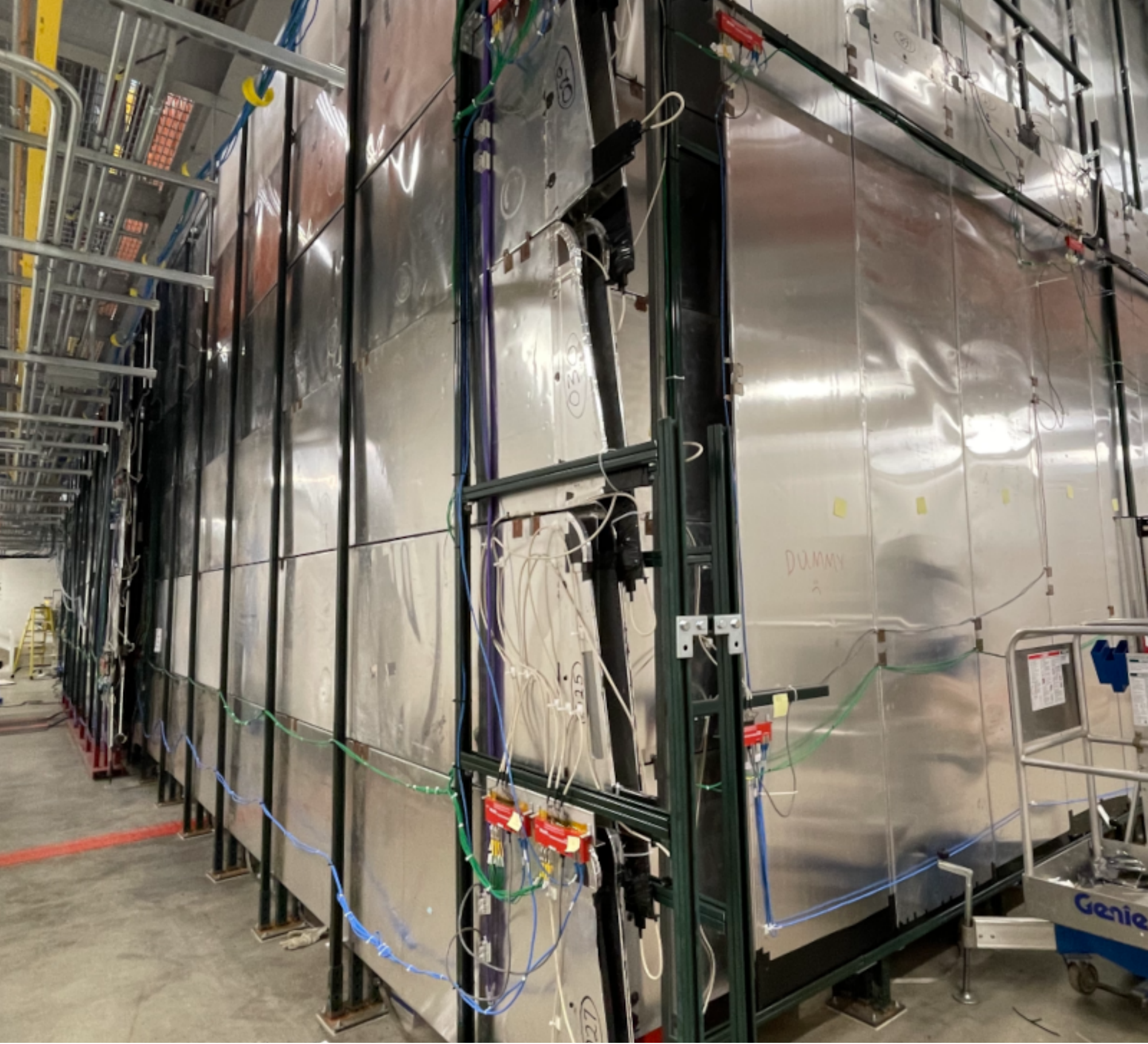}
    \includegraphics[width=0.45\textwidth]{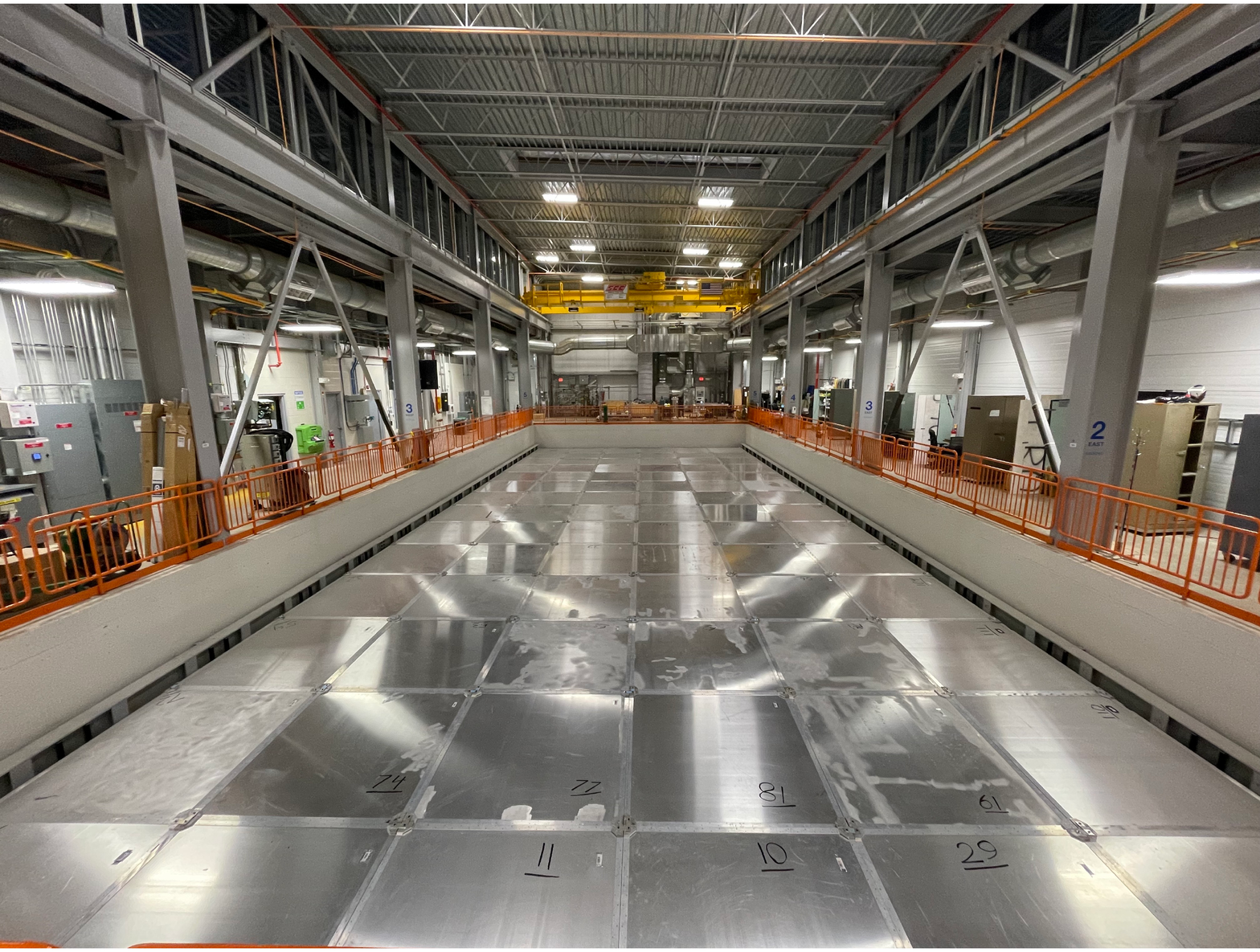}
    \caption{Left: picture of the Side CRT. Right: Top CRT horizontal modules whose installation was completed in December 2021.}
    \label{fig:TopandSideCRTs}
\end{figure*}

The Side CRT system was installed over the period from November 2019 to April 2021 (Fig.~\ref{fig:TopandSideCRTs} left).  
Following its shipping in summer 2021, the installation of Top CRT modules was carried out and completed in December 2021 (Fig.~\ref{fig:TopandSideCRTs} right). All Top and Side CRT modules were tested before and after their installation to check for electronic functionality of the channels. Data transmission to the servers is performed via ethernet cables connecting the modules in daisy chain. The distribution of a Pulse Per Second (PPS) signal 
(see Sec.~\ref{subsec:trigger}) for absolute time reference and trigger signal to the FEBs was performed with lemo cables. A voltage of 5.5~V to be provided to the FEBs is distributed via power lines assembled at FNAL during the installation. All the information on modules to cables connections, SiPM bias voltages, module positions, etc. are stored in a Fermilab SQL database.

The last ICARUS installation activity was the deployment of the 2.85-meter concrete overburden above the Top CRT. The overburden is composed of three layers of concrete blocks, each approximately 1-meter tall, 
giving a total mass of 5 million pounds. The installation of the last concrete
block was completed June 7, 2022, marking the beginning of ICARUS data taking for physics with both BNB and NuMI beams.

\section{ICARUS-T600 commissioning}\label{sec:commissioning}

After the placement of the two ICARUS modules in the pit in August 2018, all the feed-through flanges for the TPC and PMT signals and for the injection of the laser flashes used to calibrate the PMTs were installed in December 2018. The gain and the dark rate for all 360 PMTs were measured as a function of the applied voltage at room temperature. All the new TPC readout electronics in the 96 mini-crates and the low noise power supplies were installed and verified. In particular the full readout chain has been tested by injecting test pulses in wires at the far end of the chamber and reading out the signals with the A2795 boards on the other end to check the full system for noise monitoring purposes. 

In parallel all the cryogenic equipment were installed, welded and the complete system has been tested at 350 mbar over-pressure. The cold vessels were then successfully brought to vacuum, with a 10$^{-5}$ mbar residual pressure. 

The cryogenic commissioning of the ICARUS-T600 detector started on February 13, 2020 by breaking the vacuum in the two main cold vessels with ultra-purified argon gas.
Cool down started on February 14 by injecting liquid nitrogen in the cold shields. It took about four days to bring the temperature on the wire chamber below 100 K. The cooling process was continuous and the maximum temperature gradient on the wire chambers was about 35 K. On February 19, the gas recirculation units were put into operation to purify the argon gas before the start of the liquid filling. 

The continuous filling with ultra-purified liquid argon started on February 24. The filling was interrupted at around 50\% to regenerate the filling filter. The filling was stopped again when the liquid reached the $- 6$~cm LAr level probes (6 cm below the nominal level) to perform the final pressure test of the two cold vessels. After the test, the gas recirculation units were put into operation. 

\begin{figure*}[ht]
    \centering
    \includegraphics[width=\textwidth]{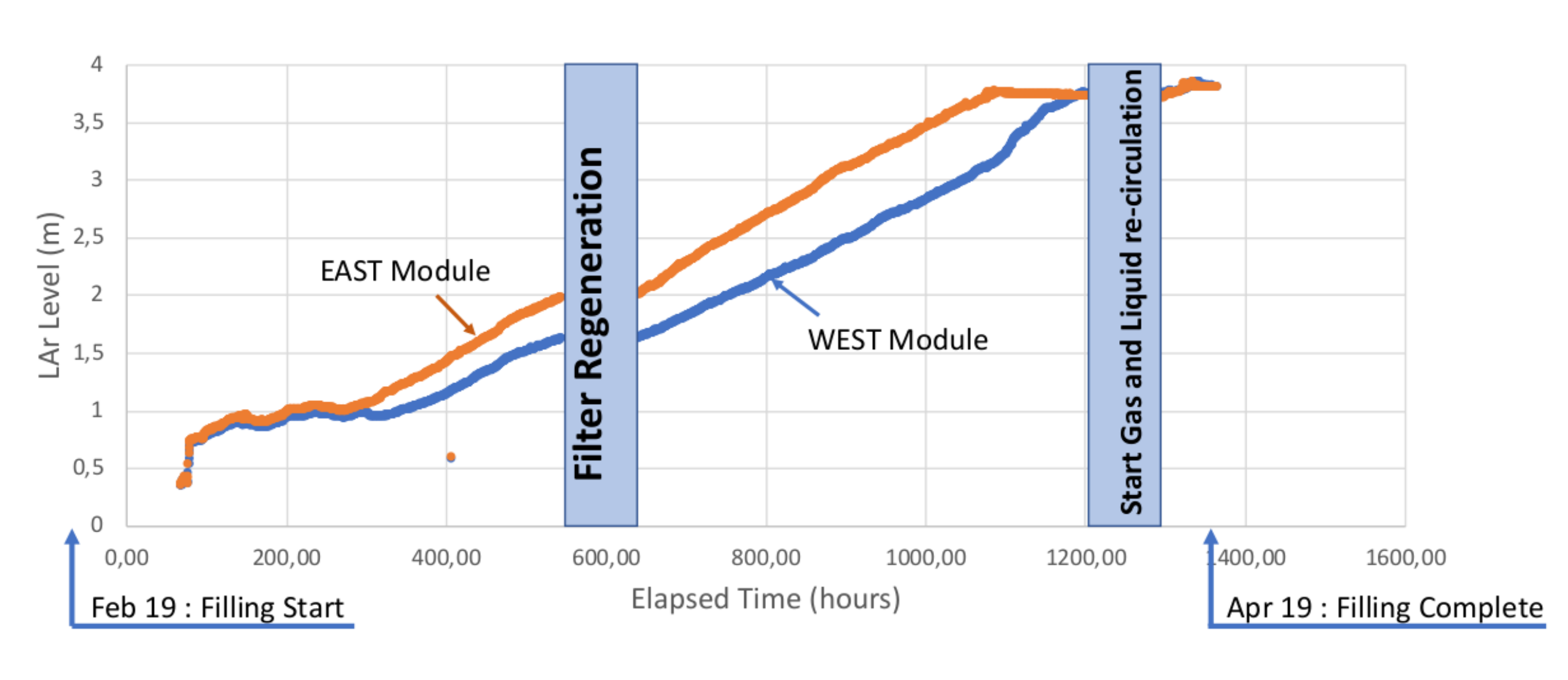}
    \caption{Trend of the liquid argon level inside the two ICARUS cryostats during the filling phase.}
    \label{fig:filling}
\end{figure*}

The filling was completed on April 19, see Fig.~\ref{fig:filling}. On April 21 the liquid recirculation was started. The recirculation rates were 1.85~m$^3$/h in the West module and 2.25~m$^3$/h in the East module.

The cryogenic stabilization phase was completed around the end of May 2020. Pressures and temperatures in the two modules were stable and no cold spots were observed on the external surface of the Warm Vessel. At the start of the cryogenic commissioning, all activities in the detector building not related to cryogenics were suspended and the building was put in a high safety condition, with strong limitations to the presence of people onsite. At the end of the liquid argon filling, after the final pressure test, the standard safety conditions were restored and regular activities on top of the detector could be restarted to complete the installation and test of all sub-detectors.

During the cryogenic commissioning, there were several activities both related to monitoring the status of the detectors (wire chambers, wires readout electronics, PMTs, CRT) and to developments for the following detector commissioning phases.
Noise data have been continuously taken of wire readout electronics, PMTs and CRT. Effects on the noise from the activation of the cryogenic plant have been continuously monitored. Functionality and stress tests of the DAQ were conducted with several useful results.

The detector activation took place on August 27, 2020 when the TPC wire planes and the cathode high voltage (HV) were taken to nominal voltages. HV has remained stable at $- 75$ kV. Significant currents were found only on a few wire bias and were addressed. All PMTs were switched on and calibrated with the laser system. 

Cosmic-ray interaction events were initially collected with a random 5~Hz trigger and data analyzed for calibration purposes (i.e. electron lifetime, space charge, drift velocity measurements). Dedicated runs were also carried out for specific commissioning tasks, such as investigation of TPC noise, PMT calibration with the laser system, DAQ upgrades/longevity tests, etc.

One of the first measurements carried out was the free electron lifetime $\tau_{ele}$. This parameter is fundamental for the monitoring of the liquid argon condition in the TPCs and to obtain the precise measurement of the energy deposition from the ionization charge signal in the collected events.
The LAr purity is continuously monitored by measuring the charge attenuation along the drift path of the electron ionization signals generated by cosmic ray tracks crossing the detector. A fast procedure has been setup starting from the method developed and used during the Gran Sasso run~\cite{lngs_purity}; it has been applied to the recorded data since the detector activation. 

\begin{figure*}[ht]
    \centering
    \includegraphics[width=.8\textwidth]{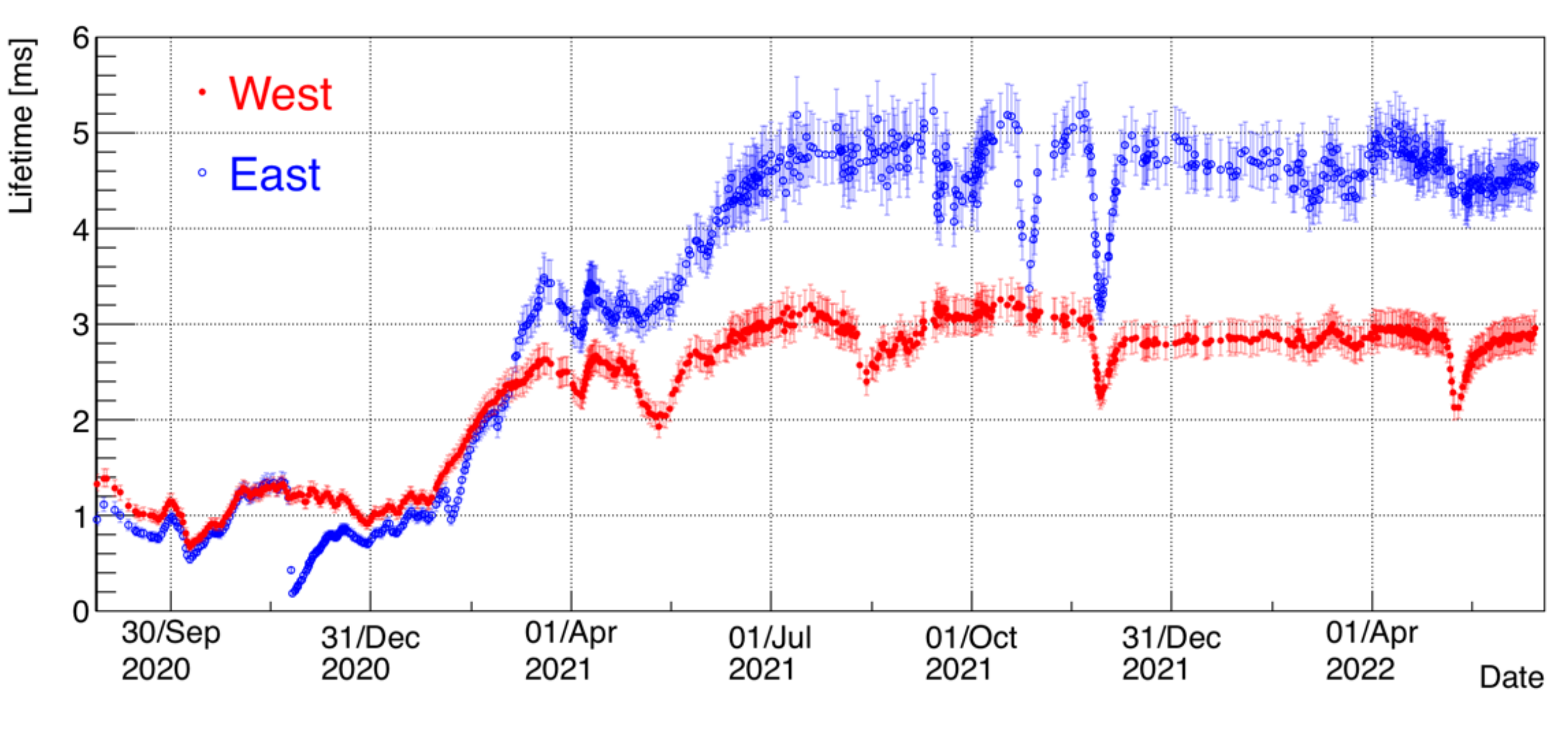}
    \caption{Trend of the drift electron lifetime in the two ICARUS cryostats during the commissioning phase. The sharp decreases of the lifetime are due to programmed interventions on the LAr recirculation pumps or on the cryogenic system. The lifetime is quickly recovered after the end of the interventions.}
    \label{fig:purity}
\end{figure*}

The $\tau_{ele}$ measurement is based on a simplified identification of the wire signals in the Collection plane and of the anode to cathode crossing muon tracks that have no indication of associated $\delta$-rays or electromagnetic showers along the track. It is used to provide a fast, real time, measurement within 5-10\% precision dominated mostly by effects related to space charge and to the electron diffusion, see Fig.~\ref{fig:purity}. 

The steady state values of $\tau_{ele}$, exceeding 3~ms in both cryostats, are high enough to allow for efficient detection and reconstruction of ionizing events inside the active volume.

\subsection{TPC commissioning}\label{subsec:TPC_comm}

After the TPC wires were biased and the cathode HV was raised to nominal operating conditions, the TPC commissioning began. With the liquid argon at a sufficient level of purity, cosmogenic activity in the detector can be used to study the detector response to ionization signals in the TPC. To characterize the performance of the ICARUS TPC, a variety of measurements were taken between August 2020 and May 2022 as summarized below.

\begin{figure*}[ht]
    \centering
    \includegraphics[width=0.95\textwidth]{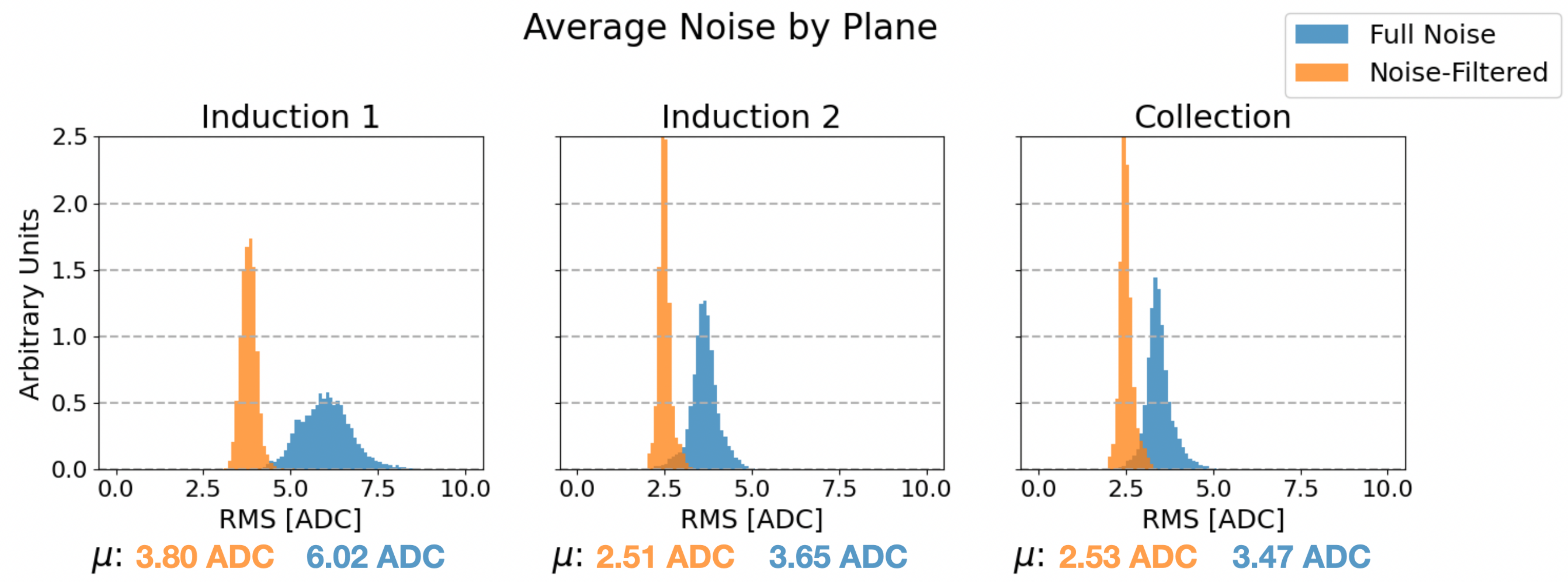}
    \caption{TPC noise levels at ICARUS before and after filtering of coherent noise, as measured by waveform RMS in ADC counts (with ENC of roughly \SI{550}{e^{-}/ADC}~\cite{Bagby_2021}). Results are shown separately for the Induction 1 plane (left), Induction 2 plane (center), and Collection plane (right). Mean values of the shown distributions are presented at the bottom of each figure.}
    \label{fig:TPC_NoiseRMS}
\end{figure*}

Noise levels in the TPC can be measured using the RMS of waveforms from the TPC readout, with an equivalent noise charge (ENC) of roughly \SI{550 }{e^{-}/ADC}~\cite{Bagby_2021}. Measured TPC noise levels at ICARUS are shown in Fig.~\ref{fig:TPC_NoiseRMS}, both before and after the filtering of coherent noise, which was performed across sets of 64 channels associated with the same front-end electronics board. 

Waveforms containing ionization signals are identified by simply applying a threshold and removing from consideration to ensure there is no bias to the noise measurements. The measurements were repeated with the cathode HV off and consistent results were obtained, validating the ionization signal identification methodology and indicating that a negligible amount of TPC noise is caused by interference from the cathode HV system.  

The noise levels after coherent noise filtering shown in Fig.~\ref{fig:TPC_NoiseRMS} are consistent with previous noise measurements of the TPC electronics in a test setup~\cite{Bagby_2021}.

\begin{figure}[ht]
    \centering
    \includegraphics[width=0.45\textwidth]{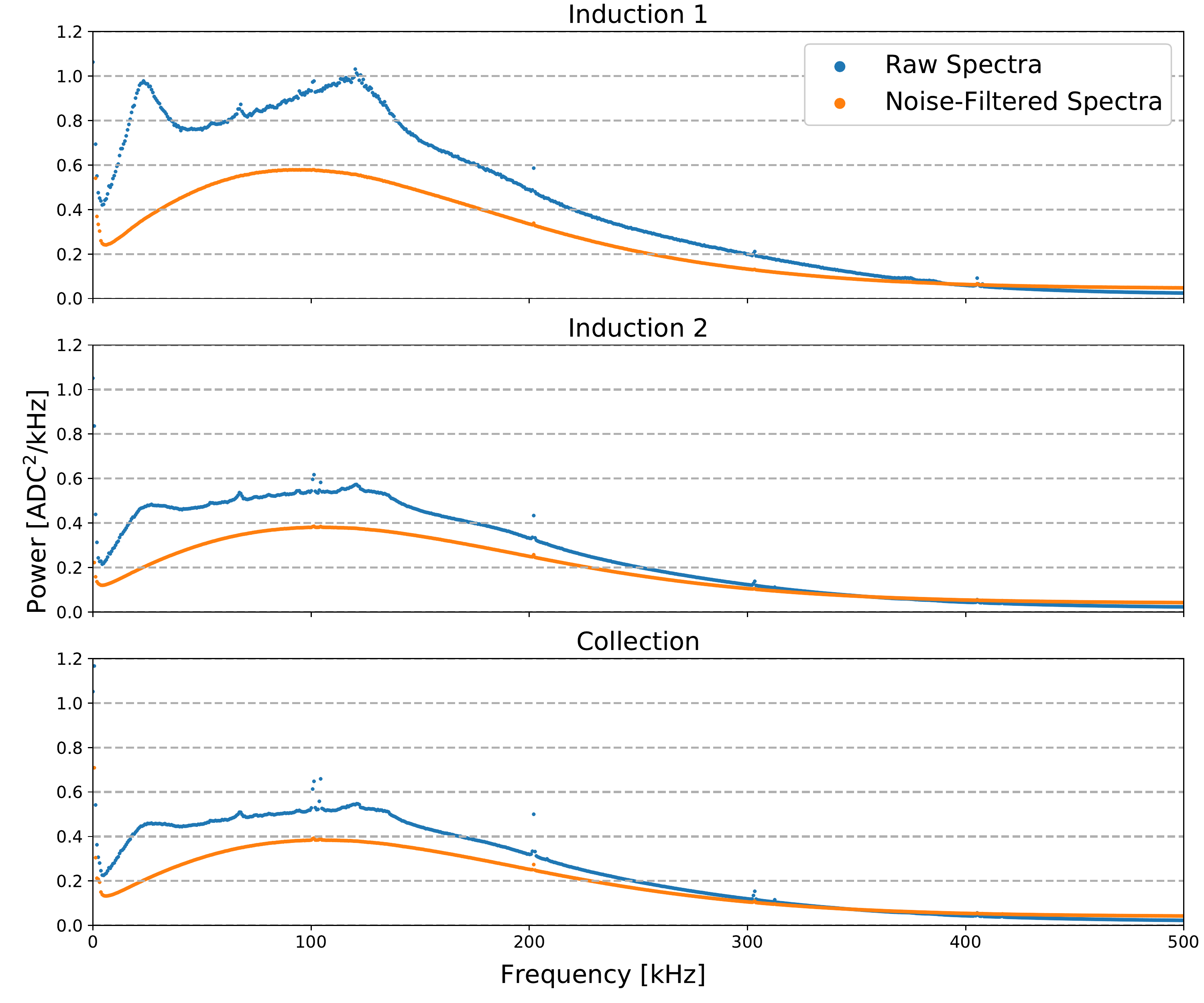}
    \caption{Fast Fourier transforms (FFTs) of noise waveform data collected by the ICARUS TPCs, before and after filtering of coherent noise.  Results are shown separately for the Induction 1 plane (top), Induction 2 plane (middle), and Collection plane (bottom).}
    \label{fig:TPC_NoiseFFT}
\end{figure}

Fast Fourier transforms (FFTs) of the same noise waveforms used in the results shown in Fig.~\ref{fig:TPC_NoiseRMS} are calculated for each of the three wire planes and averaged across the entire detector; these results are shown in Fig.~\ref{fig:TPC_NoiseFFT}. FFTs are shown both before and after coherent noise removal, showing the expected approximate Rayleigh distribution of the intrinsic noise spectrum~\cite{MicroBooNE-Noise} on all three planes after coherent noise is removed. This provides strong evidence of extrinsic noise being almost completely removed from the TPC waveform data by the noise filtering algorithm. 

The Induction 2 plane and Collection plane spectra show a similar normalization, which is expected given the same length of the wires of these two planes. The Induction 1 plane spectrum has instead a larger normalization given the longer wires and thus a higher capacitance, increasing the intrinsic noise levels. Further work is being carried out to understand the source of the coherent noise.

\begin{figure}[ht]
    \centering
    \includegraphics[width=0.45\textwidth]{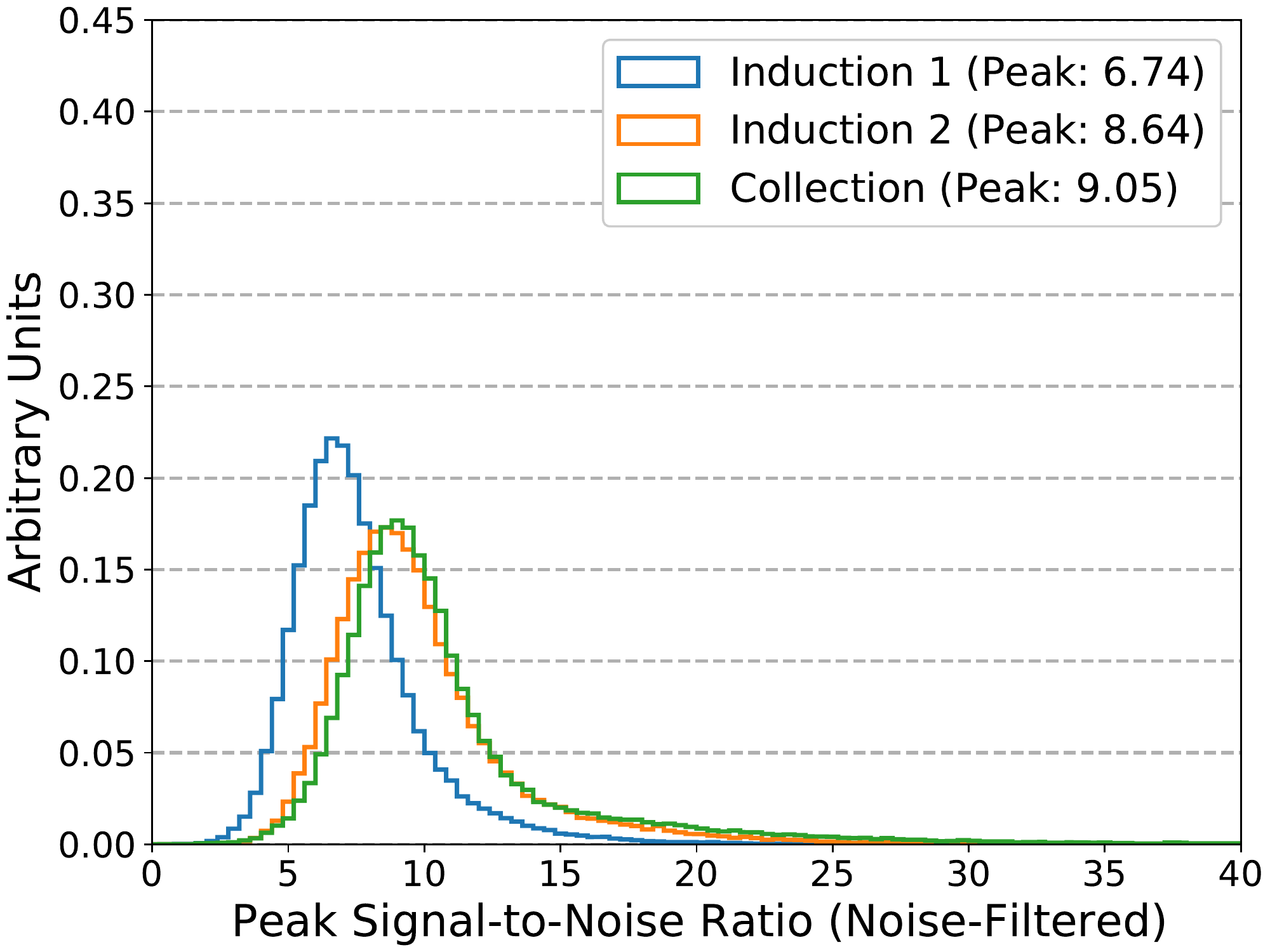}
    \caption{Peak signal-to-noise ratio (PSNR) of ionization signals for each of the three TPC wire planes using cosmic muons in ICARUS data.  Coherent noise is removed from the TPC waveforms prior to identification and measurement of the ionization signal amplitude.  See text for details on the cosmic muon data selection.}
    \label{fig:TPC_PSNR}
\end{figure}

In runs with sufficiently high electron lifetime (most runs after the very beginning of commissioning in 2020), ionization signals from anode-cathode-crossing cosmic muons are used to evaluate the peak signal-to-noise ratio (PSNR) for minimum-ionizing particles (MIPs) in the TPC.  Anode-cathode-crossing cosmic muon tracks traverse the full drift length of the detector and therefore allow for knowledge of the drift coordinate of each ionization signal along the track. Fig.~\ref{fig:TPC_PSNR} shows the PSNR of ionization signals for each plane using a large sample of cosmic muons in ICARUS data with coherent noise removed.  In this study, the peak signal (numerator in the ratio) is defined as the maximum signal ADC value minus the baseline ADC value for the unipolar signals of the Collection plane and the absolute value of the maximum signal ADC value minus the minimum signal ADC value for the bipolar signals of the two induction planes. The noise level (denominator in the ratio) is the RMS of signal-removed waveforms from the same TPC channel in units of ADCs, as shown in Fig.~\ref{fig:TPC_NoiseRMS}.  
Cosmic muon tracks used in the PSNR measurement are required to be oriented at an angle of 20 degrees or less with respect to the anode plane, and have a "3D pitch" (track segment length corresponding to the ionization signal from a single wire) of \SI{4}{mm} or less for the wire plane of interest. These selection criteria probe the phase space most relevant for beam neutrinos interacting in the detector, which have interaction products that travel mainly in the forward direction. Furthermore, only parts of the track within \SIrange{2}{10}{cm} of the anode are used in order to minimize impact from charge attenuation due to impurities in the liquid argon. Fig.~\ref{fig:TPC_PSNR} illustrates the performance of the TPC.

The detector enables robust identification of ionization signals embedded within electronics noise background, with more than 99\% of the MIP ionization signals having a PSNR greater than four.

Anode-cathode-crossing cosmic muon tracks are also used to make a measurement of ionization drift velocity in the detector.  The distance between the anode and cathode, \SI{148.2}{cm}, is divided by the maximum ionization drift time, or the difference in time between the first and last ionization signals associated with the cosmic muon tracks. The latter measurement should yield the time it takes for ionization to drift from the cathode (one end of the track) to the anode (other end of the track), so the ratio should provide the drift velocity of the ionization electrons in liquid argon at the nominal drift electric field of roughly \SI{500}{V/cm} and temperature of roughly \SI{87.5}{K}.  

\begin{figure}[ht]
    \centering
    \includegraphics[width=0.46\textwidth]{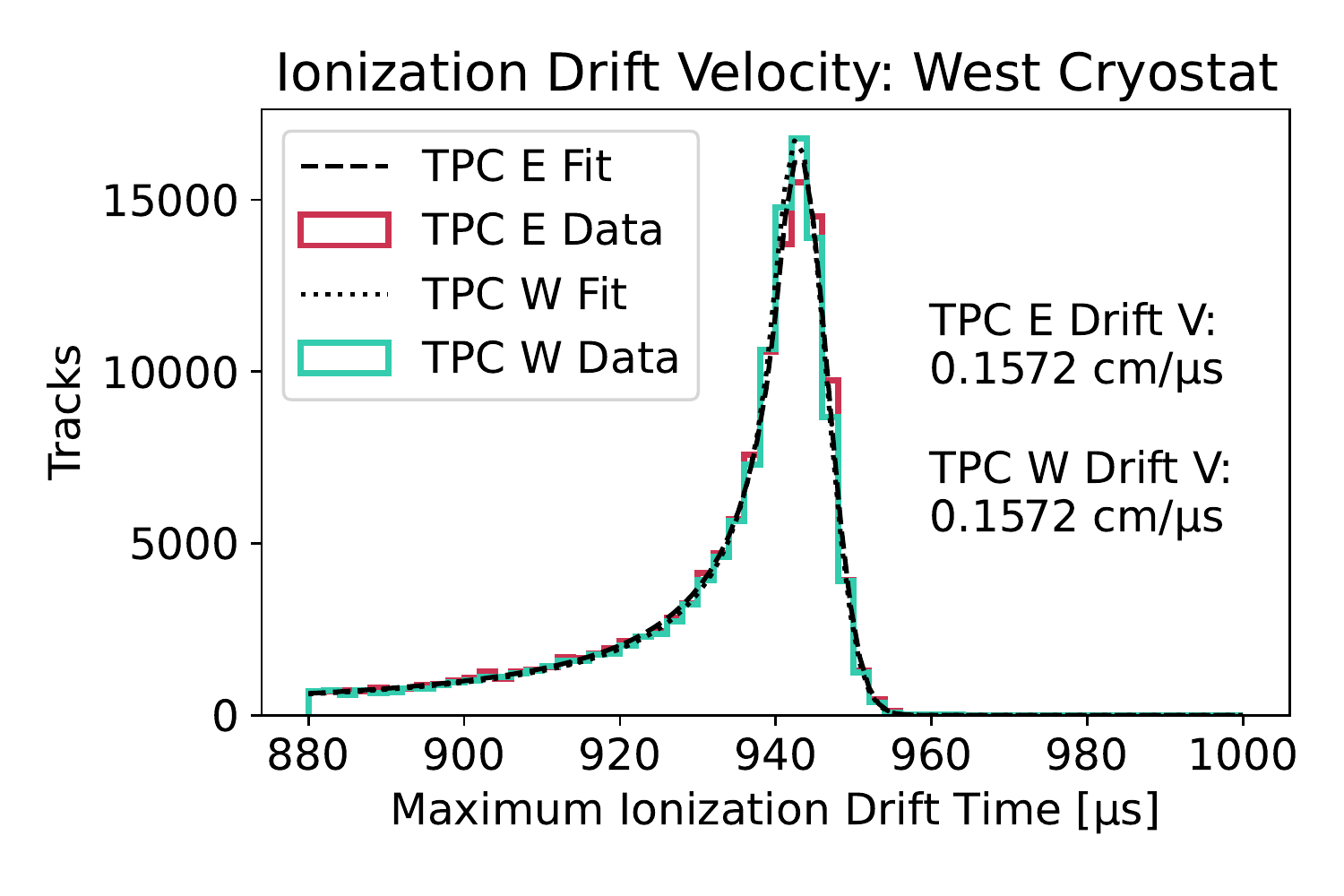}
    \hspace{1.5mm}
    \caption{Results of the ionization drift velocity measurement using ICARUS cosmic muon data.  Shown are Crystal Ball fits to the maximum ionization drift time distributions associated with anode-cathode-crossing cosmic muons in the two TPCs in the west cryostat.}
    \label{fig:TPC_DriftVel}
\end{figure}

A correction is made to account for a small bias in precisely reconstructing the drift times associated with the track end points, derived from Monte Carlo simulation.  A Crystal Ball function\footnote{The Crystal Ball function, named after the Crystal Ball Collaboration, is a probability density function commonly used to model various lossy processes in high-energy physics. It consists of a Gaussian core portion and a power-law low-end tail, below a certain threshold.} 
is then fit to the maximum ionization drift time distribution associated with cosmic muon tracks in each TPC volume (two per cryostat), with the peak value of each fit used in the ionization drift velocity calculation.  The results of the ionization drift velocity measurements in the west cryostat are shown in Fig.~\ref{fig:TPC_DriftVel}.  The results of the measurements, roughly \SI{0.1572}{cm/{\micro}s} for both TPC volumes in the west cryostat, agree with the predicted value of \SI{0.1576}{cm/{\micro}s} to within 0.3\%~\cite{Walkowiak_2020, MicroBooNE-SCE-Cosmics}.

Electric field distortions in near-surface LAr-TPCs can arise due to the accumulation of space charge, i.e. slow-moving positively-charged argon ions originating from cosmic muon ionization within the detector~\cite{Mooney-SCE}. These argon ions, which drift slowly toward the cathode at a drift velocity of several millimeters per second at a drift electric field of \SI{500}{V/cm}~\cite{MicroBooNE-SCE-Cosmics}, linger around long enough to create substantial electric field distortions that pull ionization electrons toward the middle of the TPC volume as they drift toward the anode.  These electric field distortions lead to biases in reconstructing the point of origin of ionization within the detector, a secondary effect referred to as "spatial distortions" in LAr-TPC detectors; collectively, these two related distortions are referred to as space charge effects (SCE).

\begin{figure}[ht]
    \centering
    \includegraphics[width=0.48\textwidth]{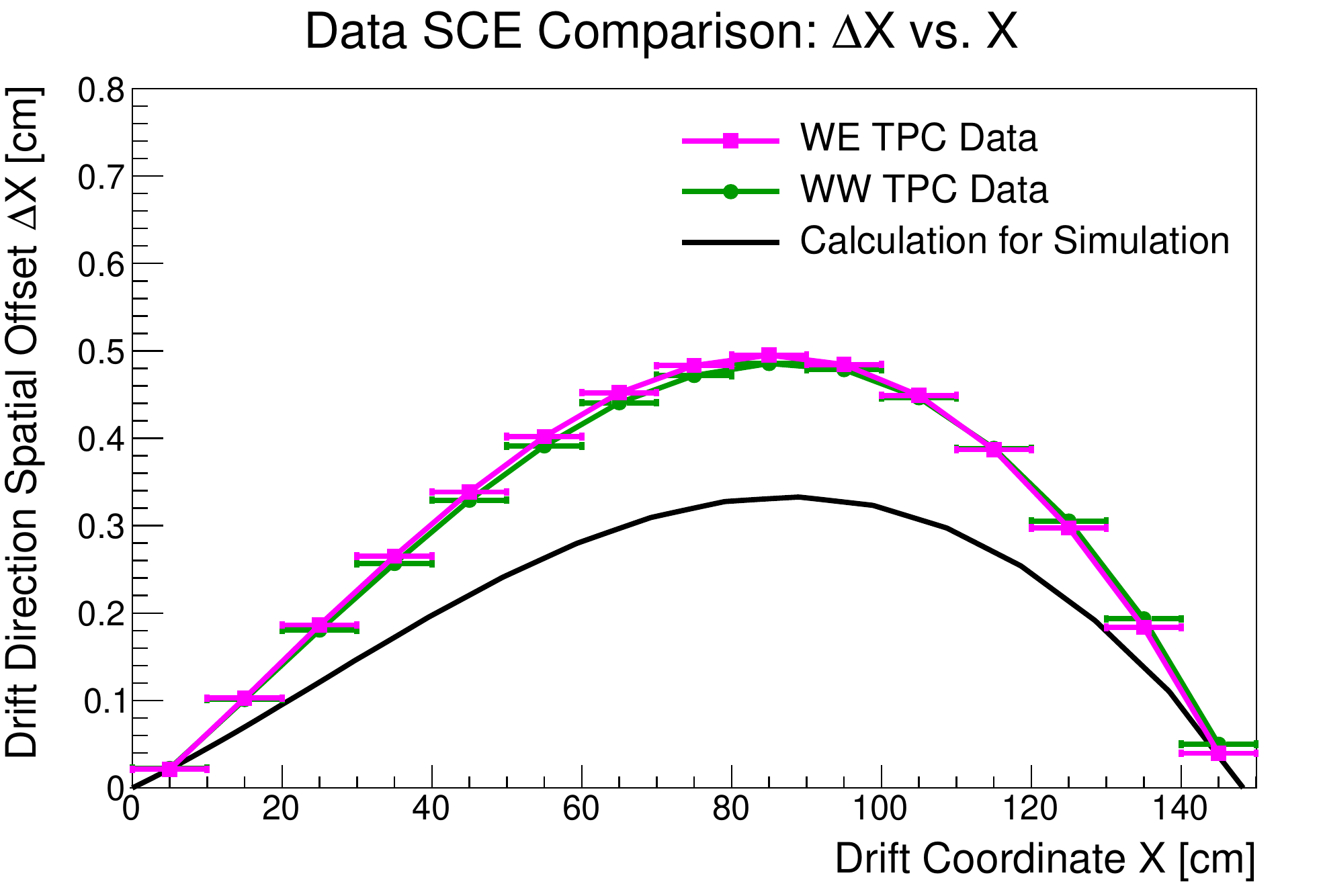}
    \caption{Measured spatial offsets in the drift direction as a function of ionization drift distance for the two TPCs in the west cryostat, evaluated using anode-cathode-crossing cosmic muon tracks in ICARUS data. The results are compared with predictions of spatial distortions from a calculation of space charge effects (SCE) presently used in ICARUS Monte Carlo simulations (to be updated with data-driven SCE measurement).}
    \label{fig:TPC_SCE}
\end{figure}

Using anode-cathode-crossing cosmic muon tracks, the magnitude of SCE in the ICARUS detector is estimated by utilizing methodology developed to measure SCE in previous near-surface running of the ICARUS detector~\cite{Antonello_2020}. The results of measurements in the two TPC volumes of the west cryostat are shown in Fig.~\ref{fig:TPC_SCE}, where they are compared to a calculation of SCE~\cite{MicroBooNE-SCE-Cosmics} used in ICARUS Monte Carlo simulations prior to measuring the magnitude of SCE in ICARUS data. The magnitude of SCE is observed to be very similar in the two TPC volumes, though underestimated by roughly 30\% in simulation.  

The energy scale of MIPs can be probed with cosmic muons that stop in the ICARUS detector, as done in similar calibrations performed at other LAr-TPC neutrino experiments~\cite{MicroBooNE-Calib}.  The known profile of muon energy loss per unit length ($dE/dx$) in liquid argon as a function of kinetic energy~\cite{PDG_2020} can be used to predict the value of $dE/dx$ versus residual range, the distance from the end of a stopped muon track in reconstructed TPC data.  After accounting for prompt electron-ion recombination~\cite{ArgoNeuT-Recomb} and charge attenuation during ionization drift due to electro-negative impurities in the detector, one can compare the most-probable value (MPV) of $dE/dx$ versus residual range from a sample of stopping muons in ICARUS data (evaluated by fitting the data with a Landau distribution convolved with a Gaussian, performed in bins of residual range) to the MPV $dE/dx$ curve expected from theory. 

\begin{figure*}[ht]
    \centering
    \includegraphics[width=0.500\textwidth]{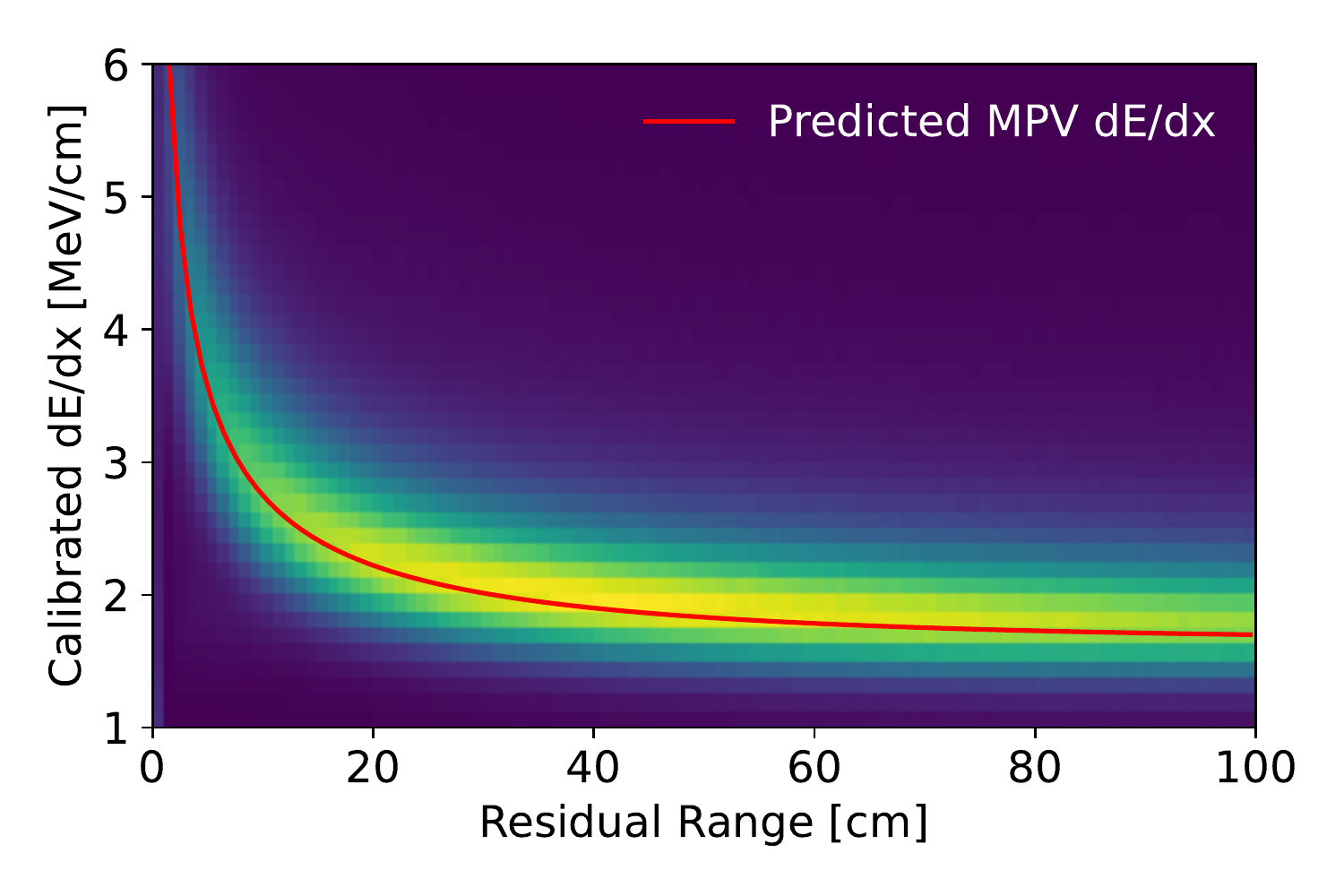}
    \includegraphics[width=0.490\textwidth]{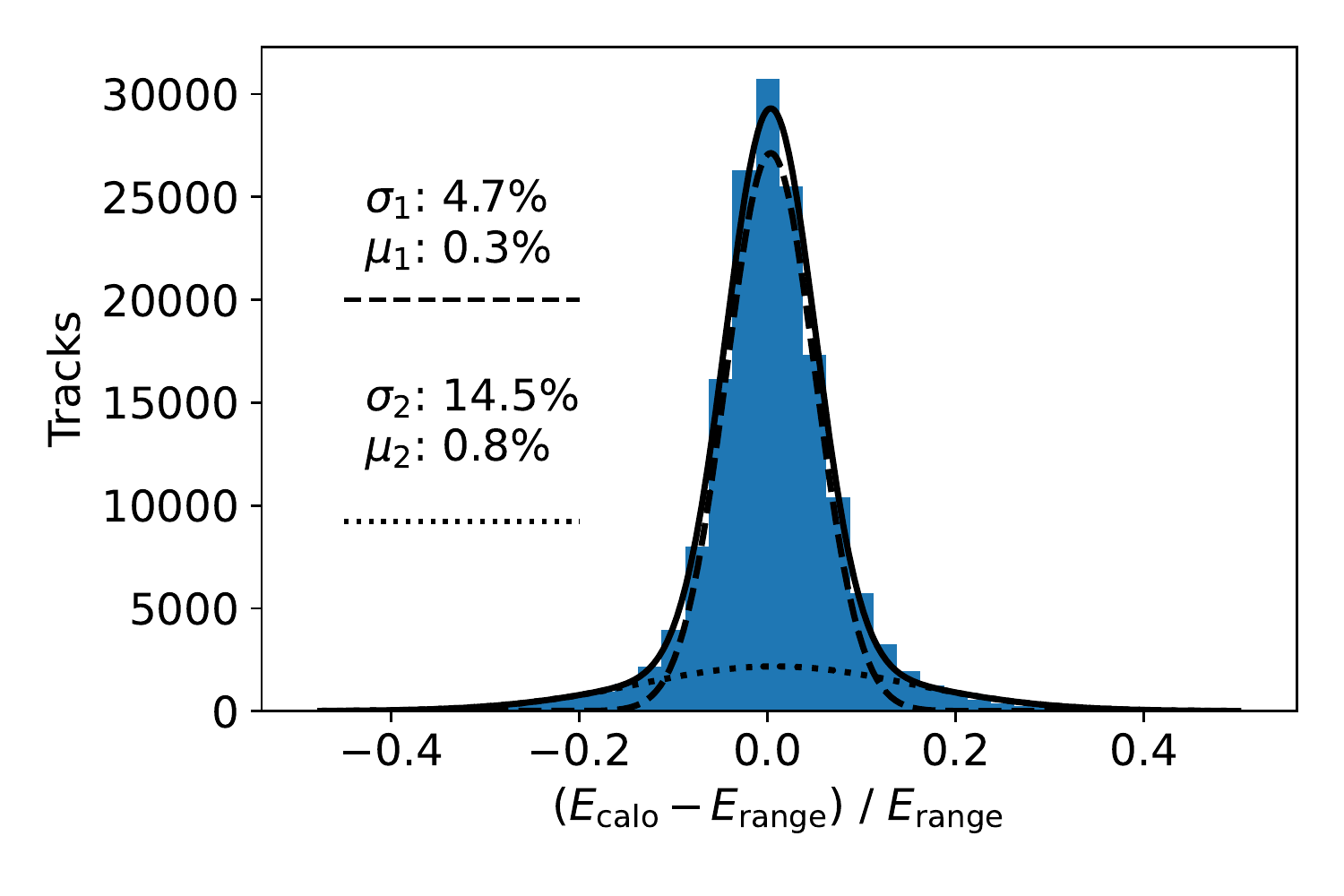}
    \caption{Calibrated Collection plane $dE/dx$ as a function of residual range for a selection of stopping muons in ICARUS cosmic muon data, including a comparison to the most-probable value (MPV) of $dE/dx$ from stopping muons predicted from theory~\cite{PDG_2020} (left); comparison of cosmic muon kinetic energy reconstruction by calorimetry, $E_{\mathrm{calo}}$, and by range, $E_{\mathrm{range}}$, showing little bias between the two methods for stopping muons in ICARUS cosmic muon data after the energy scale calibration is applied (right).}
    \label{fig:TPC_StopMu}
\end{figure*}

The result of the Collection plane energy scale calibration for the east TPC of the west cryostat is shown in Fig.~\ref{fig:TPC_StopMu} (left). Good agreement between calibrated data and predictions from theory is found for all values of stopping muon residual range after this calibration has been performed, with sub-percent agreement for values of $dE/dx < \SI{4}{MeV/cm}$; similar levels of agreement are observed for the other three TPCs as well. Additionally, the energy scale calibration is further scrutinized by comparing two different methods of stopping muon kinetic energy reconstruction: one by calorimetry (summing up charge associated with energy deposition along the track), $E_{\mathrm{calo}}$, and another by range (converting distance from end of stopping muon track to kinetic energy by use of a look-up table~\cite{PDG_2020}), $E_{\mathrm{range}}$.  
The result of this cross-check is presented in Fig.~\ref{fig:TPC_StopMu} (right), showing little bias between the two methods for stopping muons in ICARUS cosmic muon data after the energy scale calibration is applied.  Future measurements will include protons from ICARUS data, allowing for probing of the energy scale of highly-ionizing particles in the detector. 

\subsection{PMT commissioning}\label{subsec:PMT_comm}

The whole light detection system was tested at Fermilab before the cooling of the detector, once the dark condition inside the cryostats was guaranteed. 
A total of 357 (out of 360) PMTs were found to be working with performances consistent with the tests performed at CERN~\cite{Babicz:2018svg}. The same number of working PMTs were found after the filling of the detector with liquid argon, demonstrating the ability of this PMT model to withstand low temperatures. 

\begin{figure}[ht]
    \centering
    \includegraphics[width=0.45\textwidth]{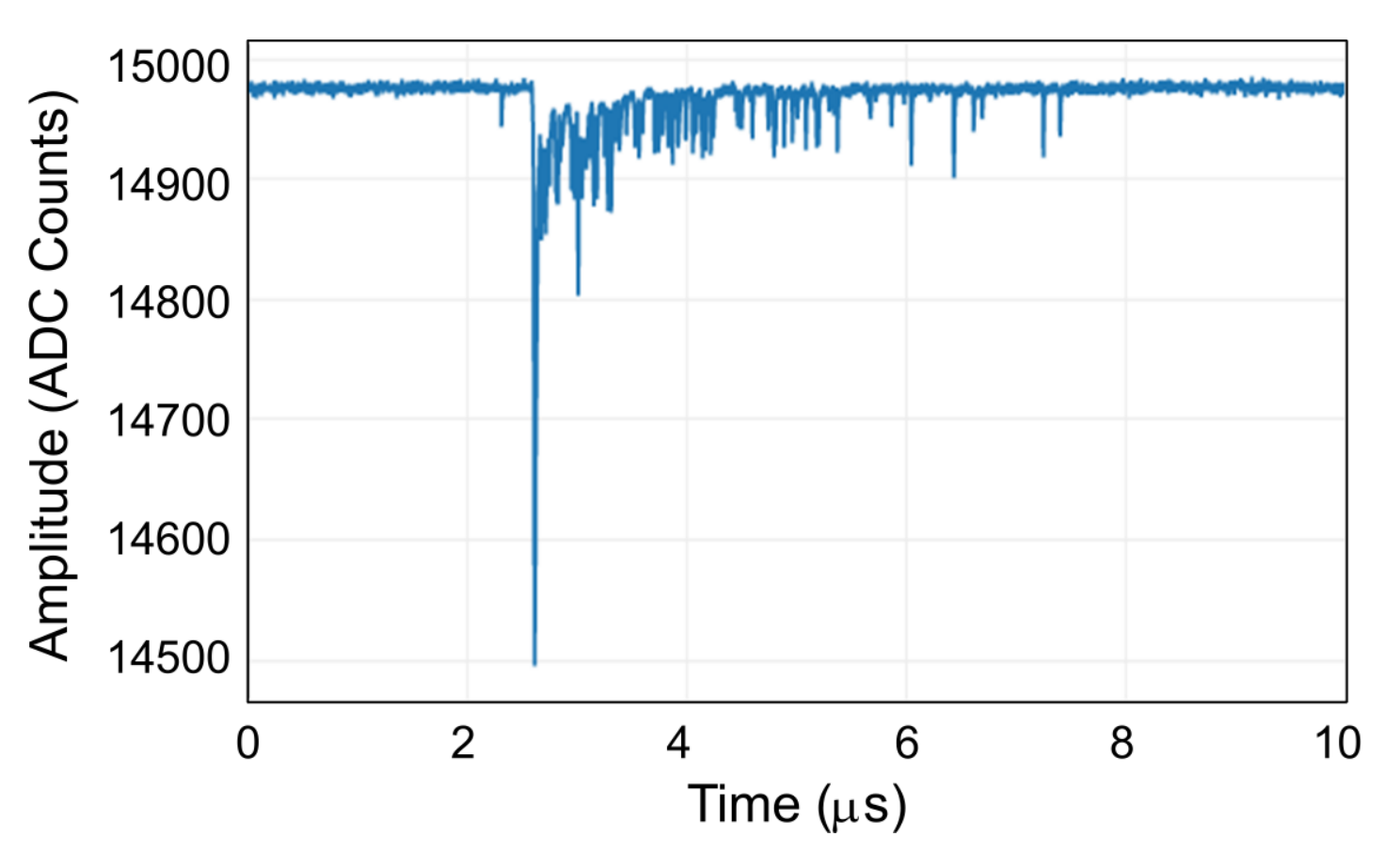}
    \caption{PMT signal as recorded by the light detection system electronics.}
    \label{fig:PMT_signal}
\end{figure}

A PMT signal, recorded by the light detection system electronics, is shown in Fig.~\ref{fig:PMT_signal}.
A gain calibration/equalization campaign was carried out during the PMT commissioning. 
At first, external fast laser pulses focused on each PMT window by means of dedicated optical fibers were used to obtain a coarse gain curve for each PMT as a function of the applied voltage around the expected values.
Laser pulses were also used to characterize, to within \SI{1}{ns} precision, the delay response of each PMT channel, which can differ due to different PMT and cable transit times. Voltages were set to values corresponding to a gain of 5 $\cdot$ 10$^6$, resulting in an equalization within 16\%, as a first approximation. 

\begin{figure}[ht]
    \centering
    \includegraphics[width=0.49\textwidth]{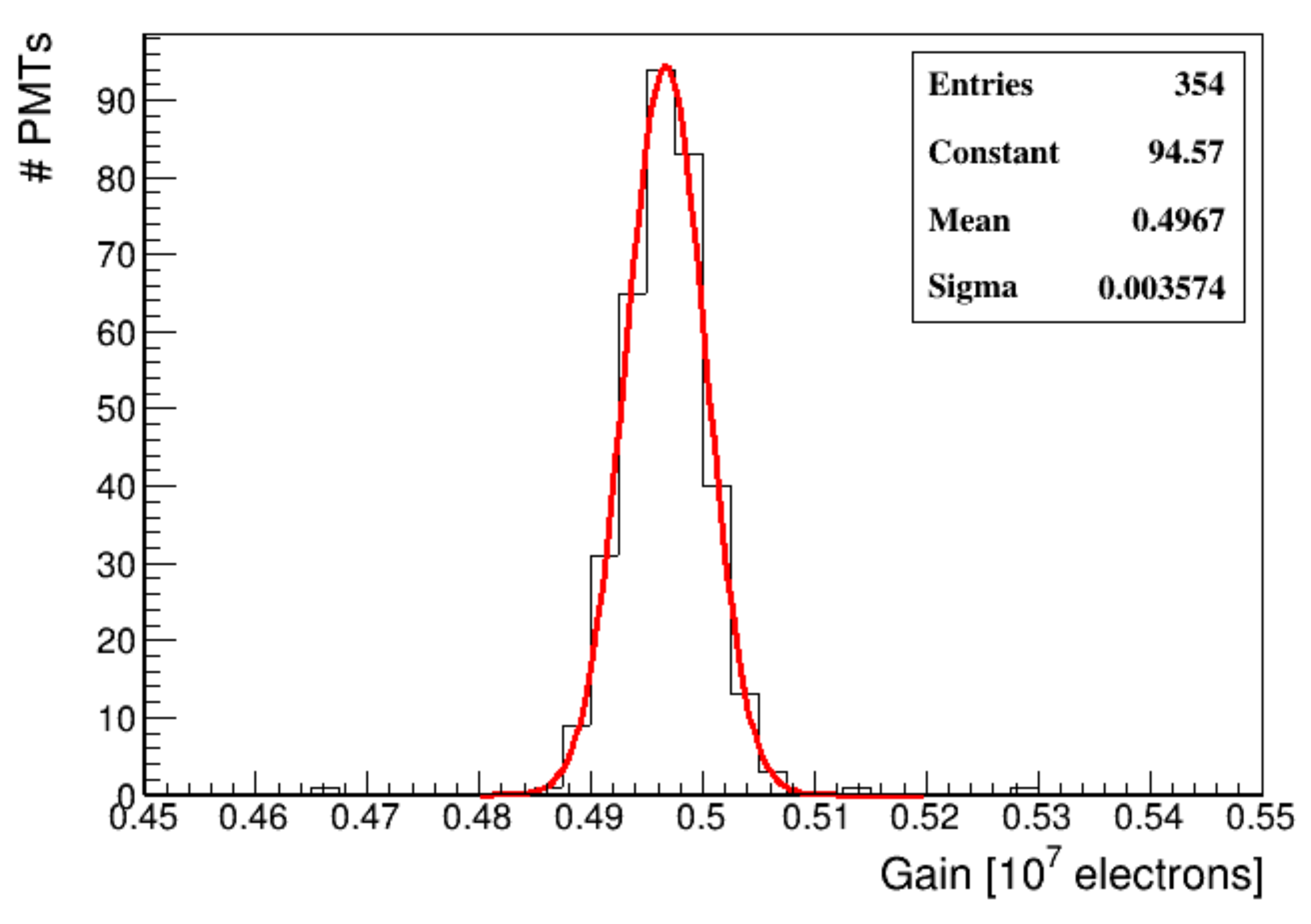}
    \caption{Gain distribution for 354 PMTs after the fine tuning equalization. The automatic procedure was not applied on 6 PMTs (not present in the plot) that were manually calibrated.}
    \label{fig:gain_equalization}
\end{figure}

Fine tuning was carried out to improve the gain equalization by means of an automatic procedure. To this purpose the response of each PMT to background single photons ($\approx$~\SI{250}{kHz}) was measured, and the voltages were adjusted according to the gain curves. This procedure led to a final equalization with a spread less than 1\%, as shown in Fig.~\ref{fig:gain_equalization}.

\subsection{CRT commissioning}\label{subsec:CRT_comm}

\begin{figure}[ht]
    \centering
    \includegraphics[width=0.49\textwidth]{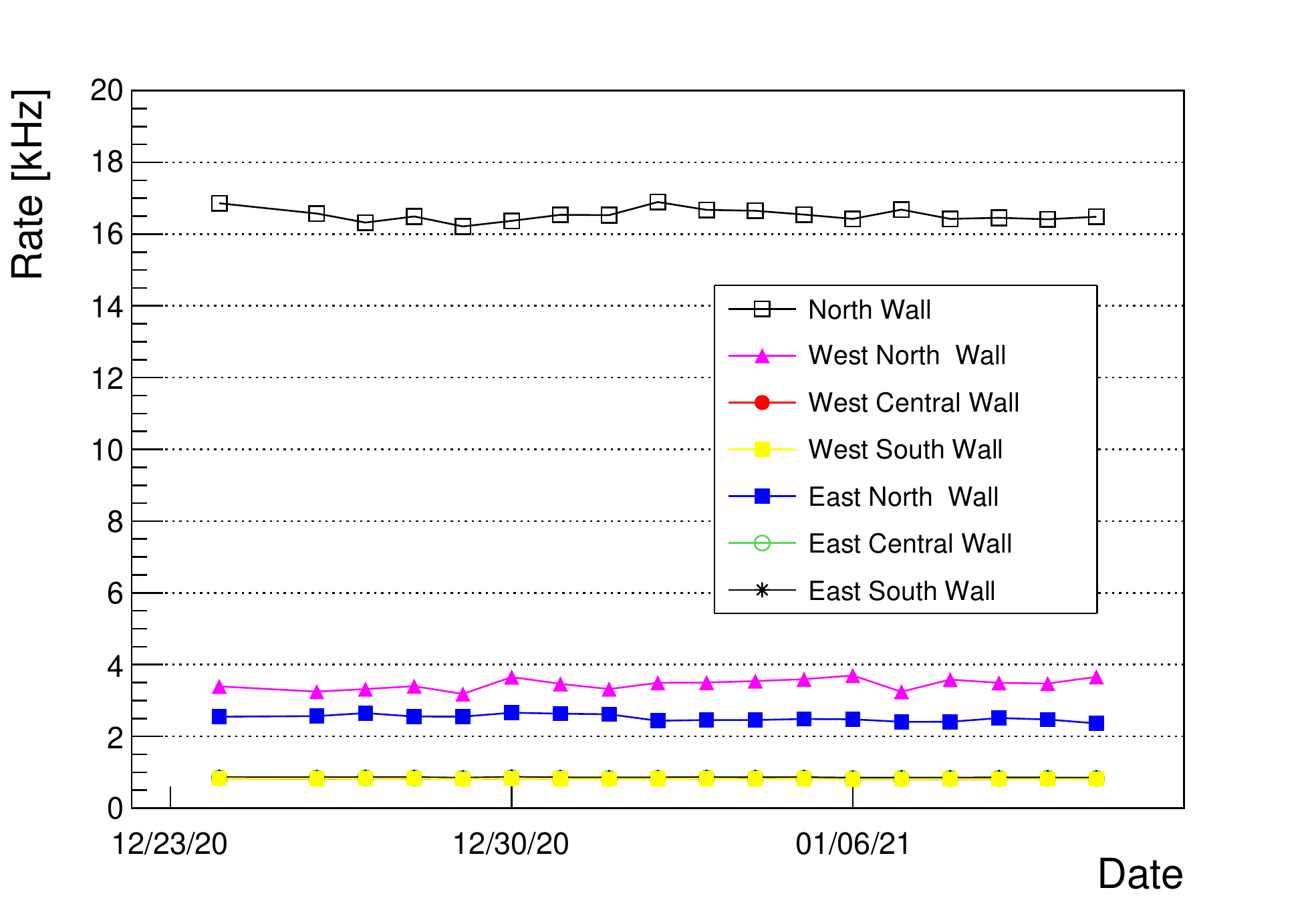}
    \caption{Side CRT cosmic event rates as a function of time. The black points corresponds to the rates from the north side CRT wall, the pink and blue points corresponds to East and West north walls, and the remaining walls are at 1 kHz rate.}
    \label{fig:crtrates}
\end{figure}

The side and top CRT modules were tested before the installation at ICARUS using a test stand. After the installation of all CRT modules, the cosmic rate over time was obtained. 
The event rates for each wall of the side CRT as a function of time are constant, as shown in Fig.~\ref{fig:crtrates}. 
The higher rates on north wall (black) are due to the proximity with the cryogenic pumps, with these modules experiencing higher electrical noise rates in addition to cosmic rates on the surface. In addition, the rates from the west north and east north walls are slightly higher from being closer to the cryogenics. Following work to characterize and mitigate the noise, electrical chokes (inductors) were installed along all Side CRT FEB power cables to reduce noise rates. 

\begin{figure*}[ht]
    \centering
    \includegraphics[width=0.49\textwidth]{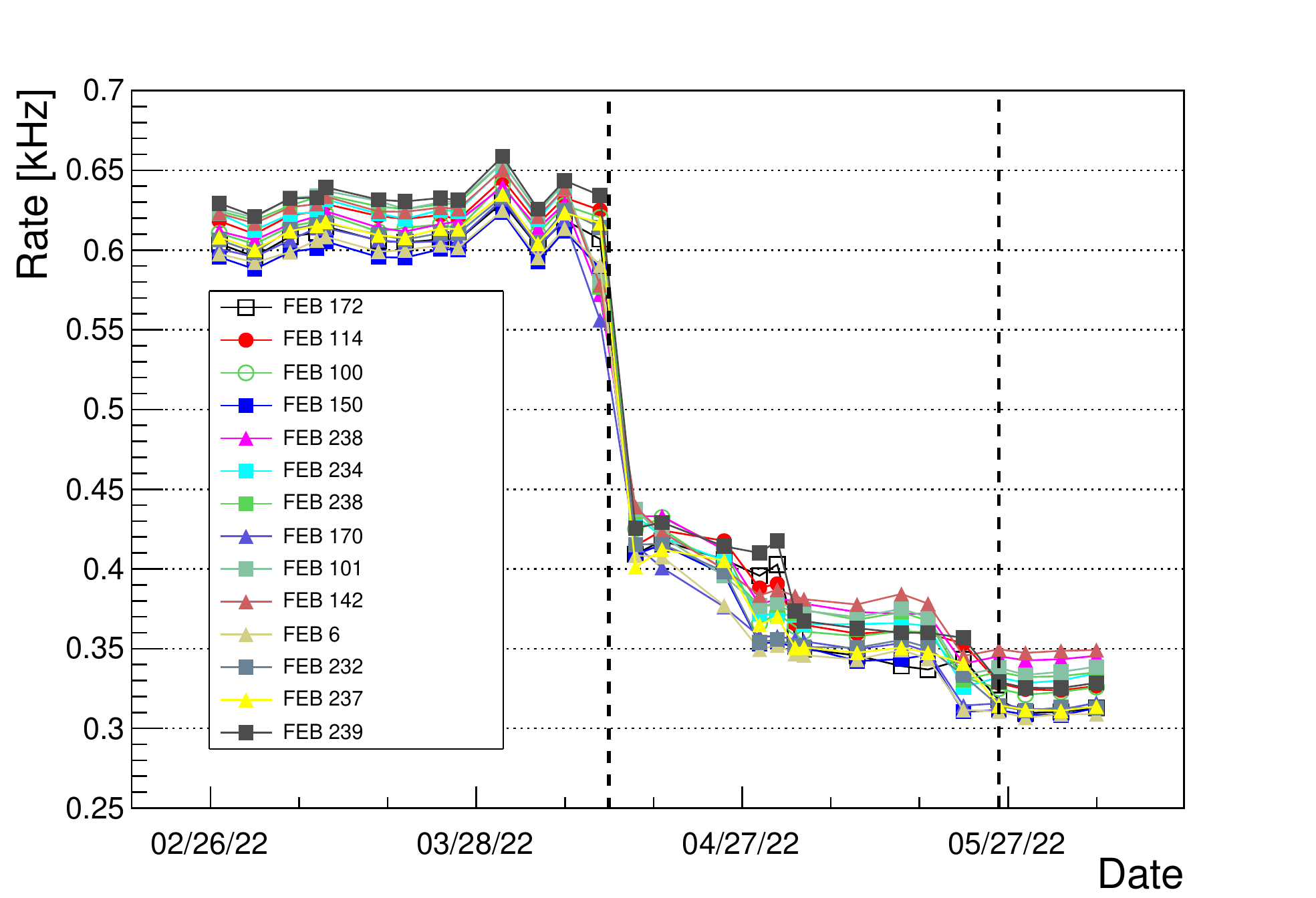}
    \includegraphics[width=0.49\textwidth]{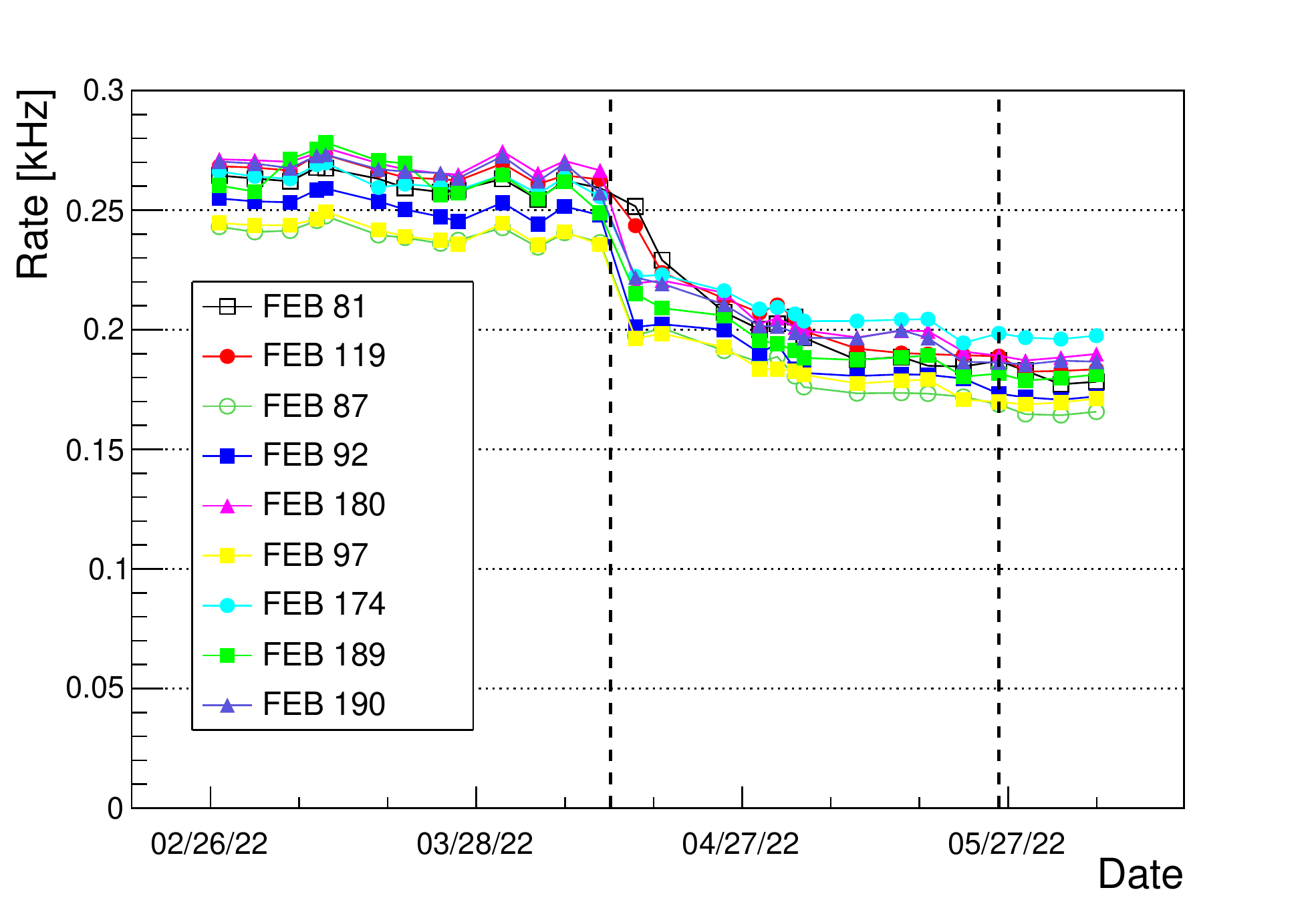}
    \caption{Cosmic ray rates as a function of time for a set of Top CRT horizontal (left) and vertical (right) modules. Numbers in the legend indicate the module's Front End Board and the black dot lines indicate the beginning and the end of 3 m overburden installation over the displayed modules: the rates reduced from $\sim$ 610 (260) Hz before to 330 (180) Hz after the installation of the overburden for the horizontal (vertical) modules.}
    \label{fig:TopCRTrates}
\end{figure*}

Top CRT cosmic event rates before and after the installation of concrete overburden are shown in Fig. \ref{fig:TopCRTrates} for horizontal (left) and vertical (right) modules. Before the installation of the overburden the mean rate was $\sim$~\SI{610}{Hz} and \SI{260}{Hz} for horizontal and vertical modules, respectively. After the installation of the overburden the rates reduced to \SI{330}{Hz} and \SI{180}{Hz} for horizontal and vertical modules, respectively. Except for variation due to concrete blocks placement above the detector, the rates are stable on a time scale of months. 
 
\subsection{Triggering on the BNB and NuMI neutrinos}\label{subsec:trigger}

The initial ICARUS trigger system exploits the coincidence of the BNB and NuMI beams spills, \SI{1.6}{\micro s} and \SI{9.6}{\micro s} respectively, with the prompt scintillation light detected by the PMT system installed behind the wire planes of each TPC~\cite{trigger2022}.

The generation of the beam spill gates is based on receiving the “Early Warning” (EW) signals for BNB and NuMI beams, 35 and 730~ms in advance of protons on target, respectively. LVDS signals from the PMT digitizers, in terms of the OR signal of adjacent PMTs, are processed by programmable FPGA logic boards to implement trigger logic for the activation of the ICARUS read-out. Additional trigger signals are generated for calibration purposes in correspondence with a subset of the beam spills without any requirement on the scintillation light (Min-Bias trigger) and outside of the beam spills to detect cosmic ray interactions (Off-Beam trigger).

\begin{figure}[ht]
    \centering
    \includegraphics[width=0.495\textwidth]{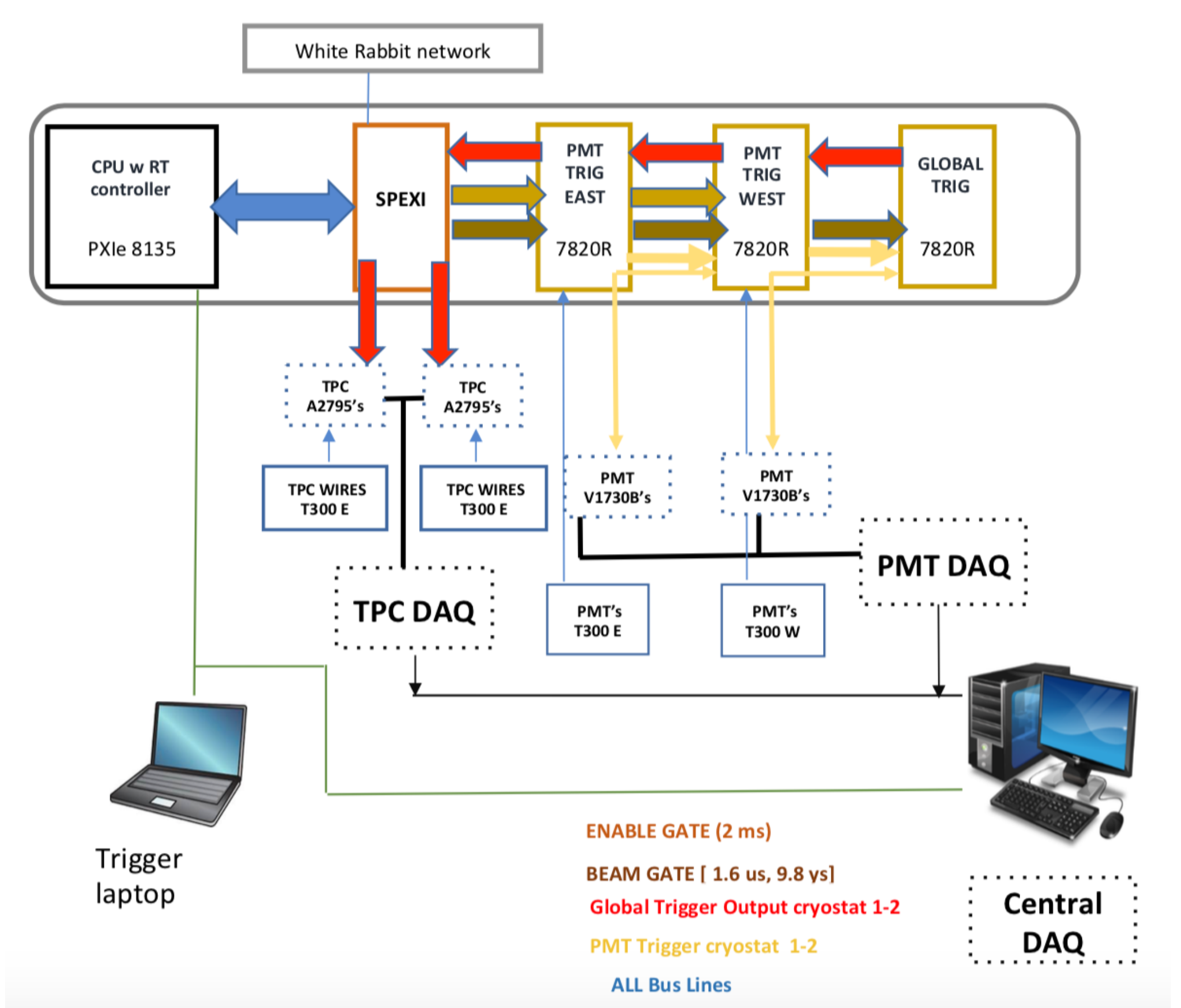}
    \caption{Layout of the trigger system. SPEXI board: synchronizes the whole ICARUS detector, generates clocks and readout signals, handles beam extraction messages; 7820 FPGA boards: generate a Global Trigger in coincidence with beam extraction (Early Warning) on the basis of selected PMT signal majorities to recognize an event interaction in the LAr, to start the PMT activity recording; RT Controller implements all the features for communication with DAQ.}
    \label{fig:trigger_layout}
\end{figure}

To synchronize all detector subsystems' read-outs with the proton beam spill extraction at the level of few nanosecond accuracy, a White Rabbit (WR) network~\cite{whiterabbit} has been deployed for distributing the beam extraction signals. An absolute GPS timing signal, in the form of PPS, is used as a reference for generating phase locked digitization clocks (62.5~MHz for the PMT and 2.5~MHz for the TPC) and for time-stamping the beam gates and trigger signals. In addition, the signals of Resistive Wall Monitor detectors (RWM) at 2~GHz sampling frequency are also recorded to precisely measure the timing and the bunched structure of protons on target, see Fig.~\ref{fig:trigger_layout}. 

In the presence of a global trigger signal, 1.5~ms and \SI{30}{\micro s} acquisition windows are activated for the TPC and PMT signal recording, respectively. In addition, PMT waveforms are collected inside a 2~ms time window around the beam spill to record all cosmic muons crossing the ICARUS TPCs during the electron drift time.

\begin{figure*}[ht]
    \includegraphics[width=\textwidth]{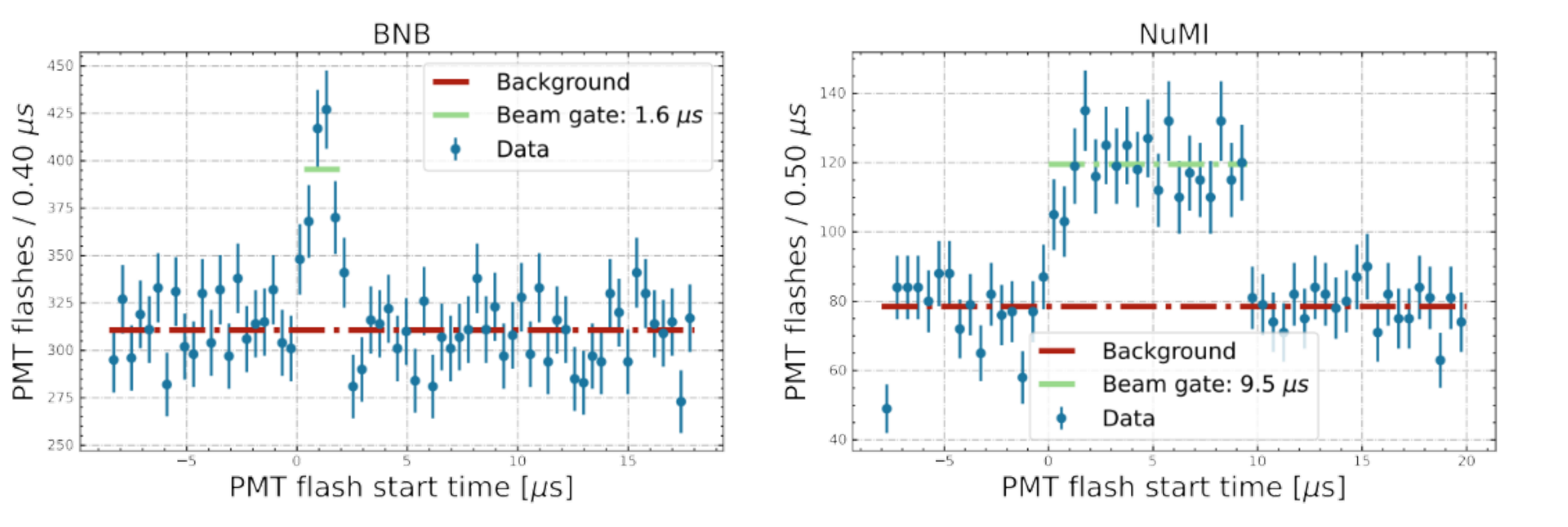}
    \caption{Time distribution of the recorded PMT light flashes ($\geq$ 5 fired PMT pairs in the left and right TPCs within 150 ns):  the beam event excess is observed for BNB (left) and NuMI beam (right). The \SI{1.6}{\micro s} and \SI{9.6}{\micro s} spills duration of the beams are well recognized.}
    \label{fig:time_distribution}
\end{figure*}

The timing of the beam spills was first approximately determined by measuring with an oscilloscope the difference between the EW signals arrival time and the actual proton extraction signal by RWM counters at the target. Then neutrino interactions were identified and associated with the muons of the beam spill in excess to cosmic rays that were clearly identified inside the time profile of the scintillation light signals (\emph{flashes}) by requiring at least 5 fired PMT pairs in the left and right TPC (Fig.~\ref{fig:time_distribution}).

Due to the energy range of BNB and NuMI neutrino beams, neutrino interactions are expected to be contained in an $\sim$~4~m section of ICARUS along the beam direction, suggesting the implementation of a trigger logic based on the recognition of fired PMTs inside a limited TPC region. The logic for processing the PMT LVDS signals has been initially determined with Monte Carlo calculations, and then it has been refined by analyzing a sample of events collected with a beam spill signal only (Min-Bias trigger), i.e. without any requirement on the scintillation light. The 18-m long TPC walls have been subdivided in 3 consecutive longitudinal slices of 6-m length including 30 PMTs each. In each of opposite facing slices a majority of 5 LVDS signals, with 8 photo-electron (phe) discrimination threshold and an OR of two adjacent PMTs, has been required to produce a PMT trigger primitive signal. The same logic with a majority of 10 LVDS PMT signals is applied to 
generate a PMT trigger primitive in time period prior to and after a beam spill. This trigger provides collection of data sampling the ~15 kHz of cosmic rays crossing the detector during the drift time. 

With trigger gates of duration 4 ms and 14 ms for BNB and NuMI, respectively, a trigger rate of $\sim$ 0.7 Hz has been obtained (0.3 and 0.15 Hz from the BNB and NuMI components, respectively, and 0.25 Hz for the Off-Beam). This is in a manageable data read-out bandwidth with good operational stability. The trigger efficiency for neutrino interactions is under study with data; expectations based on the Monte Carlo simulations indicate a $>$ 90\% efficiency for neutrino CC interactions with $>$100 MeV energy deposition.

\subsection{DAQ implementation}\label{subsec:daq}

The ICARUS data acquisition (DAQ) system 
utilizes the general \textit{artdaq} data acquisition software development toolkit ~\cite{artdaq}, providing customizable applications for reading data from detector elements (\textit{BoardReaders}), and configurable applications for performing event-building, data-logging, and data-dispatch to downstream online data quality monitoring processes. 

Customized BoardReaders acquire data fragments from the TPC, PMT, and CRT readout electronics, and from the trigger and White Rabbit timing systems. They then assign appropriate event counters and 
timestamps to each fragment and then queue that data for transfer to a configurable number of \textit{EventBuilder} applications.
For each triggered event, the ICARUS trigger BoardReader sends its data fragment to an EventBuilder, triggering a request for data from all other configured BoardReaders in the DAQ system. 
Events are written using the \textit{art} event-processing framework~\cite{art}. Data are written on separate file streams using simple filters on trigger type. 
Each event in ICARUS, after lossless data compression, is approximately 160 MB, with the majority of data corresponding to the TPCs.
The DAQ system is capable of stably supporting trigger rates in excess of 5 Hz, though typical operational trigger rates are of roughly 1 Hz or below.

The \textit{BoardReader} for the trigger system sends a single fragment containing the trigger and beam-gate timing, the type of beam gate, a global trigger counter, and a counter for the number of beam gates of each type in that DAQ run. The global trigger counter and time are used for collection of data from other subsystems; the latter derives from the common White Rabbit timing system, and is checked for validity against the network  protocol time of the trigger \textit{BoardReader} server. The number of beam gates of each type in the run is used offline for proper accounting of the total number of POT and detector exposure within a run.

In order to handle large data volumes stored on tape, the Fermilab based SAM (Serial Access to Metadata) system is exploited. 
For this purpose, a set of metadata is associated to each data file using Python scripts.
The metadata allow users to create large data sets for the analysis by requiring matching with data’s relevant information such as run number, data type (raw or reconstructed), run configuration, date, etc.

\subsection{First operations with the BNB and NuMI}\label{subsec:operations}

\begin{figure*}[ht]
    \centering
    \includegraphics[width=\textwidth]{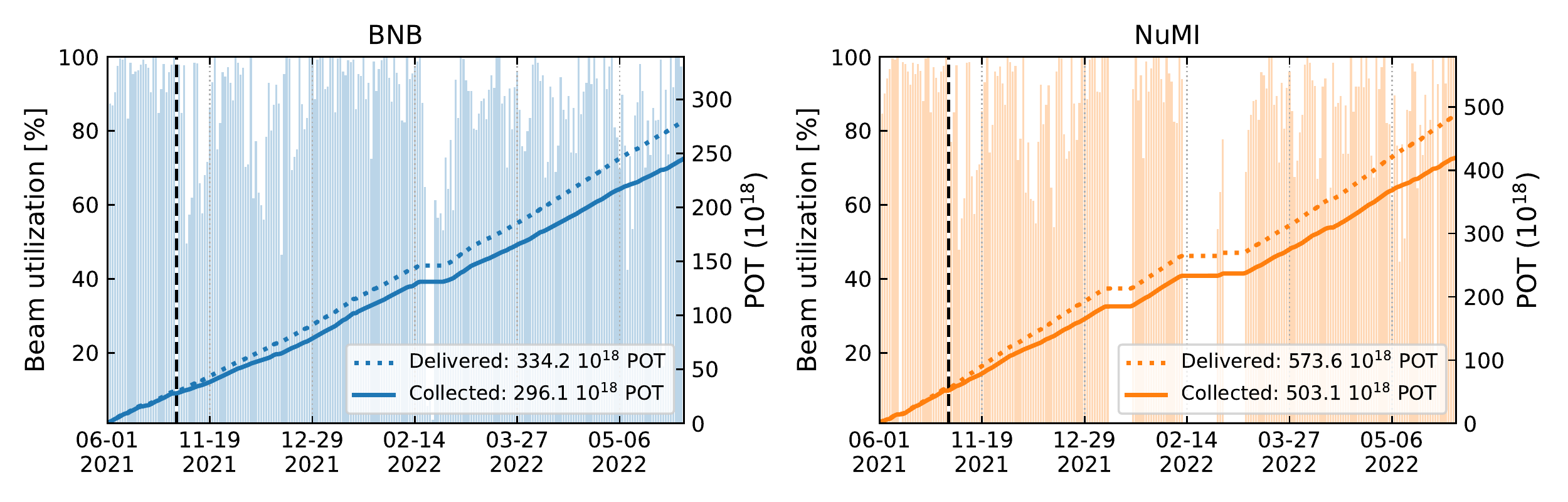}
    \caption{Cumulative sum of POT delivered by the accelerator and collected by the detector and daily beam utilization coefficient as a function of the operation time for BNB (NuMI) on the left (right). The dotted black line marks the separation between the two operation periods of the detector: the full month of June 2021 and between November 5, 2021 and June 1, 2022 (the long break between the two periods is hidden in the plot).}
    \label{fig:beam_pot}
\end{figure*}

The ICARUS-T600 detector was first fully operational in June 2021 before the summer shutdown. It restarted data collection when beam returned November 5, 2021. Figure \ref{fig:beam_pot} shows the amounts of POT delivered by the accelerator and collected by the detector during its commissioning phase, concluded in June 2022, for a total of $296\cdot10^{18}$ and $503\cdot10^{18}$ POT collected for BNB and NuMI, respectively. Beam utilization - defined as the amount of POT collected divided by the delivered - of $89\%$ for BNB and $88\%$ for NuMI. In Fig.~\ref{fig:beam_pot}, 
daily variations of the beam utilization are also visible: periods with low utilization (less than 60\%) correspond to days where the data acquisition was suspended in order to proceed with detector commissioning activities. Apart from this, the utilization is an average over 91\% per day for both beams, which corresponds to a downtime of less than two hours per day. The most frequent causes of operation downtime are data acquisition issues and less commonly hardware problems. The detector and data collection status are continuously supervised with fully-remote shifts staffed by collaborators and with the support of on-call experts for each of the main detector subsystems. 

\section{Observation and reconstruction of neutrino events}
\label{sec:reco}

The data collected by the detector are processed by offline software to obtain information necessary for reconstruction and analysis of events. The procedure to reconstruct the TPC wire and PMT signals is briefly described in the following Sec.~\ref{subsec:WireSignalReco}, \ref{subsec:PMTreco} and \ref{subsec:CRTreco}.

The detector behavior was first investigated by a visual selection of neutrino interactions in the active liquid argon, as described in Sec.~\ref{subsec:VisualScanning}. 
These sample were an important component of the development and validation of an automated event selection scheme.

\subsection{Wire signal reconstruction}
\label{subsec:WireSignalReco}

The ICARUS wire signal processing chain follows a logic similar to other LAr-TPC experiments, based on the \emph{deconvolution} of the wire signal waveform. This procedure, explained in more detail in \cite{uboone_wsp}, has the goal to recover the original time structure of the current of drift electrons generating the signal on each wire, upstream of the distortions produced by the electric field in the wire region and the shaping by the front-end electronics. Mathematically, this is obtained by inverting the response functions describing both the electric field and the electronics effects; the resulting deconvolved signal shape is approximately Gaussian for all wire planes. 

After the removal of the coherent noise (described in \ref{subsec:TPC_comm}), the deconvolution is performed on each wire waveform. Segments of waveforms corresponding to physical signals (\emph{hits}) are searched for in the deconvolved waveform with a threshold-based hit finding algorithm. Each hit is then fit with a Gaussian, whose area is proportional to the number of drift electrons generating the signal.

\begin{figure}[htb]
    \centering
    \includegraphics[width=0.5\textwidth]{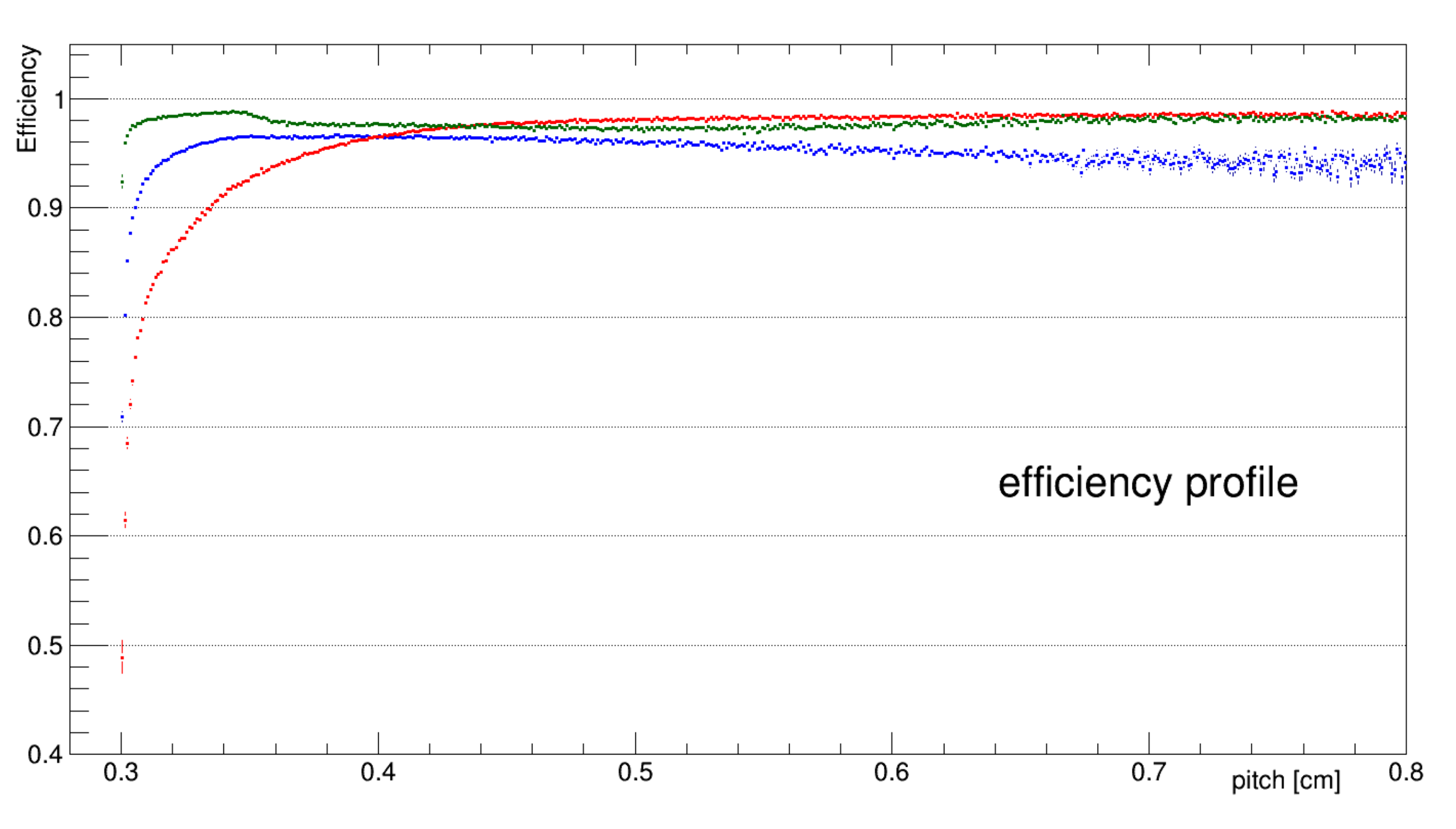}
    \caption{Hit efficiency as a function of wire "pitch": blue, red and green points correspond to Induction 1, Induction 2 and Collection wires respectively. Measurement made by means of a sample of cosmic muon tracks crossing the cathode. }
    \label{Fig:WIRE_eff_profile}
\end{figure}

Globally, the efficiency for identifying a wire signal and associating it with the corresponding track that generated is exceeding 90\% for all three wire planes when the 3D track segment length contributing to each hit (\emph{pitch}) is larger than 3.4 mm (Fig. \ref{Fig:WIRE_eff_profile}).

\subsection{PMT signal reconstruction}
\label{subsec:PMTreco}

The reconstruction of the scintillation light associated with the event of interest is based on the recorded PMTs signals in the event, sampled at 500 MHz. For any event triggered in coincidence with the beam spill, all 360 PMTs digitized signals are recorded in \SI{30}{\micro s} long time intervals. In addition, for cosmic rays crossing the detector in $\pm$1 ms around the beam gate and identified by the trigger logic, all 180 PMTs belonging to the ICARUS module containing the event are recorded in \SI{10}{\micro s} long time intervals.

\begin{figure}[ht]
    \centering
    \includegraphics[width=0.5\textwidth]{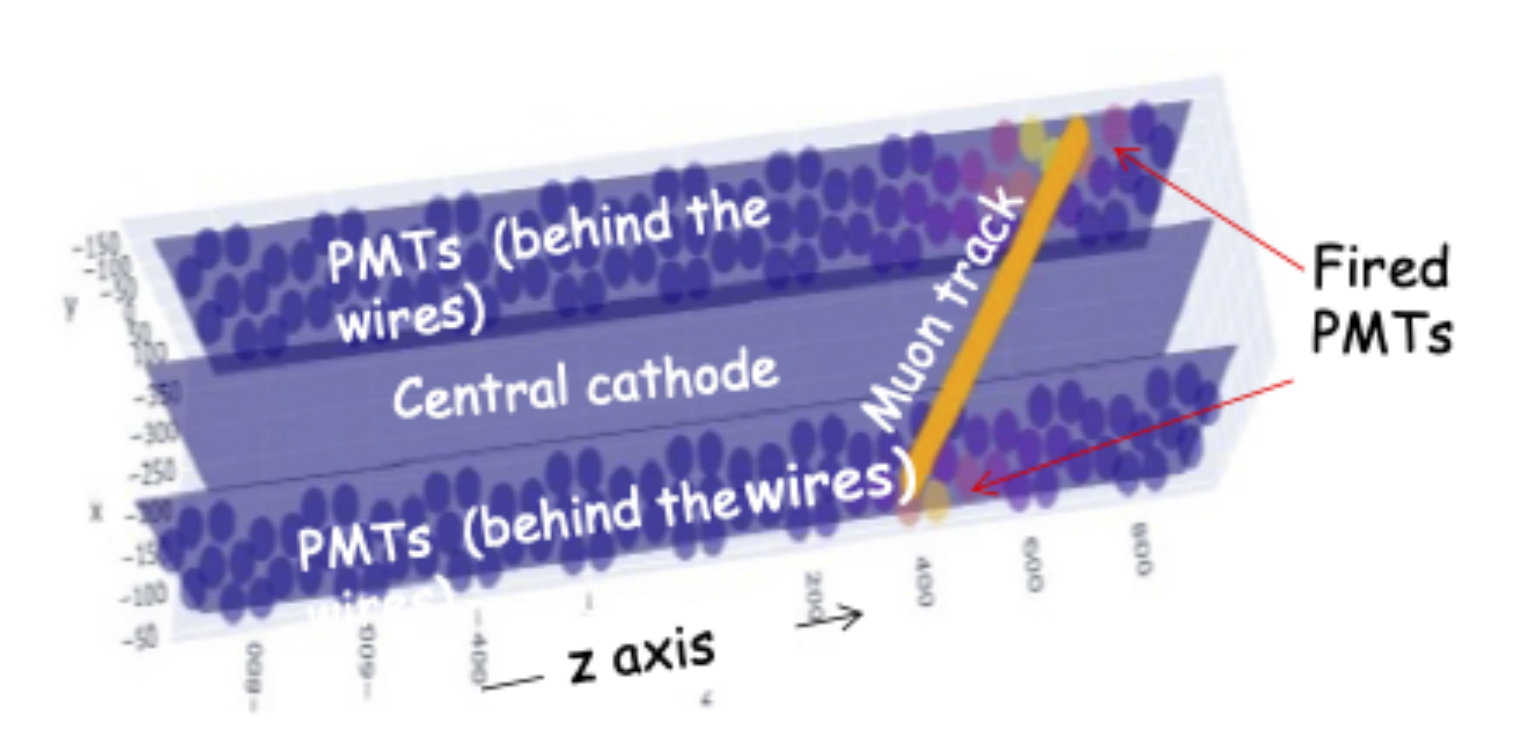}
    \caption{The PMTs associated with a cosmic ray muon crossing the cathode.}
    \label{Fig:SketchFiredPMTs}
\end{figure}

A threshold-based algorithm is applied to each recorded signal, to identify fired PMTs and to reconstruct the characteristics of the detected light to be used in the event analysis. Whenever a PMT signal exceeds the baseline by 0.5~phe, a new {\em OpHit} object is created, characterized by a start time, a time interval for the signal to return back to baseline, a maximal amplitude, and an integral of the signal over the baseline. As a second stage all OpHits in coincidence within 100 ns are clustered together into an {\em OpFlash} object. The Opflash is then expanded to include also OpHits within \SI{1}{\micro s} after the first OpHit time. Nominally, an OpFlash should correspond to the total detected light associated to each interaction, either due to cosmic rays or to a neutrino interaction.
The distribution of the PMT signals in an OpFlash (time, amplitudes, integrals and geometrical positions) is clearly determined by the associated interaction in the TPC (Fig.~\ref{Fig:SketchFiredPMTs}). 

\begin{figure}[ht]
    \centering
    \includegraphics[width=0.5\textwidth]{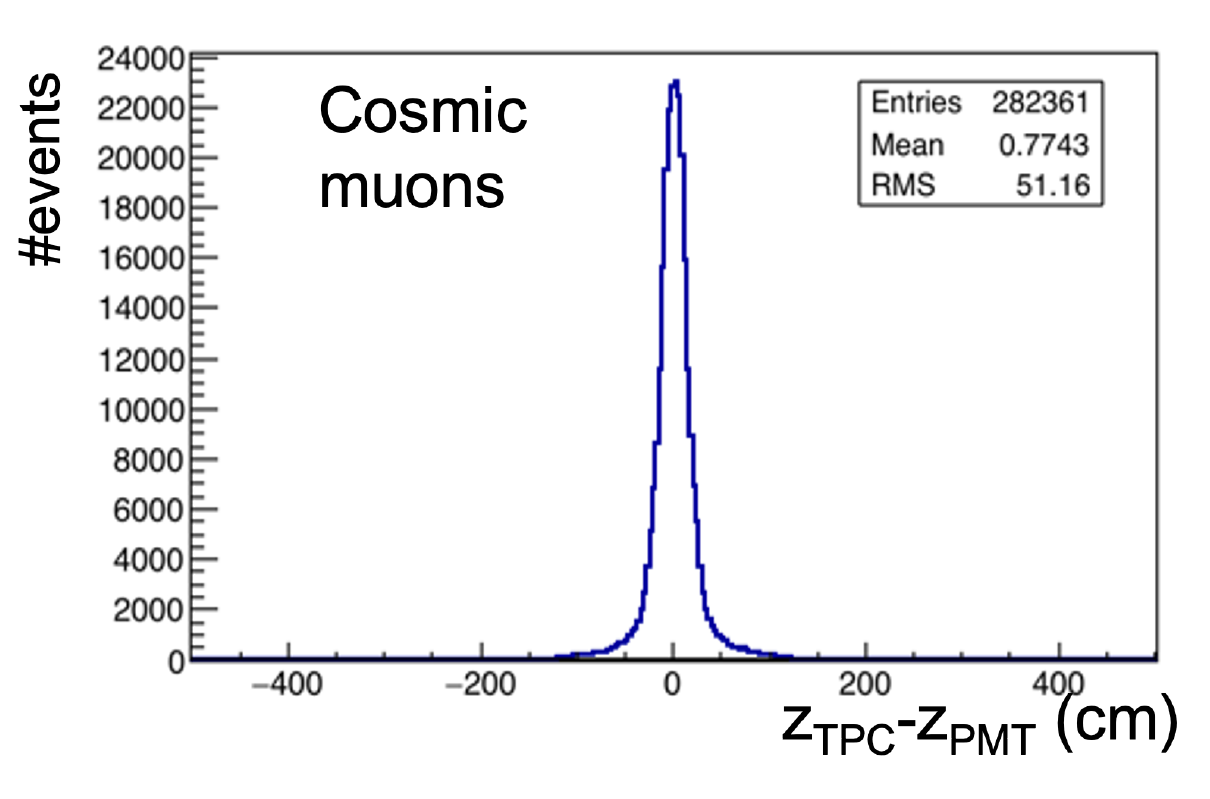}
    \caption{Distribution of $\rm \Delta z=z_{TPC}-z_{PMT}$ for a sample of cosmic ray muons crossing the cathode.}
    \label{fig:TPCPMTmuons}
\end{figure}

\begin{figure}[ht]
    \centering
    \includegraphics[width=0.5\textwidth]{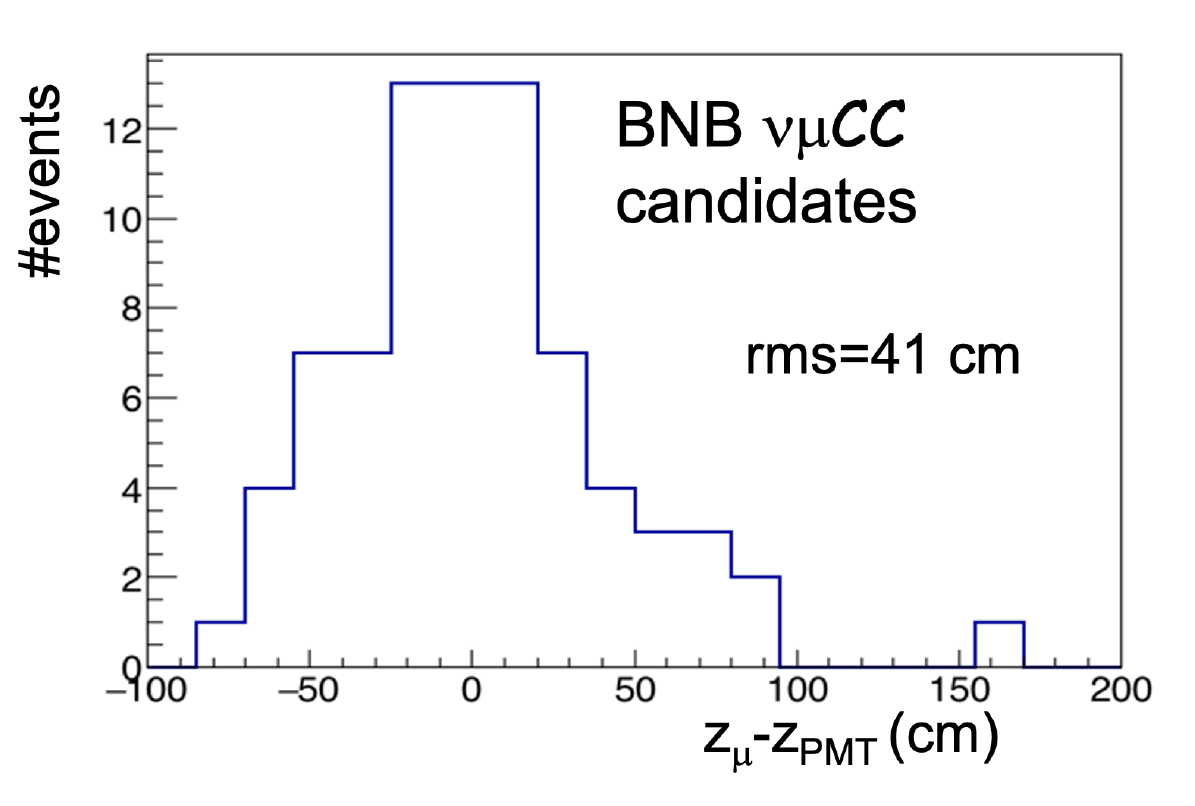}
    \caption{Distribution of $\rm \Delta z=z_{TPC}-z_{PMT}$ for a sample BNB $\nu$ interactions identified by visual scanning.} 
    \label{fig:BNBNUPMT}
\end{figure}

Initially, a very simple association between the event in the TPC and the corresponding detected light that is based on the comparison of the track and the light barycentre along the longitudinal z axis (z$_{\rm TPC}$, z$_{\rm PMT}$) has been adopted.
A correlation within few tens of centimeters was observed for the TPC and light barycentre ($\rm \Delta z=z_{TPC}-z_{PMT}$) for both cosmic muons crossing the cathode (Fig.~\ref{fig:TPCPMTmuons}) and for a sample of BNB neutrino interactions (Fig.~\ref{fig:BNBNUPMT}) selected by visual scanning.

By requiring $ \lvert \Delta$z$\rvert < $ 100 cm it is possible to restrict the analysis of the event to a detector slide that is approximately 5\% of the total active LAr, with a corresponding reduction of randomly overlapping cosmic rays.


\subsection{CRT reconstruction}
\label{subsec:CRTreco}

The CRT hit reconstruction algorithm was validated
during the commissioning phase~\cite{Firstcrtdata}. The first step in the reconstruction chain is to construct CRT hits defined as points in space and time corresponding to a muon track crossing the CRT volume. CRT data coming from Front End Board (FEB) read-outs in a given event are ordered in time and grouped by CRT region. Due to the differences in design of the side and top CRT systems, the Side and Top CRT Hits have to be handled differently.

The coincidence logic in the Side CRTs is performed offline in the reconstruction stage due to the inner and outer CRT modules being connected to FEBs in adjacent layers, whereas each top CRT module is a self-contained coincidence unit. In order to identify a coincident grouping of CRT data objects, a software-based coincidence gate is performed (the hardware-based coincidence gate width is 150 ns and this value is the minimum for the software gate). The reason for not making the coincidence window too large is to avoid introducing fake coincidences from low energy events. Studies are underway to establish a gate width that optimizes the tagging efficiency while avoiding introducing fake coincidences with low energy events if the gate is too wide. 

After the creation of coincident groupings of CRT data, the spatial information is extracted to reconstruct the position of the crossing track. The channel with the largest amplitude is the channel that generated the FEB trigger signal.
The channel position is identified and extracted from the geometry based on the global coordinates of the ICARUS building. The hit position is taken as the mean strip position where a track crosses multiple strips in each layer. 

When the charge amplitude exceeds the discriminator threshold, a CRT hit is acquired by the front-end electronics recording the values of two different time counters. The first counter, T0, is reset every second by means of the PPS signal (see Sec.~\ref{subsec:CRT_inst}) and it provides the global timing of the recorded hit. The second counter, T1, is reset by the event trigger signal and is used to determine the hit relative timing with respect to the event trigger. Each CRT hit timestamp is corrected to account for cable delays and light propagation in the scintillator and in the WLS fiber.

\par The Top CRT hit is defined by the FEB internal triggering logic (see Sec. \ref{sec:CRT}) where a signal threshold of 1.5 phe is applied to each channel. The position within a module is determined by selecting the four channels with the largest amplitude and projected in the global detector coordinates.

\begin{figure}[ht]
    \centering
    \includegraphics[width=.50\textwidth]{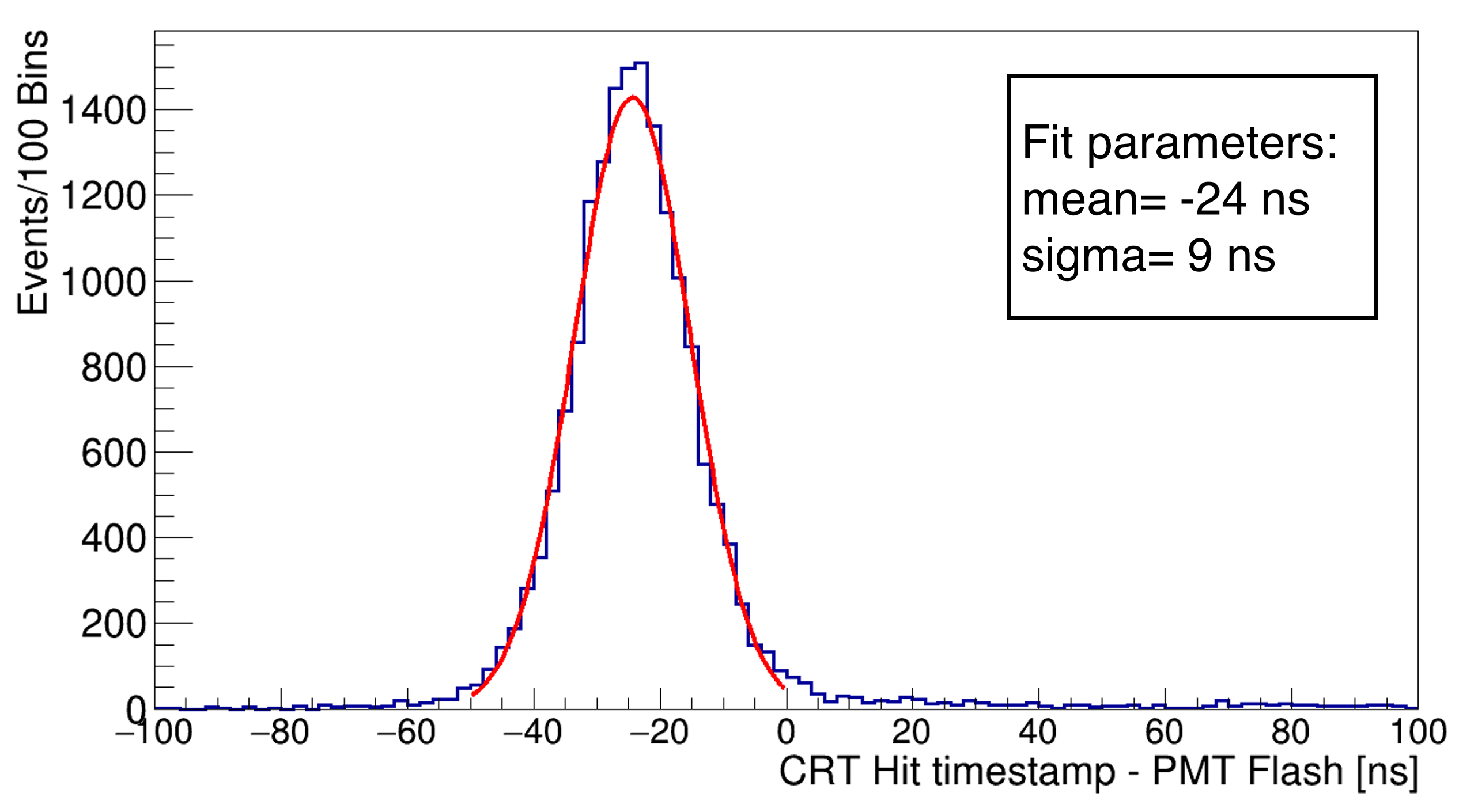}
    \caption{Time difference between matched CRT hits and PMT flashes. The plot refers to Top CRT data in time with the BNB spill.}
    \label{TopTof}
\end{figure}

The CRT timing system has been cross-calibrated with the PMT signals, using the common trigger pulse recorded by the CRT and PMT systems. A preliminary evaluation of the Time-Of-Flight (TOF) of cosmic muons has been performed by selecting particles entering the detector from the Top CRT and generating a flash in the active argon volume. The preliminary distribution of the time differences between Top CRT hits and PMT signals is shown in Fig.~\ref{TopTof}: the measured average TOF of 24$\pm$9 ns is in agreement with the expected $\sim$ 26 ns evaluated from the distance between the Top CRT plane and the first PMT row.

\begin{figure}[ht]
    \centering
    \includegraphics[width=0.45\textwidth]{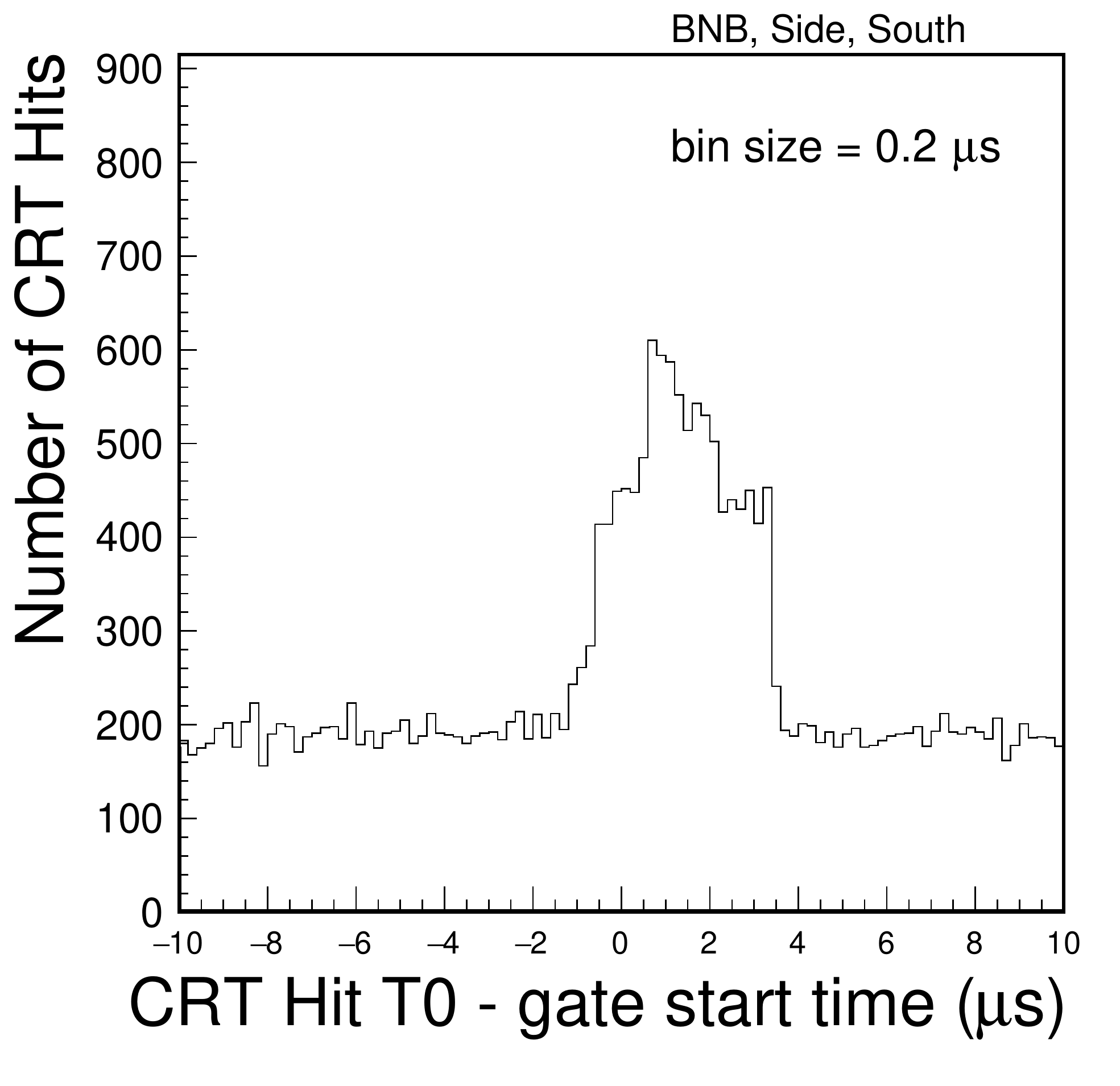}
    \caption{CRT hit time relative to the neutrino gate start time in the south wall (side CRT) for the BNB beam.}
    \label{southwallcrttiming}
\end{figure}

Figure~\ref{southwallcrttiming} shows the CRT hit time relative to the neutrino gate start time in the south side CRT wall for the BNB neutrino beam. Using 11 days of commissioning data, a clear peak can be observed, showing activity in the \SI{4}{\micro s} trigger coincidence window. Additional activity due to the beam appears inside the smaller BNB gate (\SI{1.6}{\micro s} within the \SI{4}{\micro s} window), the rest of the activity outside the \SI{1.6}{\micro s} window is due to cosmic ray triggering.

\subsection{Event display study}
\label{subsec:VisualScanning}

As a first check of the general behavior of the detector, a visual study campaign was performed to select and identify neutrino interactions in the active liquid argon using a graphical event display.

As a first step, all the events recorded in the BNB and NuMI beam for some runs were studied selecting the tracks in the cryostat where the trigger signal has been produced. An interaction was classified as a neutrino candidate if a clear vertex with more than one track was visually identified: electron neutrino CC candidate events require the presence of a clear electromagnetic shower connected to the primary vertex, while the muon neutrino CC events are selected by requiring the presence of a long track (at least 0.5 m) from the primary vertex. In addition, only events with the primary vertex at least 5 cm from top/bottom TPC sides, 50 cm from the upstream/downstream TPC wall, and 5 cm from the anode position have been initially selected.

\begin{figure*}[htb]
    \centering
    \includegraphics[width=\textwidth]{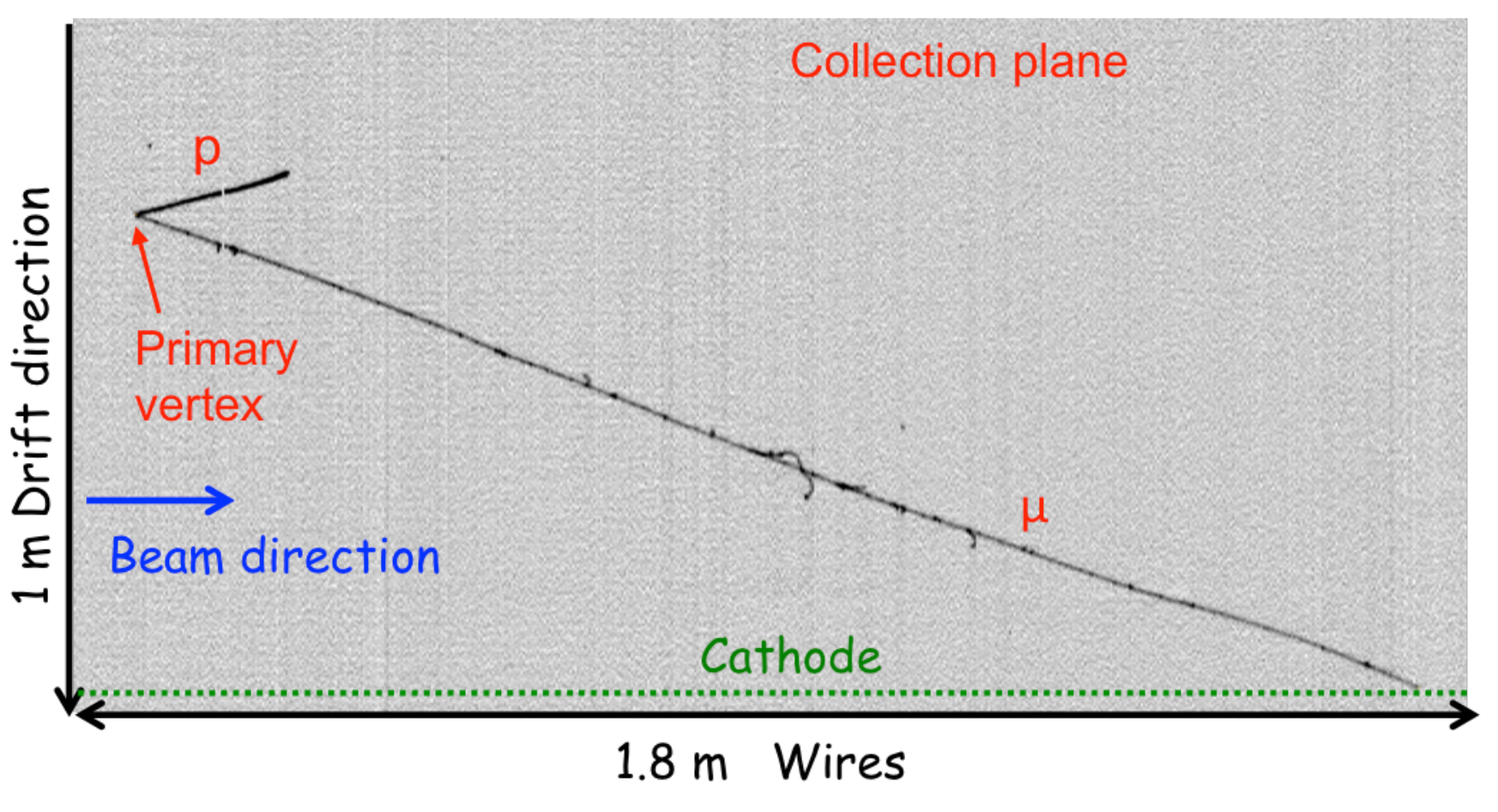}
    \caption{A visually selected $\nu_{\mu}$CC candidate from the BNB beam.}
    \label{Fig:BNB_NuMUCCQE}
\end{figure*}

\begin{figure}[htb]
    \centering
    \includegraphics[width=0.5\textwidth]{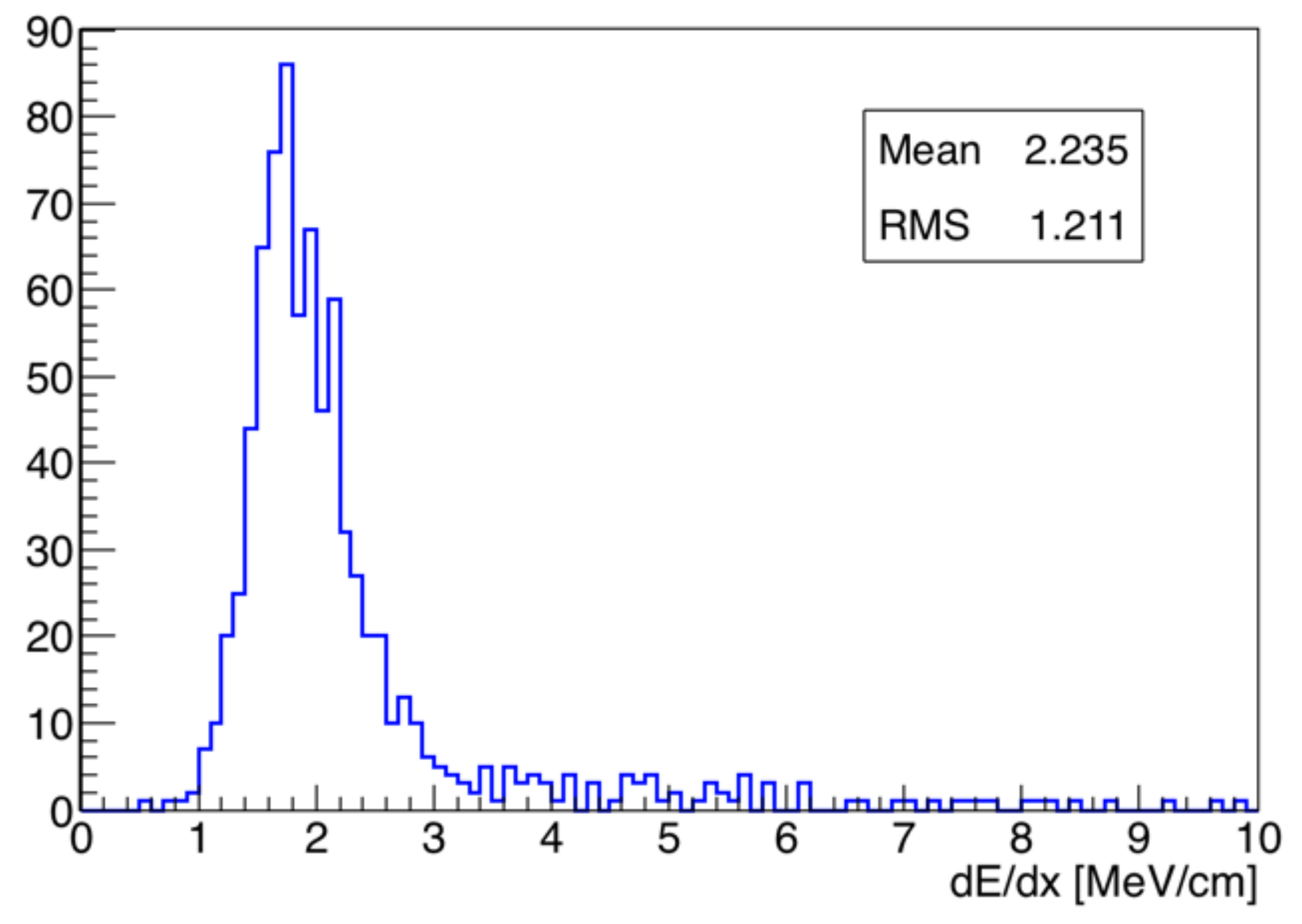}
    \caption{Distribution of the measured dE/dx of the muon candidate in the event shown in Fig.~\ref{Fig:BNB_NuMUCCQE}. dE/dx is reconstructed on each wire applying a preliminary calibration constant.}
    \label{Fig:BNB_NuMUCCQE_dEdx}
\end{figure}

An example of a $\nu_{\mu}$CC candidate is shown in Fig.~\ref{Fig:BNB_NuMUCCQE}, with an estimated total deposited energy of $\sim$ 1.1 GeV. The CC muon candidate is 3.8 m long, while the highly ionizing track from the primary vertex is identified as a 20 cm long proton. 
The full wire signal calibration is in the finalization stage, but by a very preliminary wire signal conversion to estimate the deposited energy, it is possible to reconstruct the dE/dx associated to the individual hits of the muon candidate in the same event, distributed as expected for a MIP particle  particle, as shown in Fig.~\ref{Fig:BNB_NuMUCCQE_dEdx}. 


\begin{figure}[htb]
    \centering
    \includegraphics[width=0.5\textwidth]{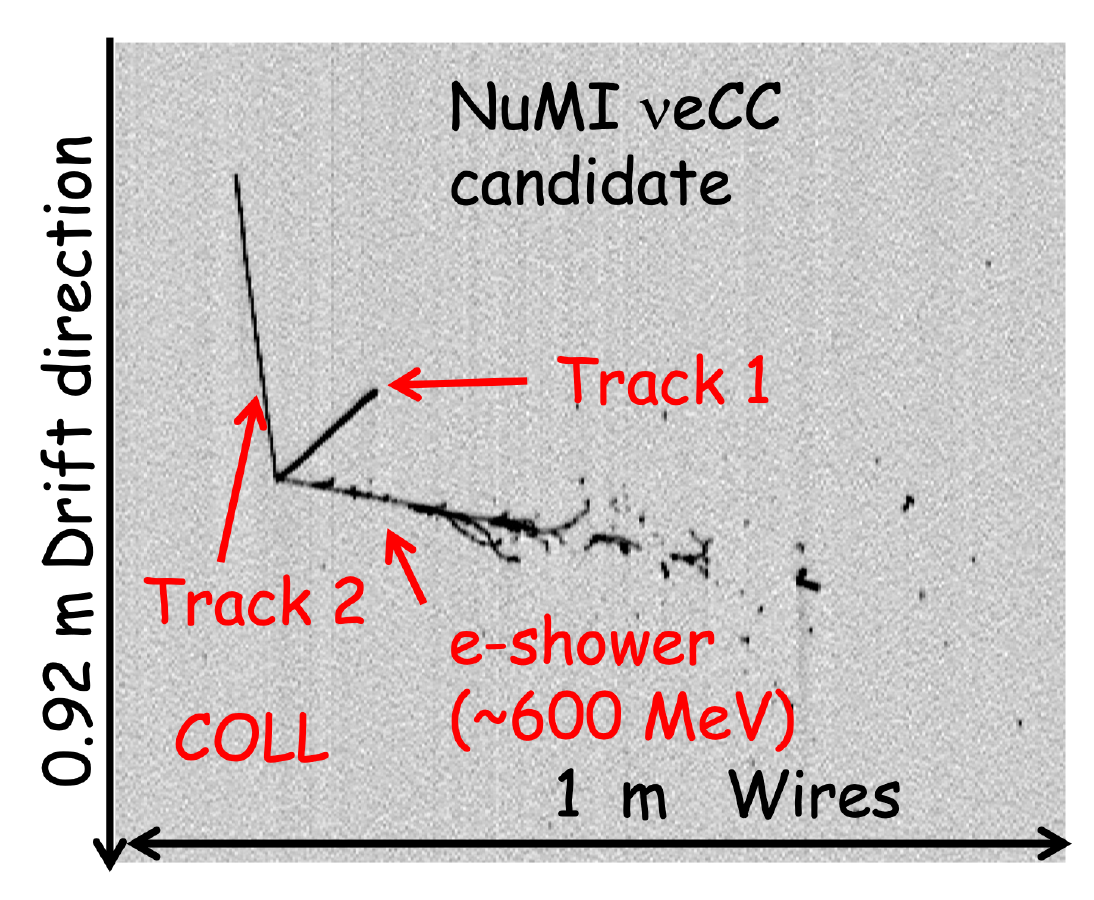}
    \caption{A visually selected $\nu_{e}$CC candidate from the NuMI beam .}
    \label{Fig:NuMI_NueCC}
\end{figure}

Visual scanning also permitted identification of $\nu_{e}$CC candidates in the NuMI beam: a remarkable example is shown in Fig.~\ref{Fig:NuMI_NueCC} for an event of $\sim$ 600 MeV deposited energy.

\subsection{Event reconstruction}
\label{sec:EventReco}

For a given cryostat, hits identified and passing a multi-plane matching algorithm are passed as input to Pandora \cite{PANDORA}: a pattern reconstruction code that performs a 3D reconstruction of the full image recorded in the collected event, including the identification of interaction vertices and of tracks and showers inside the TPC. These are organized into a hierarchical structure (called a \emph{slice}) of particles generated starting from a primary interaction vertex or particle.

The analysis uses information reconstructed in Pandora to tag and reject “clear cosmic” slices by identifying straight tracks crossing the full active liquid argon volume or that are clearly out of time with respect to the beam gate. In Monte Carlo studies, selection criteria require that the reconstructed vertex is in the fiducial volume and that PMT timing signals and the reconstructed angle of the muon track are inconsistent with that of a cosmic ray. These requirements reject 99.7\% of cosmic rays, while accepting more than 82\% of true $\nu_{\mu}$CC events in the fiducial volume. Requiring that a particle identified as a proton be reconstructed in the event further reduces background from cosmic rays. After all criteria are applied, 0.8\% of a selected $\nu_{\mu}$CC contained sample is made up of background from cosmic rays, with 0.6\% coming from intime cosmic rays and 0.2\% coming from out-of-time cosmic rays. Further tagging and rejection of cosmic rays out of time with respect to the beam spill is possible with the CRT detector, which can provide a few nanosecond absolute time measurement for the TPC tracks when they are unambiguously matched to signals on the CRT. This TPC track-CRT hit matching algorithm is still being tuned and validated with cosmic ray data collected off-beam, but is expected to facilitate improved efficiency and allow further optimization of the cosmic rejection criteria.  

Pandora and a set of algorithms to identify, measure and reconstruct tracks and showers can be exploited for the event reconstruction and analysis. These reconstruction tools represent a legacy from past efforts and made available within the LArSoft framework \cite{larsoft}, complemented by new efforts carried out within the joint SBN effort for a common near and far detector analysis. This set of algorithms is applied to tracks and showers from any slice in the event to perform particle identification and estimate the momentum from range, calorimetry and multiple Coulomb Scattering. 

\begin{figure}[htb]
    \centering
    \includegraphics[width=0.5\textwidth]{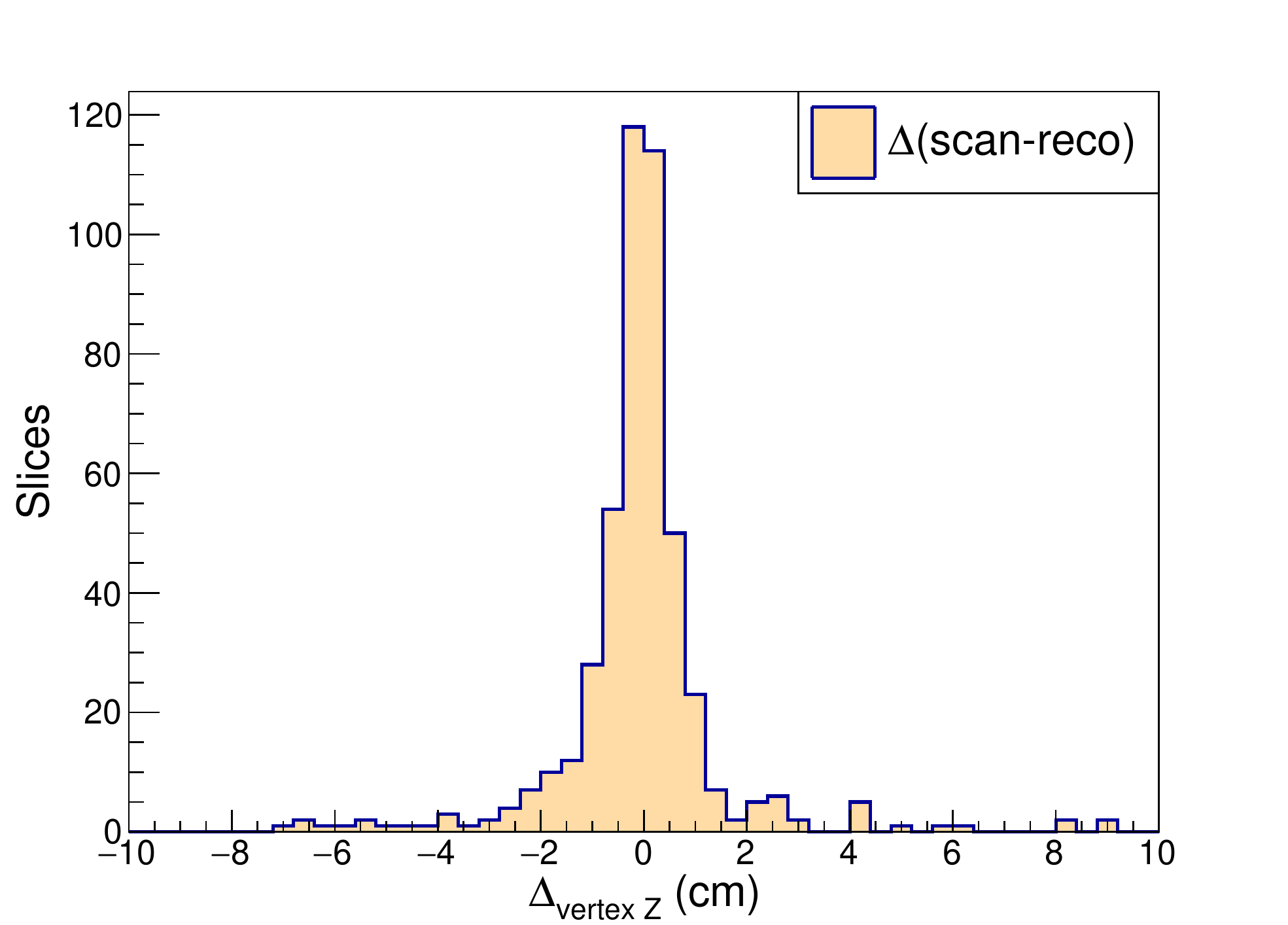}
    \caption{Difference $\Delta Z$ between the automatic and manual measured longitudinal (beam) coordinate of the neutrino interaction vertex for a sample of 476 $\nu_{\mu}$CC candidates from the BNB beam.}
    \label{Fig:Dz_BNB_NuVertex}
\end{figure}

A dedicated visual study of events was performed to select $\sim 600$ $\nu_{\mu}$CC interactions from BNB in the active liquid argon. These events have been used for validation of the Pandora reconstruction.
In order to reduce the manual effort, events to be visually studied are 
first selected by requiring, offline, the absence of signals in the CRT in coincidence with the trigger. 
In addition, full 3D reconstruction was performed for the events and only reconstructed tracks longer than 30 cm, fully contained in the detector, and whose barycenter was in agreement within 1 m with the barycenter of the light signal generating the trigger, have been visually studied. For this sample, the neutrino interaction vertex was identified and measured in 3D coordinates as well as the final point associated with the muon candidate track. Out of the full selected sample, 476 neutrino events present in the analysis files showed a reasonable match with a reconstructed object based on vertex location and were adopted as a benchmark for the validation of the reconstruction tools.
As an example, in $\sim 90$\% of these events the reconstruction reasonably identifies the neutrino interaction vertex along the beam direction, meaning the difference between the two estimates is within 3 cm, as shown in Fig.~\ref{Fig:Dz_BNB_NuVertex}. 

Comparison of the visual study to automated reconstruction, along with studies of Monte Carlo simulation, will enable further understanding of where to focus efforts and improvements in the automatic reconstruction. 
For example, in some cases inefficiencies in a wire plane for a given event reconstruction leading to loss of hits may impact some 3D steps and lead to a track broken into one or more smaller pieces; or algorithms may lead to improper clustering or determination of particle types, etc. 
Further tuning of the reconstruction is progressing, as well as the complete calibration of the detector. However the first results are quite promising, demonstrating that the basic tools for the event reconstruction and the event selection are operational and allow an initial identification and measurement of neutrino interactions. 

\section*{Conclusions}

After the successful three-year physics run at the underground LNGS laboratories
studying neutrino oscillations with the CERN Neutrino to Gran Sasso beam, the
ICARUS T600 LAr-TPC detector underwent a significant overhaul at CERN and was then
installed at Fermilab. Detector activation began in 2020 with the cryogenic
commissioning and, despite serious challenges and delays caused by prolonged
restrictions related to the COVID-19 pandemic, it started operations in 2021 and
successfully completed its commissioning phase in 2022. It collected neutrino events
from both the Booster Neutrino Beam (BNB) and the Main Injector (NuMI) beam off-axis.
Data taking started in June 2021 with the beam data acquisition, with the detector commissioning activities being conducted in parallel.
 An event sample corresponding to $\sim$~$3\cdot10^{20}$ and $5\cdot10^{20}$ POT of the Booster and NuMI beam respectively has been collected with an efficiency
exceeding 91\% during the normal operations.
This data set was used to study the single detector subsystems calibration and to test the ICARUS event selection and
reconstruction procedure and analysis algorithms.
ICARUS has already started the first year of regular data taking devoted to
a sensitive study of the claim by Neutrino-4 short baseline reactor
experiment both in the $\nu_\mu$ channel with the BNB and in the $\nu_e$ channel with NuMI. ICARUS will also address other fundamental studies such as neutrino cross sections with the NuMI beam and a number of Beyond Standard Model searches. The search for evidence of a sterile neutrino jointly with the Short-Baseline Near Detector, within the Short-Baseline Neutrino program, will follow.

\section*{Acknowledgements}

This document was prepared by the ICARUS
Collaboration using the resources of the Fermi National Accelerator Laboratory (Fermilab), a U.S. Department of Energy, Office of Science, HEP User Facility. Fermilab is managed by Fermi Research Alliance, LLC (FRA), acting under Contract No. DE-AC02-07CH11359.
This work was supported by the US Department of Energy, INFN, EU Horizon 2020 Research and Innovation Program under the Marie Sklodowska-Curie Grant Agreement No. 734303, 822185, 858199, and 101003460 and Horizon Europe Program research and innovation programme under the Marie Sklodowska-Curie Grant Agreement No. 101081478. Part of the work resulted from the implementation of the research Project No. 2019/33/N/ST2/02874 funded by the National Science Centre, Poland.
The ICARUS Collaboration would like to thank the MINOS Collaboration for having provided the side CRT panels as well as Double Chooz (University of Chicago) for the bottom CRT panels. We also acknowledge the contribution of many SBND colleagues,
in particular for the development of a number of simulation, reconstruction and analysis tools which are shared within the SBN program. Finally, our experiment could not have been carried out without the major support of CERN in the detector overhauling within the Neutrino Platform framework and of Fermilab in the detector installation and commissioning, and in providing the BNB and NuMI beams. 

\bibliographystyle{unsrt}
\bibliography{sn-bibliography}

\begin{thebibliography}{10}

\bibitem{crubbia}
C.~Rubbia.
\newblock {The Liquid Argon Time Projection Chamber: A New Concept for Neutrino
  Detectors}.
\newblock {\em CERN-EP}, 77-08, 1977.

\bibitem{lsnd}
A.A. Aguilar-Arevalo et~al. (LSND~Collaboration).
\newblock {Evidence for Neutrino Oscillations from the Observation of Electron
  Anti-neutrinos in a Muon Anti-Neutrino Beam}.
\newblock {\em Phys. Rev.}, D64:112007, 2001.

\bibitem{miniboone}
A.A. Aguilar-Arevalo et~al. (MiniBooNE~Collaboration).
\newblock {Updated MiniBooNE neutrino oscillation results with increased data
  and new background studies}.
\newblock {\em Phys. Rev.}, D103:052002, 2021.

\bibitem{lngs_nue}
M.~Antonello et~al. (ICARUS~Collaboration).
\newblock {Search for anomalies in the $\nu_e$ appearance from a $\nu_\mu$
  beam}.
\newblock {\em Eur. Phys. J.}, C73:2599, 2013.

\bibitem{sbn_proposal}
R.~Acciarri et~al. (SBND MicroBooNE ICARUS~Collaborations).
\newblock {A Proposal for a Three Detector Short-Baseline Neutrino Oscillation
  Program in the Fermilab Booster Neutrino Beam.}
\newblock arXiv:1503.01520, 2015.

\bibitem{sbn_2019}
O.~Palamara P.A.N.~Machado and D.W. Schmitz.
\newblock {The Short-Baseline Neutrino Program at Fermilab}.
\newblock {\em Annual Review of Nuclear and Particle Science}, 69:367--387,
  2019.

\bibitem{neutrino4}
A.P.~Serebrov et~al. (Neutrino-4~Collaboration).
\newblock {First Observation of the Oscillation Effect in the Neutrino-4
  Experiment on the Search for the Sterile Neutrino.}
\newblock {\em JETP lett.}, 109:213--221, 2019.

\bibitem{NP01}
G.L.~Raselli (on behalf of~the ICARUS~Collaboration).
\newblock {The upgrading of the ICARUS T600 detector}.
\newblock {\em POS (EPS-HEP2017)}, page 515, 2017.

\bibitem{t600_pavia}
S.~Amerio et~al. (ICARUS~Collaboration).
\newblock {Design, construction and tests of the ICARUS T600 detector}.
\newblock {\em Nucl. Instr. Meth.}, A526:329--410, 2004.

\bibitem{ica_lngs}
C.~Rubbia et~al. (ICARUS~Collaboration).
\newblock {Underground operation of the ICARUS T600 LAr-TPC: first results}.
\newblock {\em JINST}, 6 P07011, 2011.

\bibitem{lngs_purity}
M.~Antonello et~al. (ICARUS~Collaboration).
\newblock {Experimental observation of an extremely high electron lifetime with
  the ICARUS-T600 LAr-TPC}.
\newblock {\em JINST}, 9 P12006, 2014.

\bibitem{lngs_reco}
M.~Antonello et~al. (ICARUS~Collaboration).
\newblock {Precise 3D track reconstruction algorithm for the ICARUS T600 liquid
  argon time projection chamber detector}.
\newblock {\em Advances in High Energy Physics}, 2013:260820, 2013.

\bibitem{lngs_mcs}
M.~Antonello et~al. (ICARUS~Collaboration).
\newblock {Muon momentum measurement in ICARUS-T600 LAr-TPC via multiple
  scattering in few-GeV range}.
\newblock {\em JINST}, 12 P04010, 2017.

\bibitem{lngs_atm}
C.~Farnese (on behalf of~the ICARUS~Collaboration).
\newblock {Atmospheric Neutrino Search in the ICARUS T600 Detector}.
\newblock {\em Universe}, 5(1), 2019.

\bibitem{ica_electronics}
L.~Bagby et~al. (ICARUS~Collaboration).
\newblock {Overhaul and installation of the ICARUS-T600 liquid argon TPC
  electronics for the FNAL Short Baseline Neutrino program}.
\newblock {\em JINST}, 16 P01037, 2021.

\bibitem{Babicz:2018svg}
M.~Babicz et~al. (ICARUS~Collaboration).
\newblock {Test and characterization of 400 Hamamatsu R5912-MOD photomultiplier
  tubes for the ICARUS T600 detector}.
\newblock {\em JINST}, 13 P10030, 2018.

\bibitem{Ali_Mohammadzadeh_2020}
B.~Ali-Mohammadzadeh et~al. (ICARUS~Collaboration).
\newblock {Design and implementation of the new scintillation light detection
  system of ICARUS T600}.
\newblock {\em JINST}, 15 T10007, 2020.

\bibitem{Bonesini:2018ubd}
M.~Bonesini et~al.
\newblock {An innovative technique for TPB deposition on convex window
  photomultiplier tubes}.
\newblock {\em JINST}, 13 P12020, 2018.

\bibitem{laser}
M.~Bonesini et~al. (on behalf of~the ICARUS~Collaboration).
\newblock {The laser diode calibration system of the Icarus T600 detector at
  FNAL}.
\newblock {\em JINST}, 15 C05042, 2020.

\bibitem{ICARUSOverburden}
B.~Behera (on behalf of~the ICARUS~Collaboration).
\newblock {Cosmogenic background suppression at the ICARUS using a concrete
  overburden}.
\newblock {\em J. Phys. Conf. Ser.}, 2156(1):012181, 2021.

\bibitem{Bagby_2021}
L.~Bagby et~al. (ICARUS~Collaboration).
\newblock {Overhaul and installation of the ICARUS-T600 liquid argon TPC
  electronics for the FNAL Short Baseline Neutrino program}.
\newblock {\em JINST}, 16 P01037, 2021.

\bibitem{MicroBooNE-Noise}
R.~Acciarri et~al. (MicroBooNE~Collaboration).
\newblock {Noise Characterization and Filtering in the MicroBooNE Liquid Argon
  TPC}.
\newblock {\em JINST}, 12 P08003, 2017.

\bibitem{Walkowiak_2020}
W.~Walkowiak.
\newblock {Drift velocity of free electrons in liquid argon}.
\newblock {\em Nucl. Instrum. Methods Phys. Res. A}, 449:288--294, 2000.

\bibitem{MicroBooNE-SCE-Cosmics}
P.~Abratenko et~al. (MicroBooNE~Collaboration).
\newblock {Measurement of space charge effects in the MicroBooNE LArTPC using
  cosmic muons}.
\newblock {\em JINST}, 15 P12037, 2020.

\bibitem{Mooney-SCE}
M.~Mooney.
\newblock {The MicroBooNE Experiment and the Impact of Space Charge Effects}.
\newblock arXiv:1511.01563, 2015.

\bibitem{Antonello_2020}
M.~Antonello et~al. (ICARUS~Collaboration).
\newblock {Study of space charge in the ICARUS T600 detector}.
\newblock {\em JINST}, 15:P07001, 2020.

\bibitem{MicroBooNE-Calib}
C.~Adams et~al. (MicroBooNE~Collaboration).
\newblock {Calibration of the charge and energy loss per unit length of the
  {MicroBooNE} liquid argon time projection chamber using muons and protons}.
\newblock {\em JINST}, 15 P03022, 2020.

\bibitem{PDG_2020}
P.A.~Zyla et~al. (Particle Data~Group).
\newblock {The Review of Particle Physics}.
\newblock {\em Prog. Theor. Exp. Phys.}, 2020 083C01, 2020.

\bibitem{ArgoNeuT-Recomb}
R.~Acciarri et~al. (ArgoNeuT~Collaboration).
\newblock {A study of electron recombination using highly ionizing particles in
  the ArgoNeuT Liquid Argon TPC}.
\newblock {\em JINST}, 8 P08005, 2013.

\bibitem{trigger2022}
C.~Farnese et~al. (ICARUS~Collaboration).
\newblock {Implementation of the trigger system of the ICARUS-T600 detector at
  Fermilab}.
\newblock {\em Nucl. Instr. Meth.}, A1045:167498, 2023.

\bibitem{whiterabbit}
J.~Serrano et~al.
\newblock {The White Rabbit Project}.
\newblock {\em Proceedings of the 12$^{th}$ International Conference On
  Accelerator And Large Experimental Physics Control Systems, Kobe, Japan},
  pages 93--95, 2009.

\bibitem{artdaq}
K.~Biery et~al.
\newblock {artdaq: An Event-Building, Filtering, and Processing Framework}.
\newblock {\em IEEE Trans. Nucl. Sci.}, 60:3764--3771, 2013.

\bibitem{art}
C.~Green et~al.
\newblock {The Art Framework}.
\newblock {\em J. Phys. Conf. Ser.}, 396:022020, 2012.

\bibitem{uboone_wsp}
C.~Adams (on behalf of~the MicroBooNE~Collaboration).
\newblock {Ionization electron signal processing in single phase LArTPCs. Part
  I. Algorithm Description and quantitative evaluation with MicroBooNE
  simulation}.
\newblock {\em JINST}, 13 P07006, 2018.

\bibitem{Firstcrtdata}
B.~Behera (on behalf of~the ICARUS~Collaboration).
\newblock {First Data from the Commissioned ICARUS Side Cosmic Ray Tagger}.
\newblock {\em PoS}, NuFact2021:201, 2022.

\bibitem{PANDORA}
R.~Acciarri et~al. (MicroBooNE~Collaboration).
\newblock {The Pandora multi-algorithm approach to automated pattern
  recognition of cosmic-ray muon and neutrino events in the MicroBooNE
  detector.}
\newblock arXiv:1708.03135v1, 2017.

\bibitem{larsoft}
R.~Pordes and E.~Snider.
\newblock {The Liquid Argon Software Toolkit (LArSoft): Goals, Status and
  Plan}.
\newblock {\em PoS}, ICHEP2016:182, 2016.

\end{thebibliography}


\end{document}